\documentclass[preprint,10pt]{elsarticle}
\usepackage{amsmath}
\usepackage{float}
\usepackage{subfig}
\usepackage{array,graphicx,dblfloatfix}
\usepackage{mathrsfs}
\usepackage{amsthm}
\usepackage{amssymb}
\usepackage[linesnumbered, ruled]{algorithm2e}
\usepackage{booktabs}
\usepackage{hyperref}

\newtheorem{thm}{Theorem}
\theoremstyle{definition} \newtheorem{definition}{Definition}
\theoremstyle{plain} 
\theoremstyle{plain} \newtheorem{proposition}{Proposition}
\theoremstyle{plain} \newtheorem{lemma}{Lemma}

\begin{document}

\title{RECOME: a New Density-Based Clustering Algorithm Using Relative
KNN Kernel Density}

\author[label1]{Yangli-ao Geng}
\author[label1]{Qingyong Li}
\author[label2]{Rong Zheng}
\author[label3,label4]{Fuzhen Zhuang}
\author[label5]{Ruisi He}
\author[label6]{Naixue Xiong}

\address[label1]{School of Computer and Information Technology,
Beijing Jiaotong University, Beijing, China. \{gengyla,liqy\}@bjtu.edu.cn}
\address[label2]{Department of Computing and
Software, McMaster University, Hamilton, Canada. rzheng@mcmaster.ca}
\address[label3]{Key Lab of Intelligen Information Processing of Chinese Academy of Sciences (CAS), ICT, CAS, Beijing, 100190, China.}
\address[label4]{University of Chinese Academy of Sciences, Beijing 100049, China. zhuangfz@ics.ict.ac.cn}
\address[label5]{State Key Laboratory of Rail Traffic Control and Safety, Beijing Jiaotong University, Beijing, China. ruisi.he@bjtu.edu.cn}
\address[label6]{Department of Mathematics and Computer Science
Northeastern State University, Tahlequah, USA. xiongnaixue@gmail.com}



%

\begin{abstract}
Discovering clusters from a dataset with different shapes, densities, and scales is a known challenging problem in data clustering. In this paper, we propose the RElative COre MErge (RECOME) clustering algorithm. The core of RECOME is a novel density measure, i.e., Relative $K$ nearest Neighbor Kernel Density (RNKD). RECOME identifies core objects with unit RNKD, and {partitions} non-core objects into atom clusters by successively following higher-density neighbor relations toward core objects. Core objects and their corresponding atom clusters are then merged through $\alpha$-reachable paths on a KNN graph. We discover that the number of clusters computed by RECOME is a step function of the $\alpha$ parameter with jump discontinuity on a small collection of values. A fast jump discontinuity discovery (FJDD) method is proposed based on graph theory. RECOME is evaluated on both synthetic datasets and real datasets. Experimental results indicate that RECOME is able to discover clusters with different shapes, densities, and scales. It outperforms six baseline methods on both synthetic datasets and real datasets. Moreover, FJDD is shown to be effective to extract the jump discontinuity set of parameter $\alpha$ for all tested datasets, which can ease the task of data exploration and parameter tuning.
\end{abstract}

\begin{keyword}
density-based clustering; density estimation; K nearest neighbors; graph theory.
\end{keyword}


\maketitle

\section{Introduction}
\label{sec:intr}
Clustering, also known as unsupervised learning, is a process of discovery and exploration for investigating inherent and hidden structures within a large dataset \cite{han2011data}. It has been extensively applied to a variety of tasks \cite{KimH09,QIU2017102,He8013075Kernel,Qixiang2003Multimedia,Laohakiat2016A, Kisore2016Improving,zhang2016spectral,newman2006modularity,Yin2014A,Wang2012Graph,Kubal2010Clustering}.
Many clustering algorithms have been proposed in different scientific disciplines \cite{jain2010data}, and these methods often differ in the selection of  objective functions, probabilistic models or heuristics adopted. Nonetheless, two difficulties, how to choose appropriate clustering number and how to discover clusters of an arbitrary shape, are faced by most methods. Density-based clustering approaches are characterized by aggregating mechanisms based on density \cite{miller2001spatial}. They can handle data with irregular shapes and determine clustering number automatically. Ester {\it et al.} \cite{ester1996density} and Sander \textit{et al.} \cite{Sander1998Density} pioneered two density-based methods, Density Based Spatial Clustering of Applications with Noise (DBSCAN) and Generalizing DBSCAN, to detect clusters in a spatial database according to density differences. Although both methods can detect clusters with different shapes, they face the challenge of choosing appropriate parameter values. Subsequently, many improved methods have been proposed \cite{ankerst1999optics, liu2007vdbscan, Cassisi2013Enhancing, loh2015fast, nanda2015design}. Recently, a novel density based clustering method, named Fast search-and-find of Density Peaks (FDP) \cite{rodriguez2014clustering}, was proposed. This algorithm assumes that cluster centers are surrounded by neighbors with lower local density and that they are at a relatively large distance from any point with higher density. FDP can recognize clusters regardless of their shape and of the
dimensionality of the space in which they are embedded, but it lacks an efficient quantitative criterion for judging cluster centers. Accordingly, approaches such as 3DC \cite{Liang2016delta} and STClu \cite{Wang2016Automatic} have been proposed to improve FDP.


Density-based clustering methods have the advantages of discovering clusters with arbitrary shapes and dealing with noisy data, but they face two challenges. First, traditional density measures are not adaptive to clusters with different densities.
Second, performances of traditional methods (e.g., DBSCAN and FDP) are sensitive to parameters, and it is non-trivial to set these parameters properly for different datasets.

Aiming to address these challenges, we propose the RElative COre MErge (RECOME) clustering algorithm, which is based on two density measures: the $K$ nearest Neighbor Kernel Density (NKD) and Relative $K$ nearest Neighbor Kernel Density (RNKD). RECOME firstly identifies \textit{core objects} corresponding to objects with RNKD equal 1. A core object and its descendants, which are defined by a directed relation (i.e., \textit{higher density nearest-neighbor}) based on the NKD, form an \textit{atom cluster}. These atom clusters are then merged using a novel notion of $\alpha$-connectivity on a KNN graph. RECOME has been evaluated using both synthetic datasets and real world datasets. Experimental results demonstrate that RECOME outperforms six baseline methods. Furthermore, we find that the clustering results of RECOME can be characterized by a step function of its parameter $\alpha$, and therefore devise a fast jump discontinuity discovery (FJDD) algorithm to extract the small collection of jump discontinuity values. In summary, this work makes the following contributions.
\begin{enumerate}

\item We give a formal analysis showing that the density measure NKD enjoys some desirable properties. Furthermore, based on the NKD, we propose a new density measure RNKD, which is instrumental in detecting clusters with different densities.

\item RECOME can avoid the ``decision graph fraud'' problem \cite{Liang2016delta} of FDP and can handle clusters with different shapes, densities, and scales. Furthermore, RECOME has nearly linear computational complexity if the $K$ nearest neighbors of each object are computed in advance.

\item FJDD can extract all jump discontinuity values of parameter $\alpha$ for any dataset in $\mathcal{O}(n\log n)$ time, where $n$ is the number of objects. It will greatly benefit parameter selection in real-world applications.
\end{enumerate}

This paper is organized as follows. Section \ref{sec:relaWork} introduces the related work. Section \ref{sec:RelaKNNKern} presents the new density measure RNKD and discusses the robustness of NKD and RNKD. Section \ref{sec:RECOME} describes the proposed clustering method RECOME. Section \ref{sec:FJDD} presents the auxiliary algorithm \textit{FJDD}. Section \ref{sec:Expe} demonstrates experimental results. Finally, we conclude the paper in Section \ref{sec:conc}.

%
%
%
%

\section{Related Work}
\label{sec:relaWork}

Existing clustering methods can be categorized into partitional methods, hierarchical methods, grid-based methods, graph-based methods, density-based methods, etc \cite{han2011data}. Partitional methods such as K-means \cite{macqueen1967some} and K-medoids \cite{Kaufmann1987Clustering}, divide data to a number of partitions and a certain quantitative measure of the ``goodness" of the resulting clusters is maximized iteratively. Hierarchical clustering methods can be agglomerative (bottom-up) or divisive (top-down). An agglomerative clustering (e.g., AGNES \cite{kaufman2008agglomerative}) starts with one object for each cluster and recursively merges two or more of the most appropriate clusters. A divisive clustering (e.g., DIANA \cite{kaufman2008divisive}) starts with the dataset as one cluster and recursively splits the most appropriate cluster. The process continues until a stopping criterion is reached. Grid-based methods such as STING \cite{wang1997sting} and CLIQUE \cite{agrawal1998automatic}, divide the original data space into grids, and then group the grids according to the statistical characters of objects in each grid. Graph-based methods, such as SCAN \cite{Xu2007SCAN} and spectral clustering \cite{Shi2000Normalized}, first construct a similarity graph from a dataset, and then utilize the notion of structural-context similarity or
the eigenvalues of Laplacian matrix to generate clusters. Density-based methods (e.g., DBSCAN \cite{ester1996density} and DENCLUE \cite{Hinneburg1999An}) first estimate the distribution density of objects in a feature space, and then recognize clusters as regions of high density separated by regions of lower density. In this paper, we focus on density-based methods because they are highly relevant to the proposed algorithm.

In \cite{ester1996density}, Ester et al. proposed the first density-based method DBSCAN. In DBSCAN, a {\it cut-off density} of an object $o$ is defined
as the number of objects falling inside a ball of radius $\epsilon$
centered at $o$. If the cut-off density of $o$ is higher than a threshold, $MinPts$,
$o$ is regarded as a key object. When the distance between two key objects is
less than $\epsilon$, they are called density-reachable. Density-reachable
key objects form basic clusters. A non-key object is assigned to a basic
cluster if it is within $\epsilon$ distance to a key object in the respective
cluster; otherwise, the non-key object is treated as noise. DBSCAN is sensitive to the choice of parameters $\epsilon$ and $MinPts$, and can hardly handle clusters with heterogeneous densities. To overcome these drawbacks, Ankerst {\it et al.} \cite{ankerst1999optics} proposed an enhanced density-connected algorithm OPTICS. OPTICS provides a visual tool to help users find the cluster structure and determine the parameters. Although OPTICS reduces the subjectivity in a parameter estimation, when dealing with a complex dataset, it is also difficult to determine how many $\epsilon$'s are needed to find potential clusters \cite{Cassisi2013Enhancing}.

Kernel density \cite{terrell1992variable} is a well-known alternative to cut-off density. It is continuous and less sensitive to parameter selection. DENCLUE \cite{Hinneburg1999An} is a method based on kernel density, in which the local peaks (i.e., local density maxima) of the kernel density function are used to define clusters. Then, each object is assigned to a cluster by a hill-climbing procedure. However, traditional kernel density methods tend to give biased estimation when handling clusters with different scales. To overcome this difficulty, KNN kernel density~\cite{Loftsgaarden1965A} has been introduced in the KNN-kernel density-based Clustering (KNNC)~\cite{tran2006knn}. In our work, the proposed RNKD estimation is inspired by KNN kernel density with further improvement allowing the inclusion of low-density clusters.

Detecting clusters with heterogeneous densities is another challenge for density-based approaches \cite{Kriegel2011Density}. Some algorithms \cite{Levent2003Finding} \cite{pei2009decode} have been proposed to solve this problem. In \cite{Levent2003Finding}, a Shared Nearest Neighbor (SNN) clustering algorithm was proposed. SNN first finds $K$ nearest neighbors of each data object according to similarity, and then refines the similarity between pairs using the number of neighbors that the two objects share. Based on the new measure, a DBSCAN-like process is used to generate clusters. SNN has been shown the capacity in detecting clusters with complex distribution, whereas its excessive parameters and relatively high computational cost weaken its applicability in practice. DiscovEring c-clusters of different dEnsities (DECODE) \cite{pei2009decode} assumes that the target dataset is generated by a series of point processes and tries to find clusters as connected regions of objects whose distances to their $m$-th nearest neighbor are similar. It has shown to be capable of determining the number of density types with little prior knowledge, but the high computational cost limits its application to large data. In \cite{Breunig2000LOF}, a density-based outlier detection approach was proposed. It relies on the local outlier factor (LOF) of each object, which is equal to the average of the ratios between the local density of an object and those of its $K$ nearest neighbors. LOF can effectively distinguish outliers from normal clusters. However, it is not suitable for finding clusters with complex distribution. In this work, we introduce the novel density measure RNKD, which allows handling clusters with heterogeneous densities efficiently.

Rodriguez and Laio proposed a novel density-based clustering method by finding
density peaks called FDP \cite{rodriguez2014clustering}. FDP discovers clusters by a
two-phase process. First, local density is computed for each object according to the number of objects in
its $d_c$ neighborhood, and then a group decision method is applied to determine cluster centers, called density peaks. Second, remaining objects are assigned to the same cluster as its nearest neighbor with higher density. FDP is effective in finding clusters
with different shapes. However, reasonable cluster centers are hard to determine when several density peaks exist in a cluster. In this work, RECOME adopts an agglomerative procedure to merge atom clusters (analogous to those resulting from ``density peaks"), which is feasible even encountering  clusters with multi-peaks.

Key notations used in the paper are listed in Table~\ref{TabNotations}.

\begin{table}[!htbp]
\centering
\caption{Main notations used throughout the paper.}
\label{TabNotations}
\begin{tabular}{l|p{0.83\textwidth}}
\hline
Notation & Description \\
\hline
$|\cdot|$ & Absolute value of a scalar or cardinality of a set. \\
$V$ & Dataset. Lower case symbols $u,v,w,u_i,v_i,w_i$ denote elements of $V$.\\
$d(\cdot,\cdot)$ & The distance function on $V$. In the formal analysis, we assume it is a metric (i.e., satisfying non-negativity, symmetry and triangle inequality).\\
$N(u)$ & The $K$ nearest neighbors set of $u$ in $V$ with respect to $d(\cdot,\cdot)$.\\
$N_i(u)$ & The $i$-th nearest neighbor of $u$ in $V$.\\
$d_i(u)$ & The distance between $u$ and $N_i(u)$, i.e., $d(u, N_i(u))$.\\
\hline
\end{tabular}
\end{table}

\section{Relative KNN Kernel density estimation}
\label{sec:RelaKNNKern}
Though many density measures have been proposed, few considers the local relative density levels (will be shown in Section \ref{ssec:DensEstiIntro}), which are crucial to detect clusters with various densities. In this section, we will first illustrate the disadvantage of classical density measures used in density clustering, and then introduce the proposed measure RNKD that homogenizes the density estimation across clusters with different densities.

\subsection{Density Estimation}
\label{ssec:DensEstiIntro}
\begin{figure}[!htb]
\centering
\subfloat[$\epsilon$ cut-off density]{
  \label{fig:cutoDens}
   \includegraphics[width=0.19\textwidth]{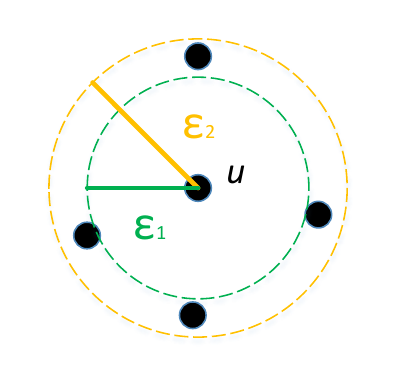}
}
\subfloat[Kernel density]{
  \label{fig:globKernDens}
   \includegraphics[width=0.19\textwidth]{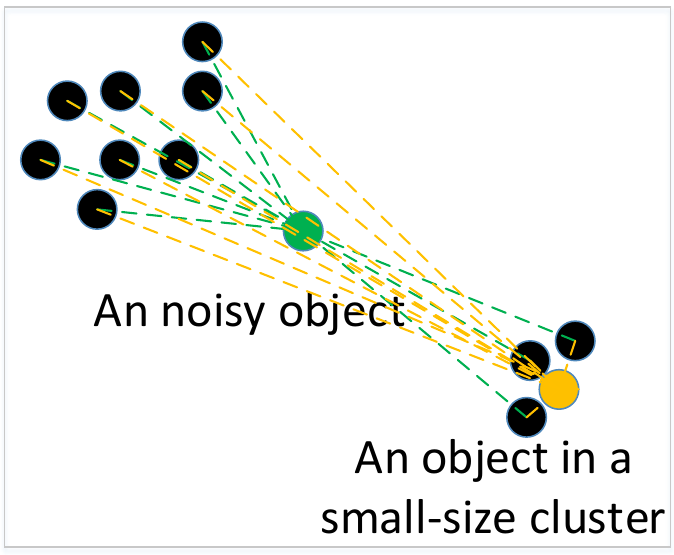}
}
\subfloat[NKD]{
  \label{fig:KNNkernDens}
   \includegraphics[width=0.19\textwidth]{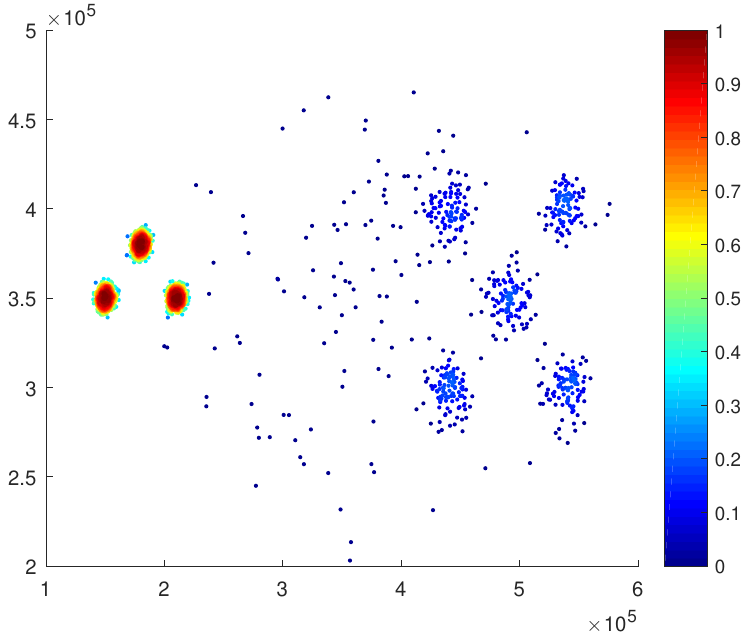}
}
\subfloat[RNKD]{
  \label{fig:RNKD}
   \includegraphics[width=0.19\textwidth]{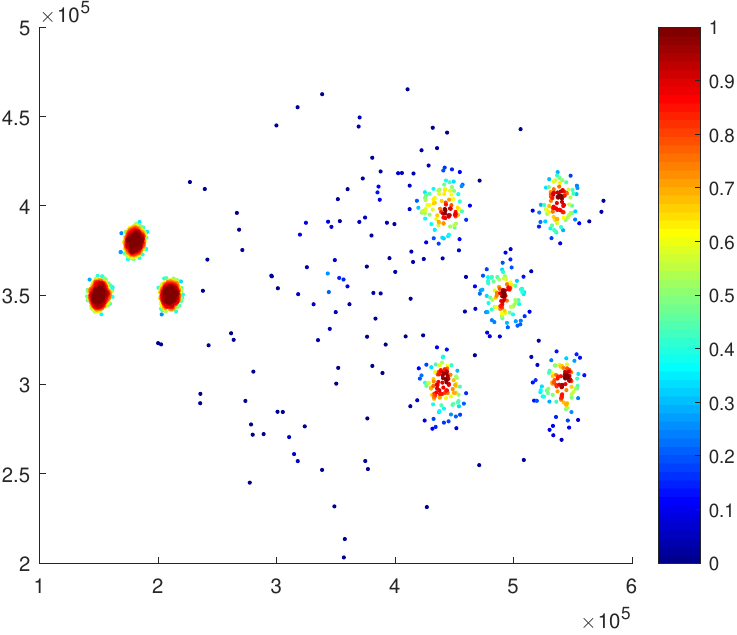}
}
\subfloat[Density gap between $u$ and $v$]{
  \label{fig:cutoDensGap}
   \includegraphics[width=0.19\textwidth]{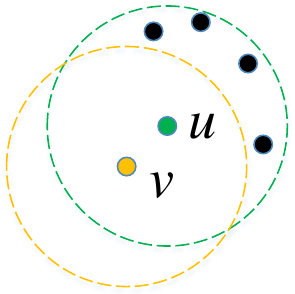}
}
\caption{Comparison of different density measures. (a) Sensitivity to
$\epsilon$ in DBSCAN. (b) Non-parametric kernel density methods are sensitive
to noise. (c) NKD estimates may mistake low-density clusters as noise. (d) RNKD
allows  discrimination of low-density clusters from noise. The temperature of
data points in (c) and (d) indicates the NKD and RNKD,
respectively.
}
\label{fig:density}
\end{figure}

The most commonly used density measure is cut-off density, which is defined as the number of objects in an
$\epsilon$-ball centered at the respective object. However, it is highly sensitive to the parameter $\epsilon$. As shown in Figure~\ref{fig:cutoDens}, a small variation in $\epsilon$ can result in drastic differences in
density estimation. Another classical measure is kernel density defined as

$$
\bar{\rho}(u)=\sum_{v\in V}ker\left(\frac{\text{d}(u, v)}{h}\right),
$$
where $ker(.)$ is usually a monotonically decreasing and continuous function, and $h$ is a constant controlling the scale. Kernel density is continuous and less sensitive to parameter selection, but it tends to give biased estimation for objects in a small-size cluster because it considers contributions of all objects in the dataset (see Figure~\ref{fig:globKernDens}).

Our proposed RNKD estimation is inspired by KNN kernel density~\cite{tran2006knn}, and only considers the objects in $N(u)$ for the density estimation of object $u$. In addition, the density of an object should be positive and has a negative relation with the distances between itself and its neighbors. Thus, the K-nearest neighbor kernel density (NKD) of object $u$ is defined as
\begin{equation}
\rho(u)=\theta\sum_{v\in N(u)}\text{exp}\left(-\frac{\text{d}(u, v)}{\sigma}\right),
\label{equ:NKD}
\end{equation}
where $\sigma=\frac{\sum_{u\in V} d_K(u)}{|V|}$ is the mean of
the distance between $v$ and its $K$-th nearest neighbors, and $\theta$ is a normalizing factor.

NKD exhibits some good properties (as will be discussed in Section~\ref{ssec:propOfNKD_AndRNKD}) and allows easy discrimination of outliers. However, it may mistake low-density clusters as outliers. One such example is shown in Figure~\ref{fig:KNNkernDens}, where low-density clusters on the right may be mistaken as noise. The key observation is that though the NKD of a low-density cluster is low
compared to high-density clusters, its NKD is still comparatively high in
its surroundings. This observation motivates us to introduce a new density
measure, called RNKD as follows.

\begin{definition}
\label{DefinitionRNKD}
Given $u\in V$, \textit{RNKD} of $u$, denoted by $\rho^*(u)$, is defined as
\begin{equation}
\rho^*(u)=\frac{\rho(u)}{\max\limits_{v\in N(u)\cup\{u\}}\{\rho(v)\}}.
\label{equ:RNKD}
\end{equation}
\end{definition}
From the definition, we can see RNKD is a normalized local relative density measure. Figure \ref{fig:RNKD} shows that it can effectively accentuate both dense and sparse clusters, but diminish noises. Though the parameter $K$ still needs to be tuned (the number of nearest neighbors), we will show that it is generally robust to $K$ in Section \ref{ssec:propOfNKD_AndRNKD}.

\subsection{Properties of NKD and RNKD}
\label{ssec:propOfNKD_AndRNKD}

As discussed in Section \ref{ssec:DensEstiIntro}, one shortcoming of cut-off density is its sensitivity to parameter $\epsilon$. Additionally, even for a fixed $\epsilon$, there can be significant discontinuity in the cut-off densities of two adjacent objects (see Figure \ref{fig:cutoDensGap}), which will lead to undesirable results. Though non-parametric kernel density can avoid this gap, it may suffer from the problem illustrated in Figure~\ref{fig:globKernDens}. Next, we show that NKD exhibits local continuity. In particular, the ratio of $\rho(u)$ and $\rho(v)$ between neighbor points $u$ and $v$ are bounded on both sides.

\begin{thm}
\label{thm:contOfNKD}
$\forall u,v\in V, u\not=v$, we have $\frac{|\rho(u)-\rho(v)|}{\rho(u)+\rho(v)}<1-\exp{\left(-\frac{d(u,v)}{\sigma}\right)}$.
\end{thm}
\begin{proof}
See appendix.
\end{proof}
 Consequently, for any $u,v\in V$, the following results hold
\begin{small}
\begin{equation}
\label{equ:NKDCont}
\frac{\exp\left(-\frac{d(u,v)}{\sigma}\right)}{2-\exp\left(-\frac{d(u,v)}{\sigma}\right)}\leq\frac{\rho(u)}{\rho(v)}\leq\frac{2-\exp\left(-\frac{d(u,v)}{\sigma}\right)}{\exp\left(-\frac{d(u,v)}{\sigma}\right)}.
\end{equation}
\end{small}
This implies that $\rho(u)$ and $\rho(v)$ are close to each other when $\frac{d(u,v)}{\sigma}$ is small. In other words, NKD changes slowly in the interior of a cluster. Furthermore, \eqref{equ:NKDCont} also shows that RNKD will be stable at a high level in the interior of a cluster.


\begin{figure}[!htb]
\centering
\begin{minipage}{0.85\linewidth}
\includegraphics[width = 0.23\linewidth]{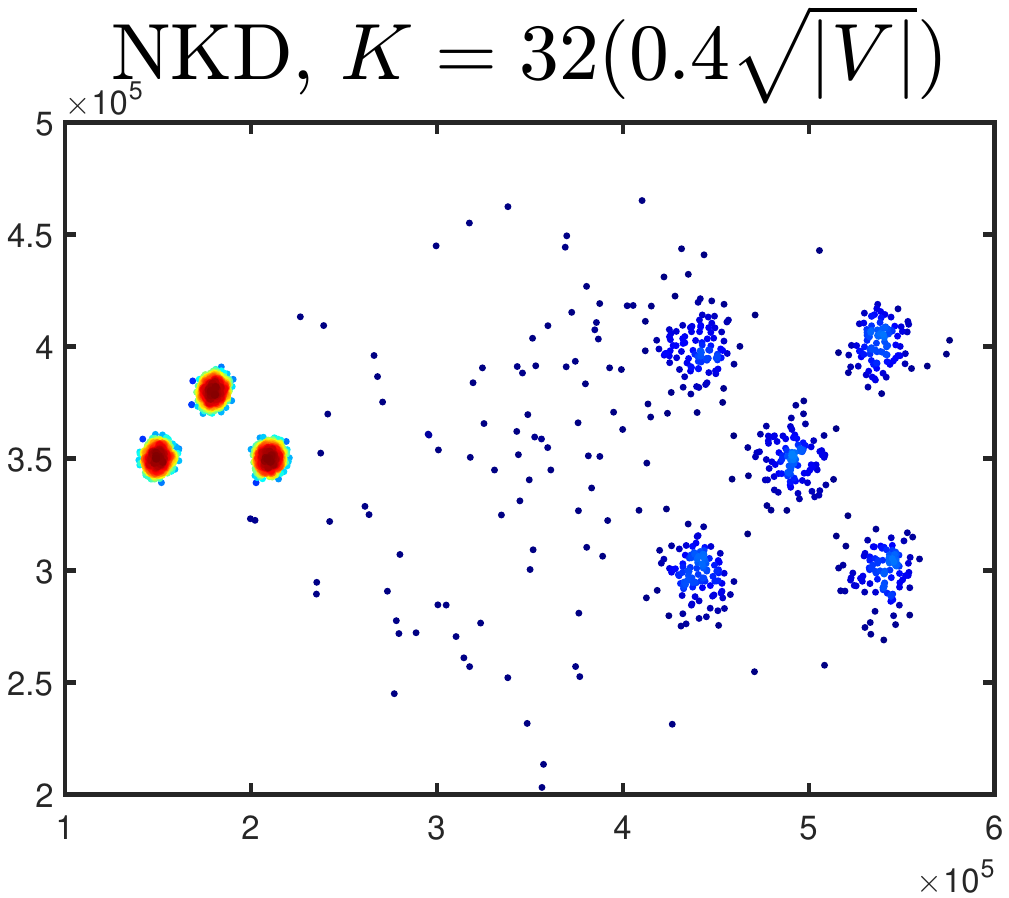}\
\includegraphics[width = 0.23\linewidth]{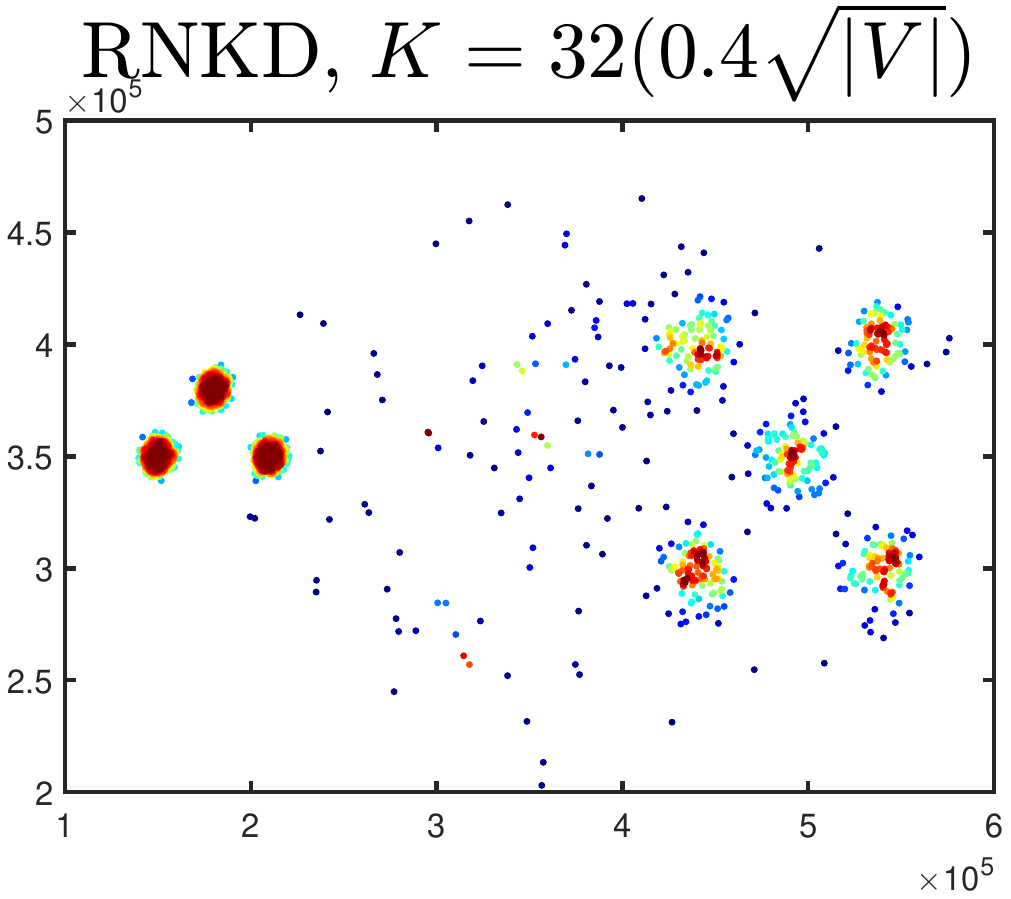}\
\includegraphics[width = 0.255\linewidth]{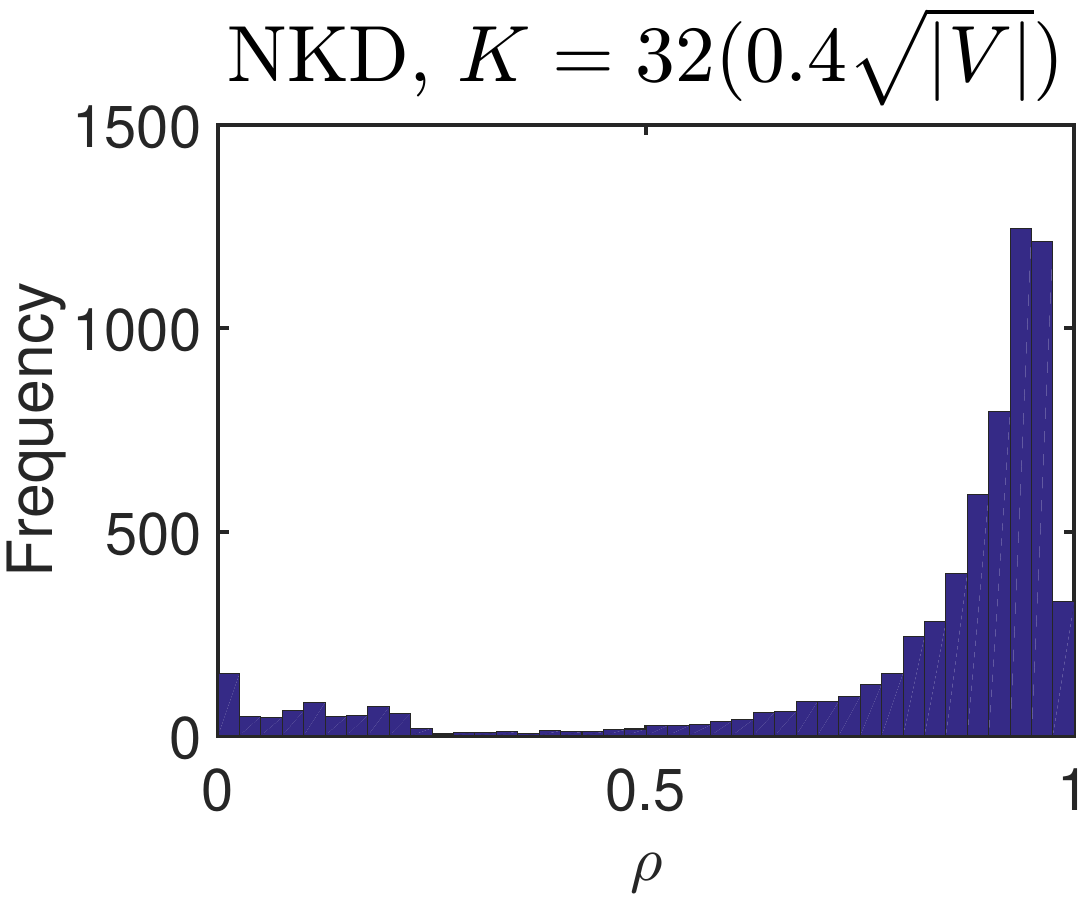}\
\includegraphics[width = 0.255\linewidth]{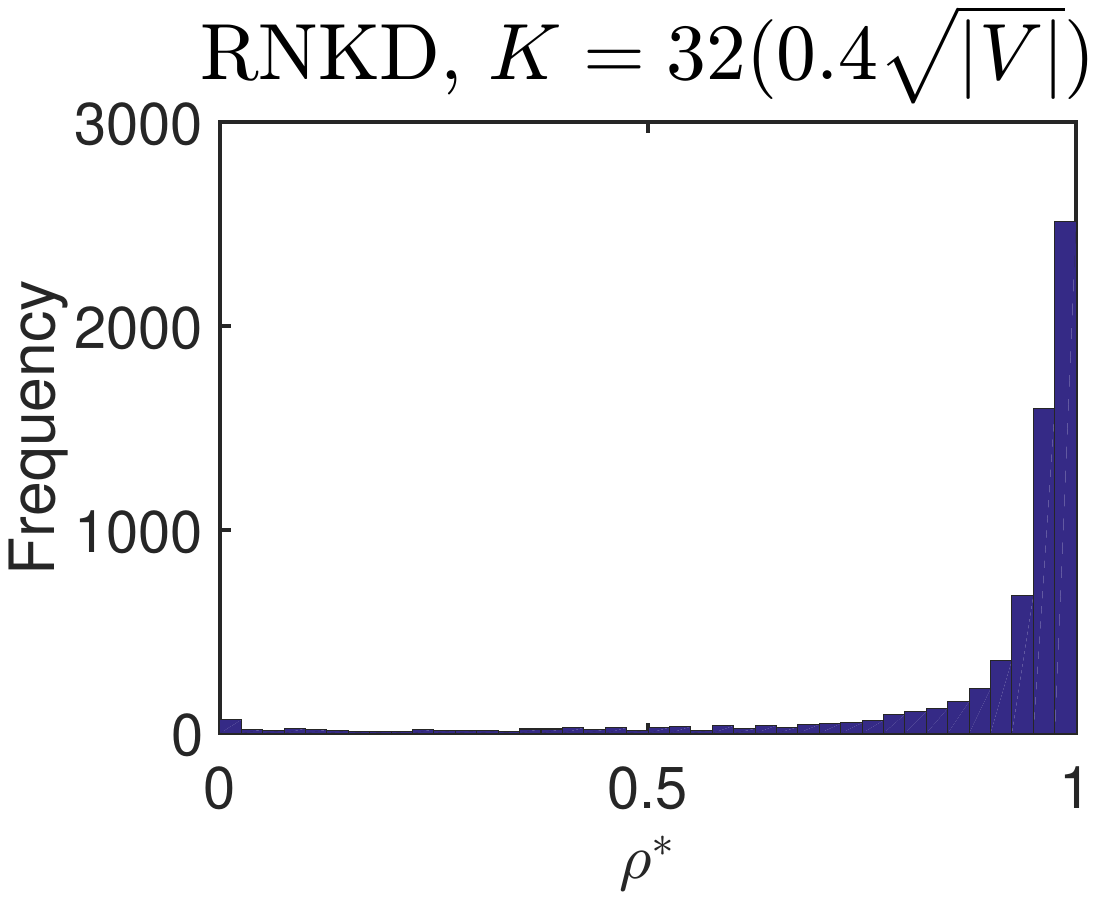}\\[1mm]
\includegraphics[width = 0.23\linewidth]{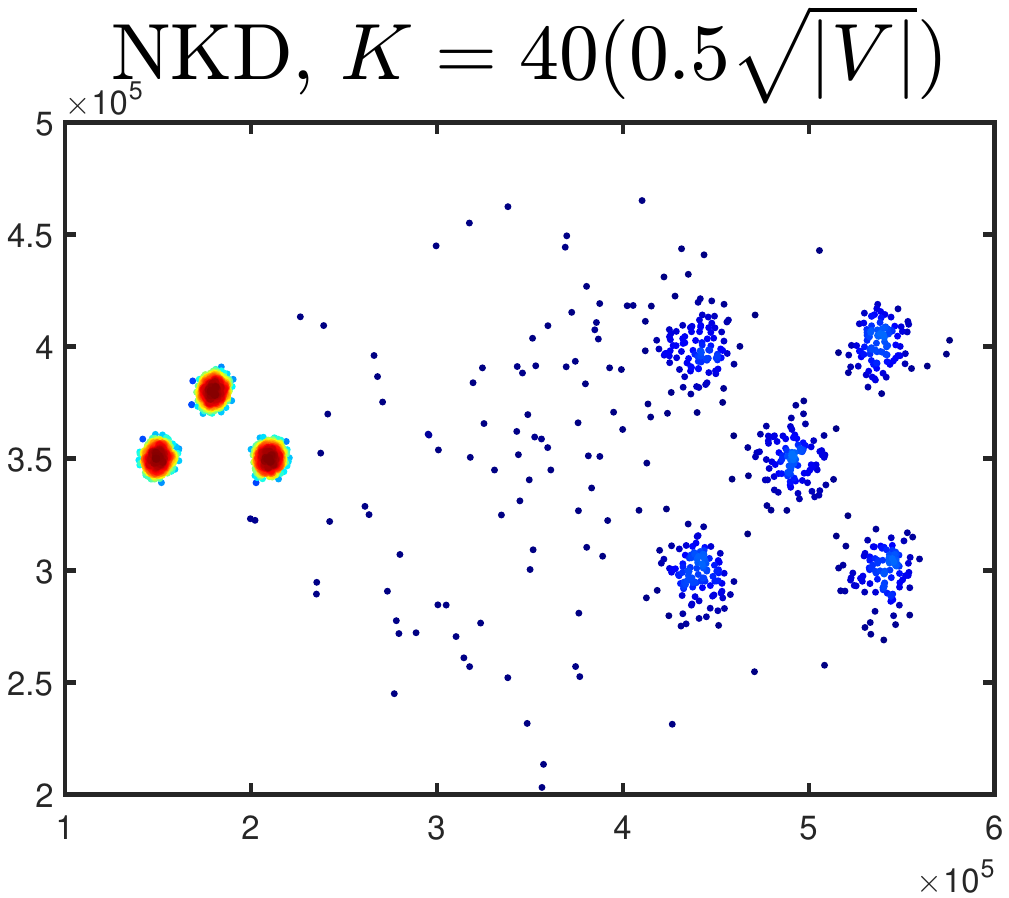}\
\includegraphics[width = 0.23\linewidth]{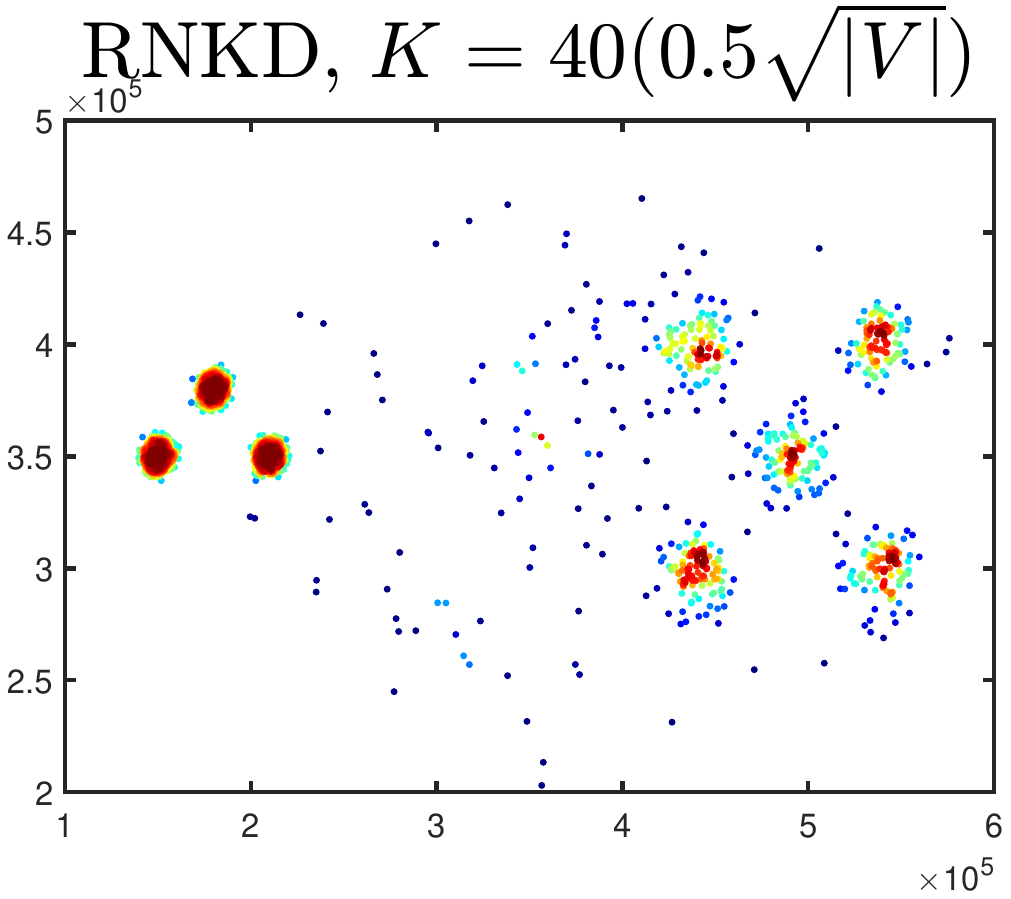}\
\includegraphics[width = 0.255\linewidth]{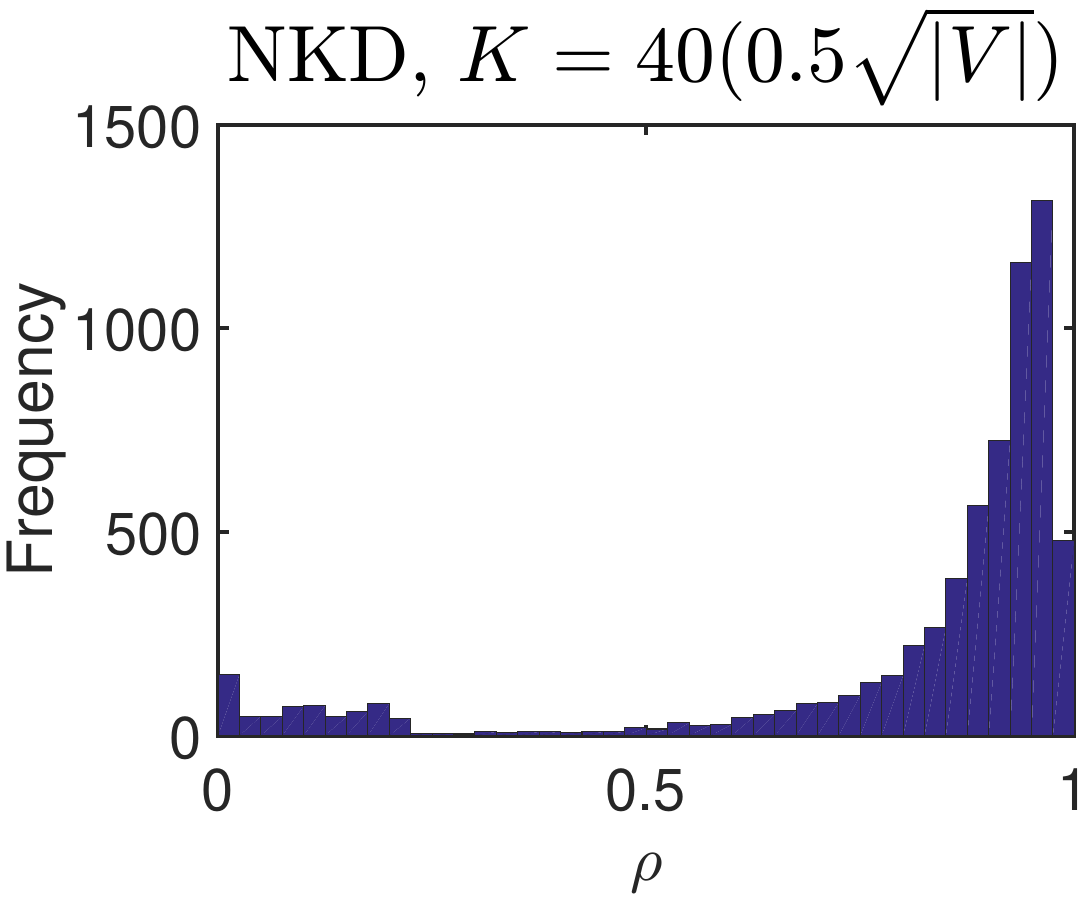}\
\includegraphics[width = 0.255\linewidth]{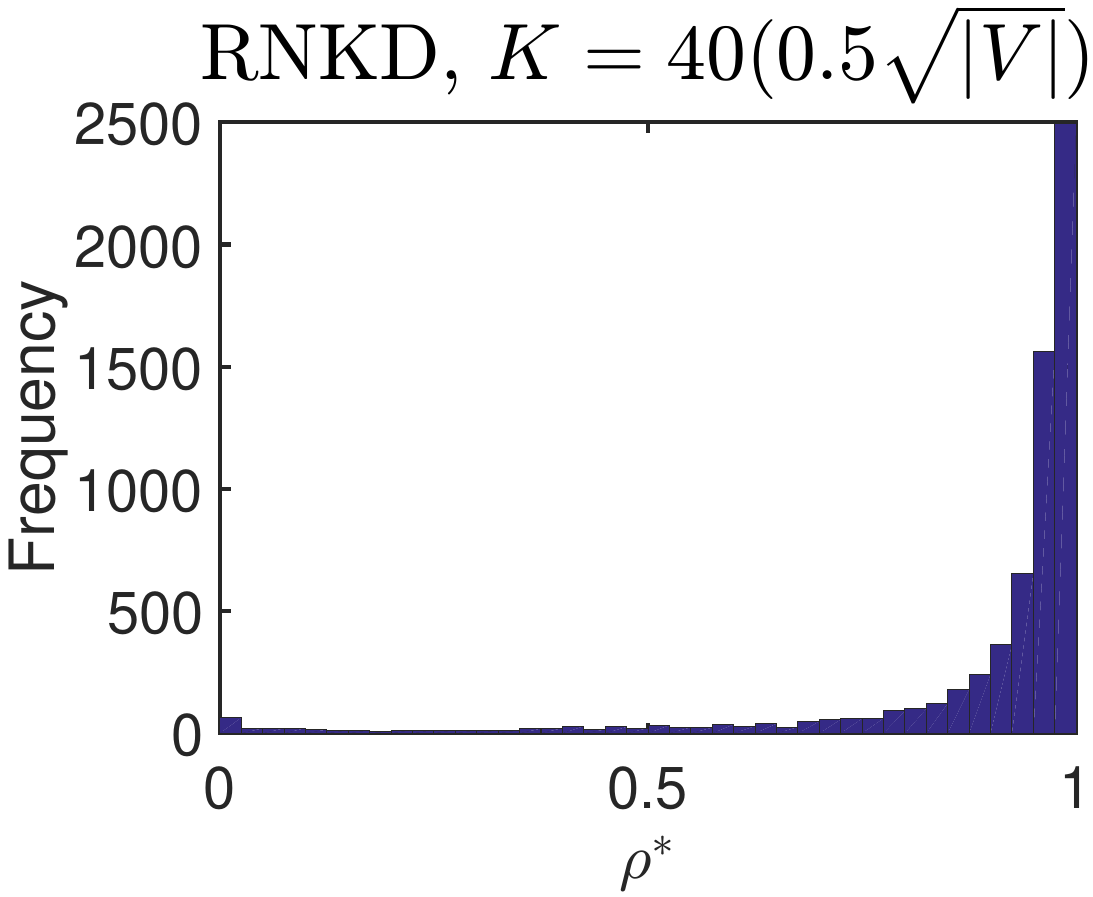}\\[1mm]
\includegraphics[width = 0.23\linewidth]{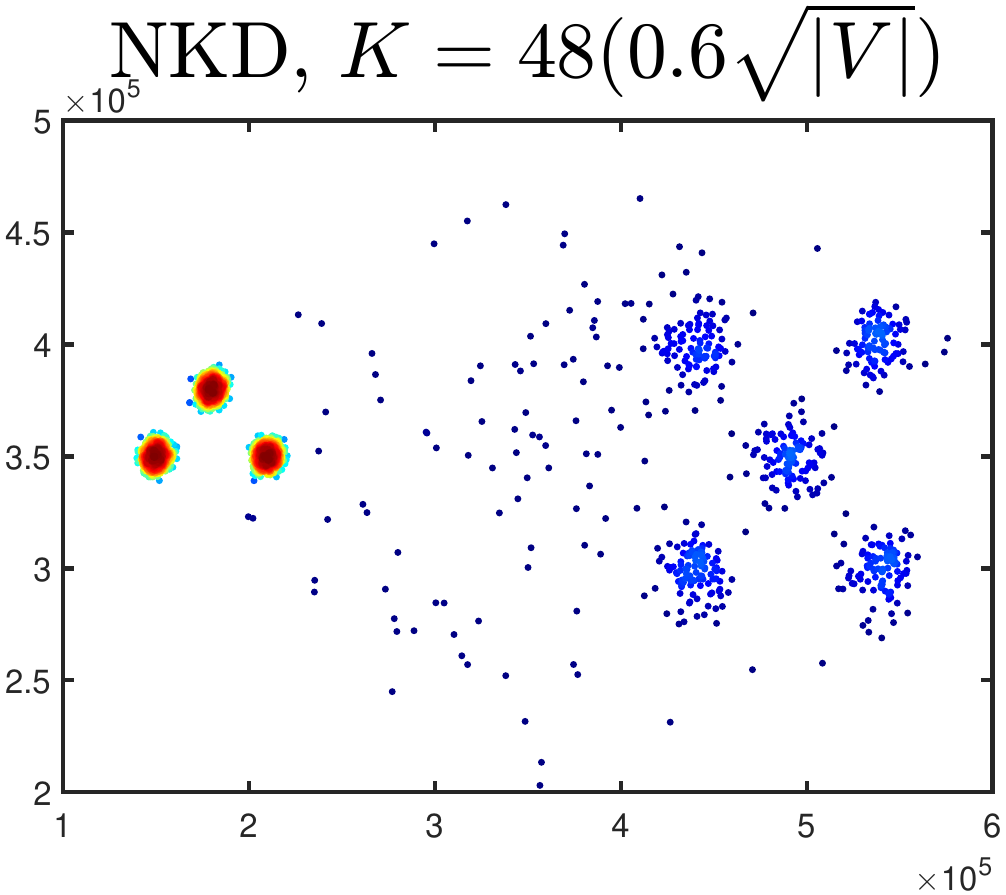}\
\includegraphics[width = 0.23\linewidth]{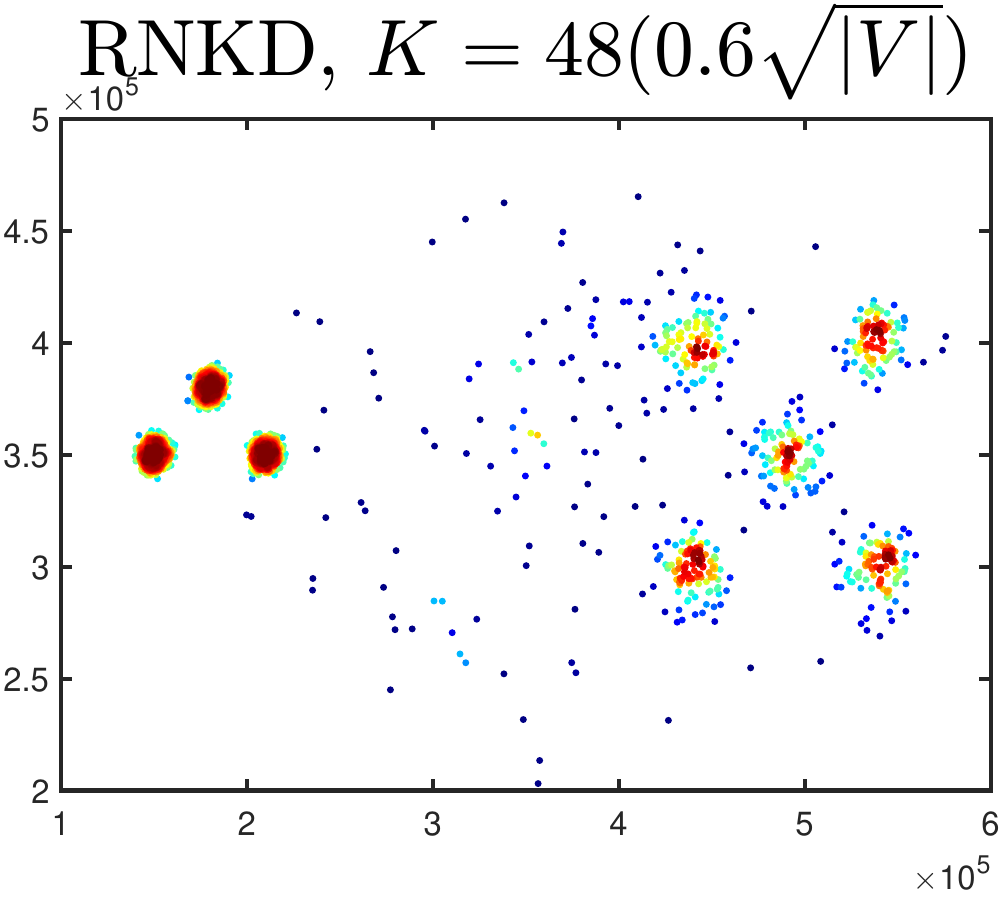}\
\includegraphics[width = 0.255\linewidth]{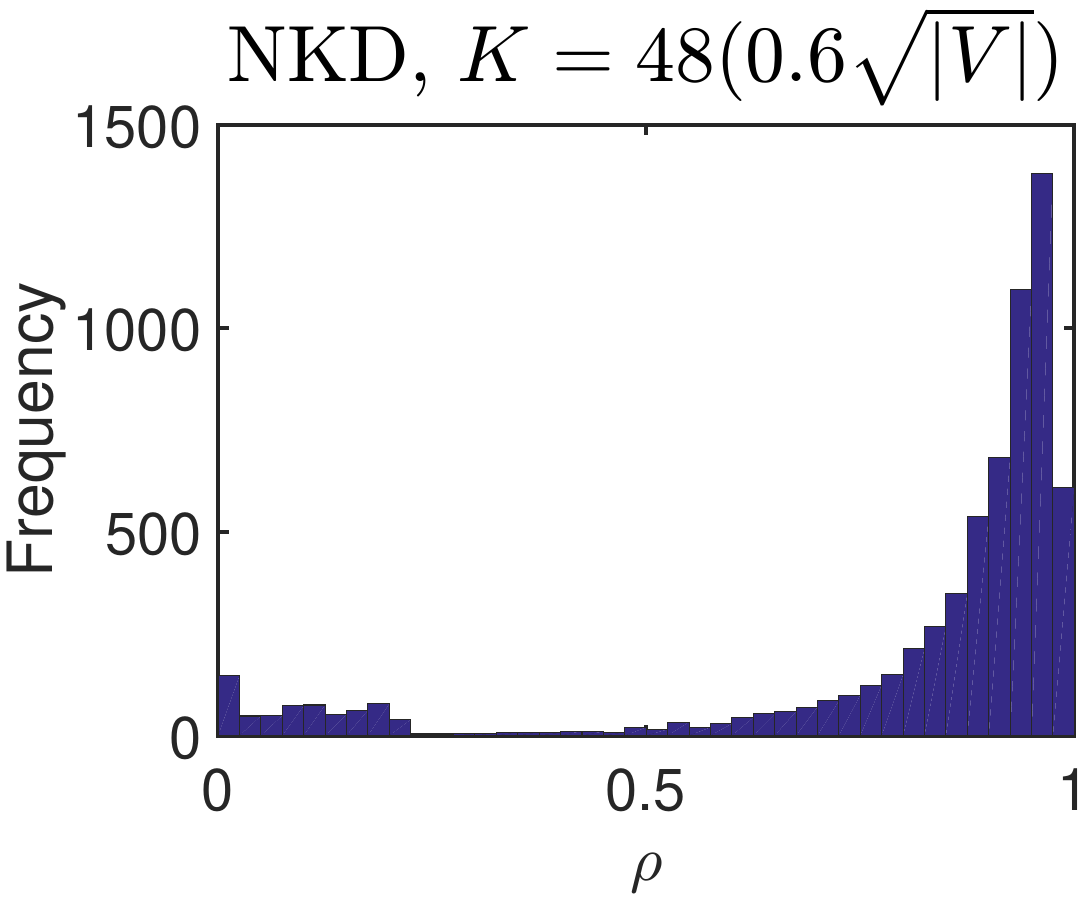}\
\includegraphics[width = 0.255\linewidth]{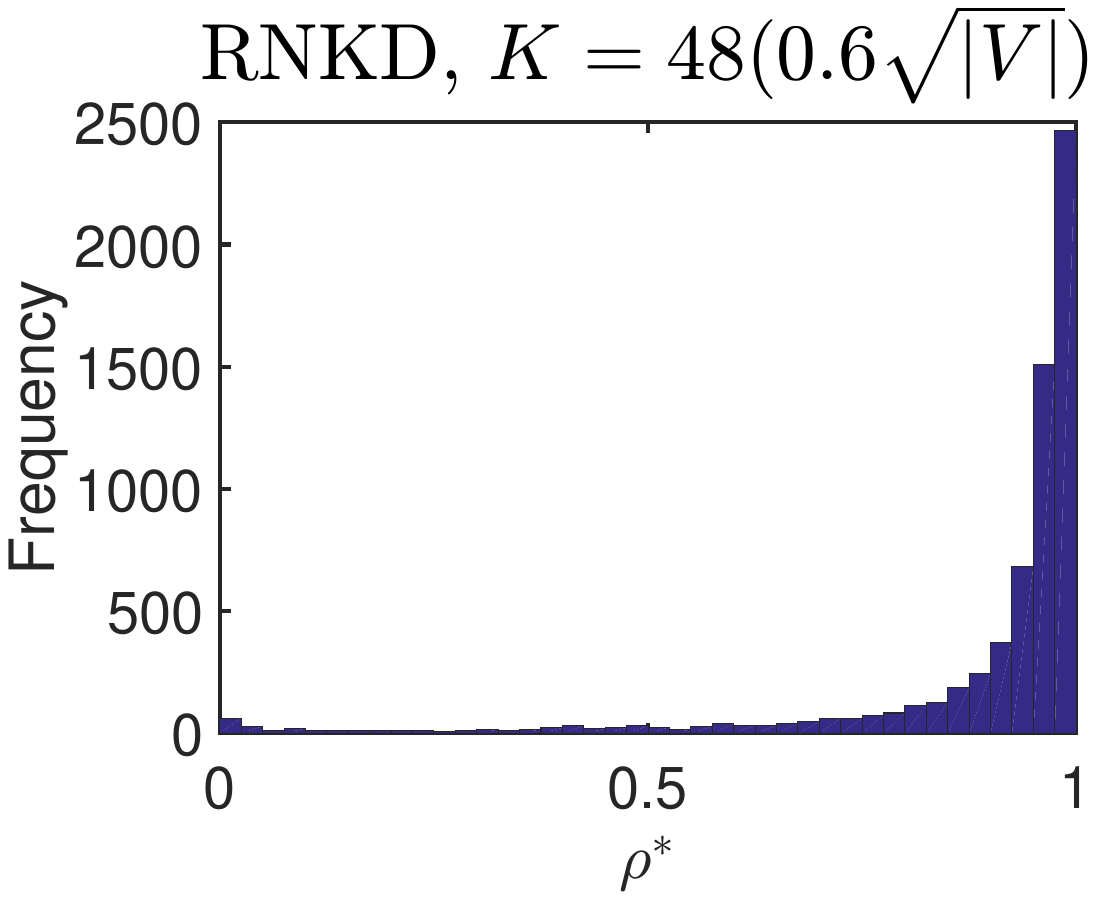}\\[1mm]
\includegraphics[width = 0.23\linewidth]{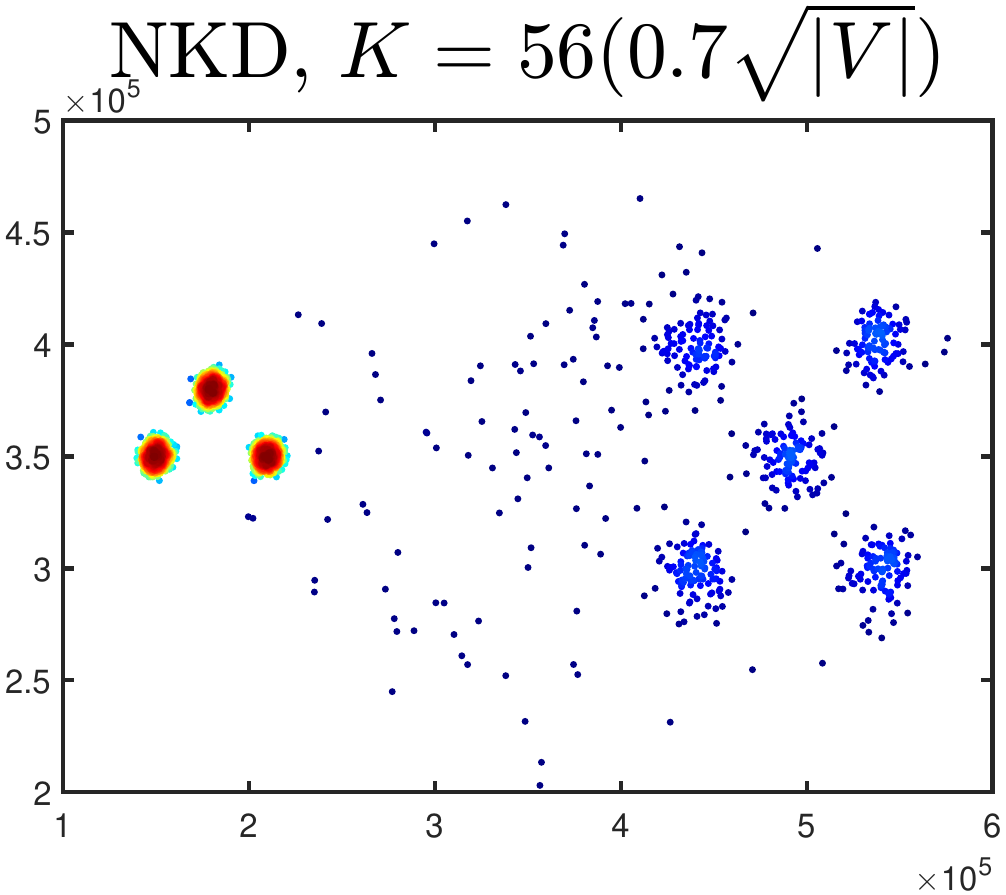}\
\includegraphics[width = 0.23\linewidth]{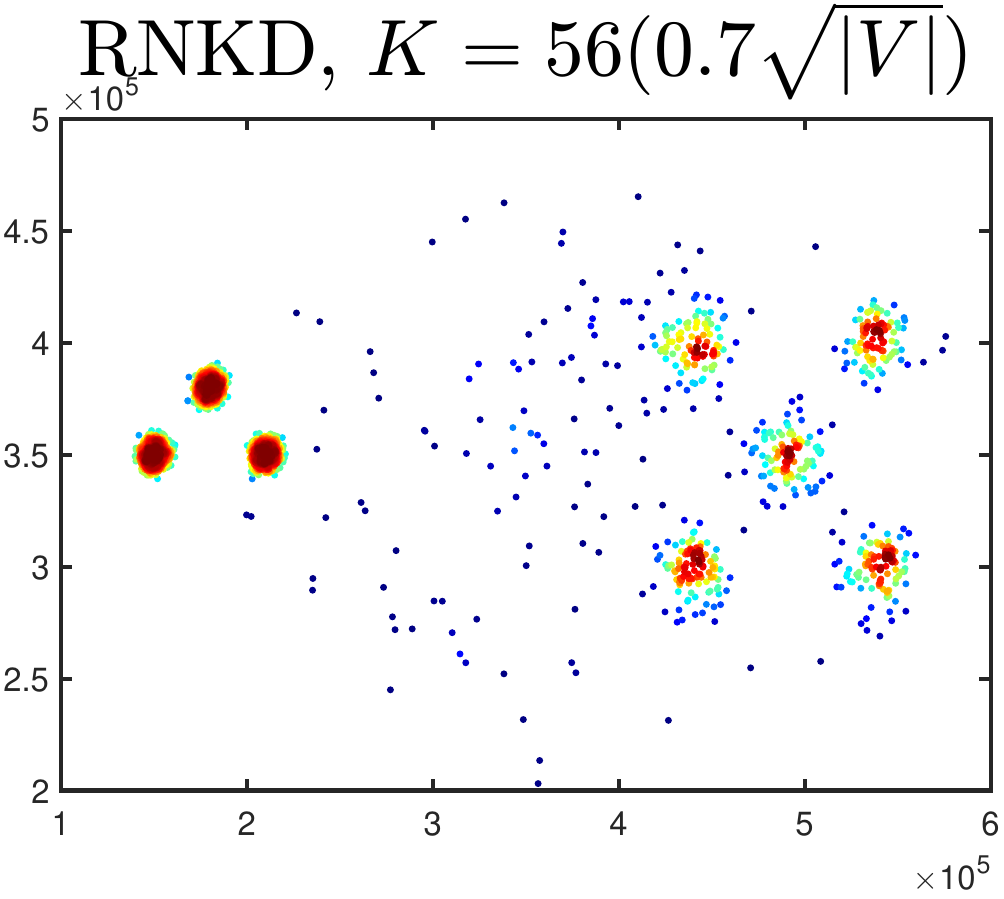}\
\includegraphics[width = 0.255\linewidth]{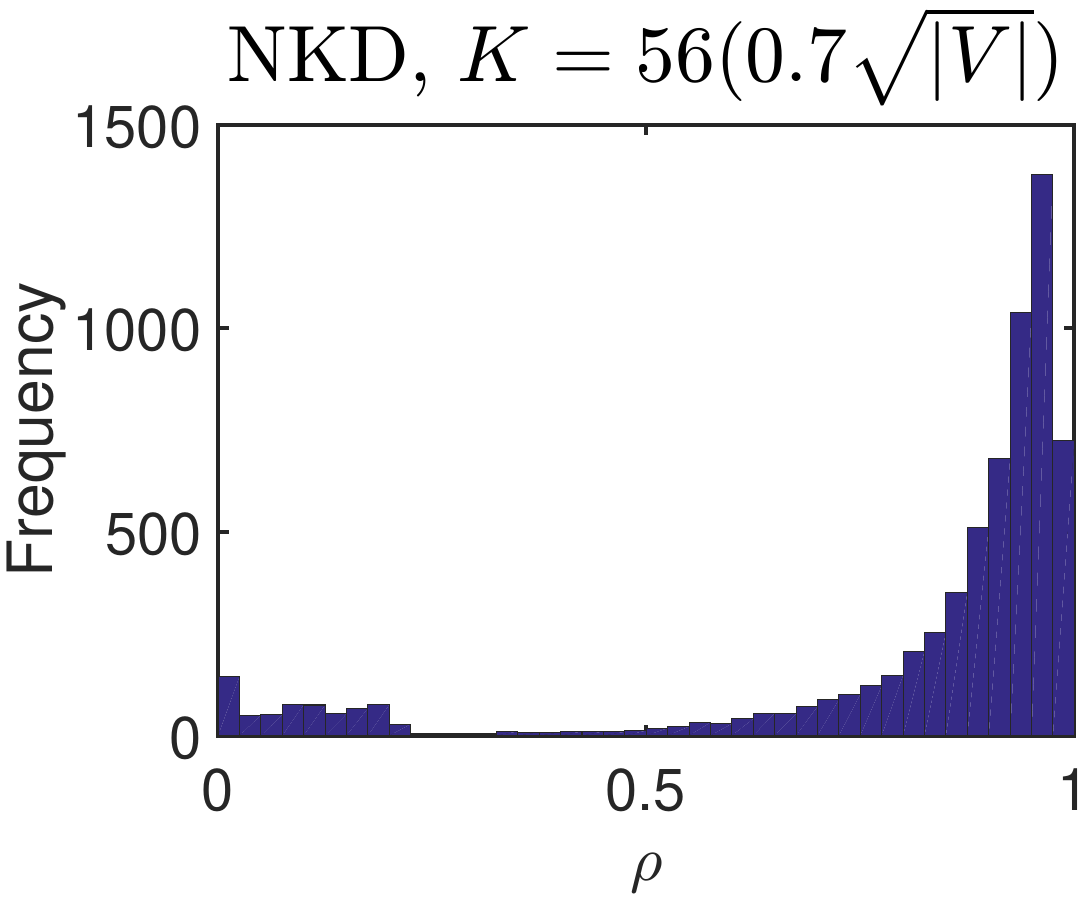}\
\includegraphics[width = 0.255\linewidth]{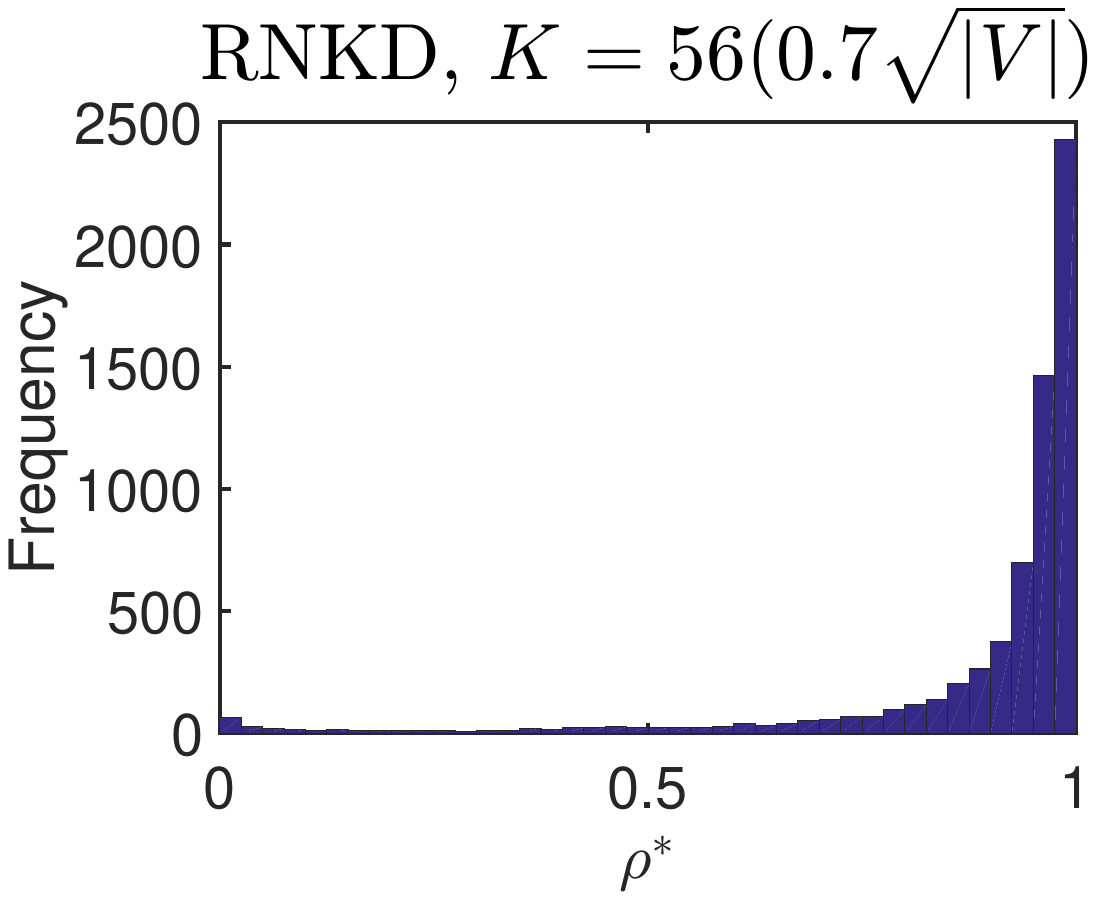}\\[1mm]
\includegraphics[width = 0.23\linewidth]{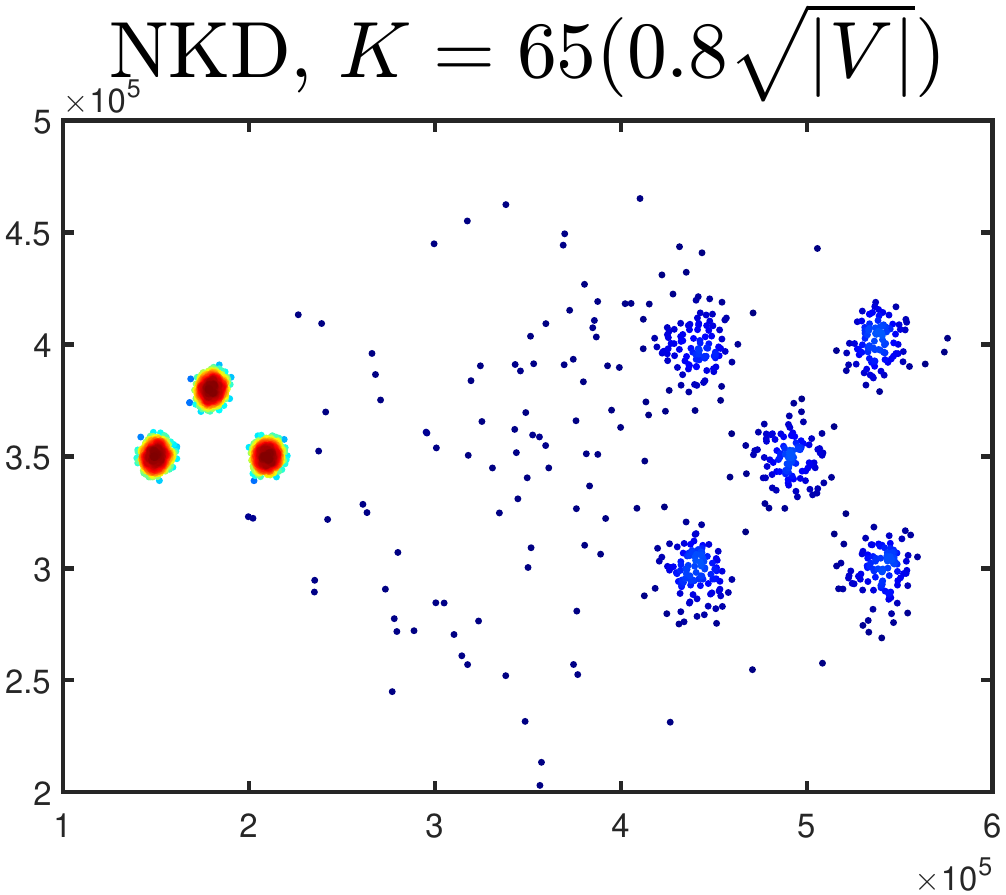}\
\includegraphics[width = 0.23\linewidth]{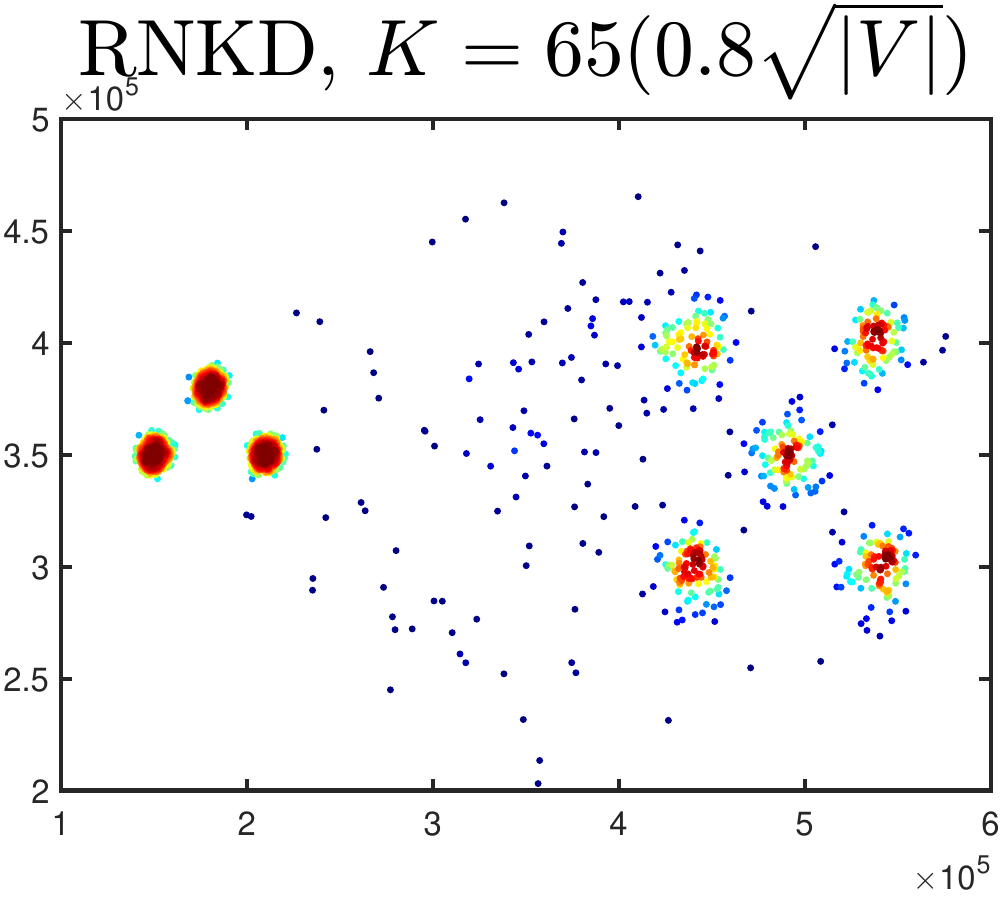}\
\includegraphics[width = 0.255\linewidth]{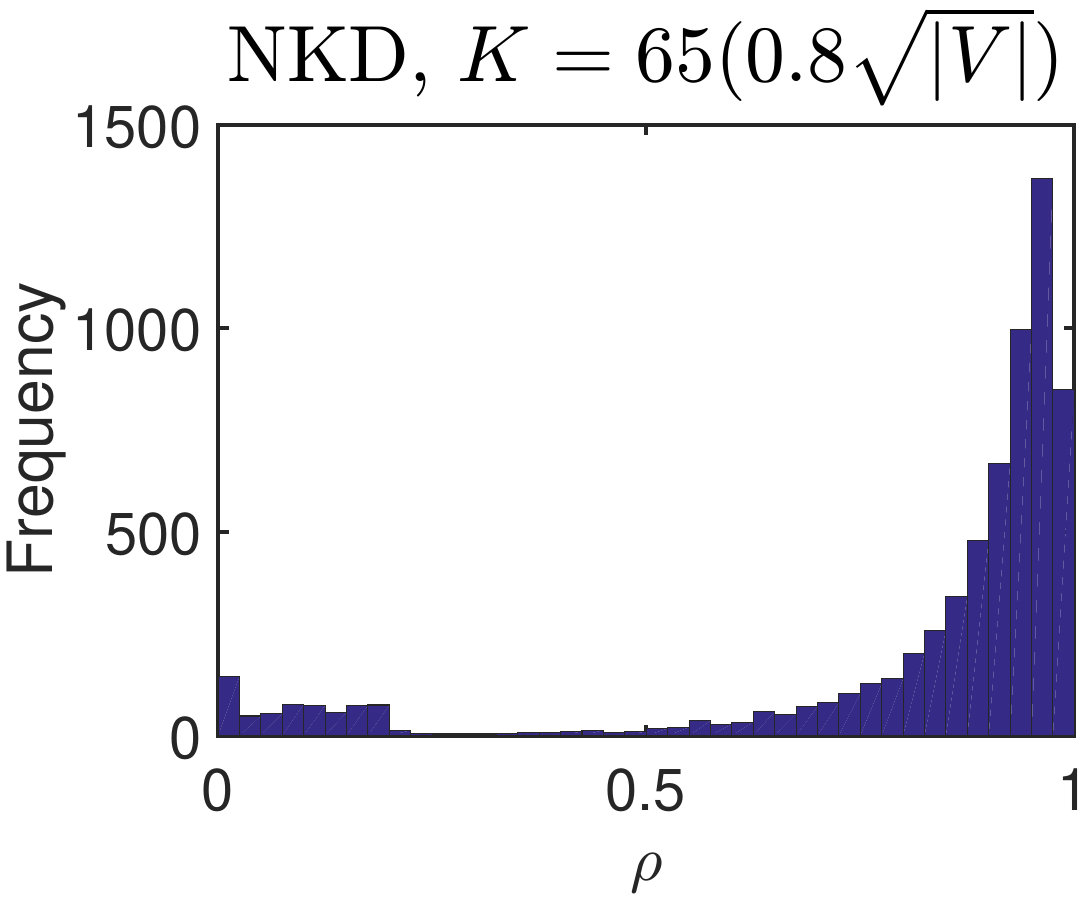}\
\includegraphics[width = 0.255\linewidth]{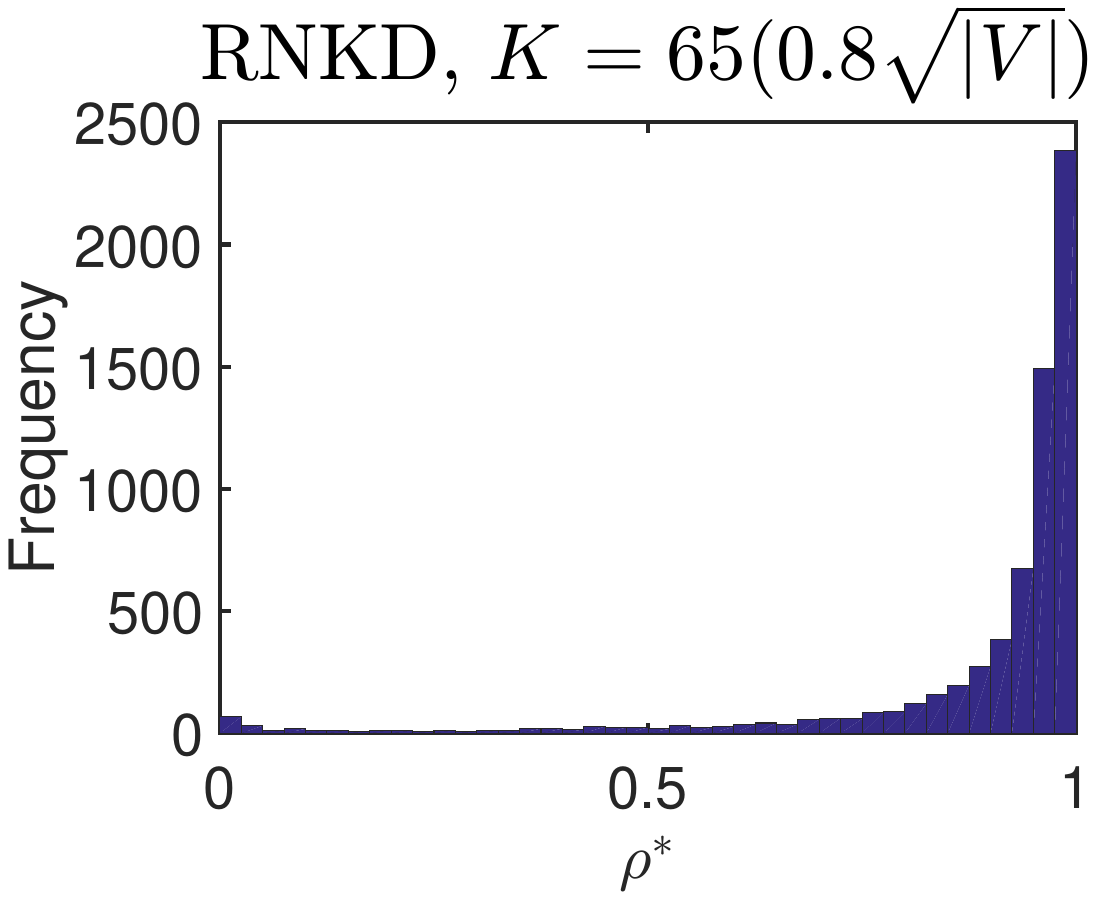}\\[1mm]
\includegraphics[width = 0.23\linewidth]{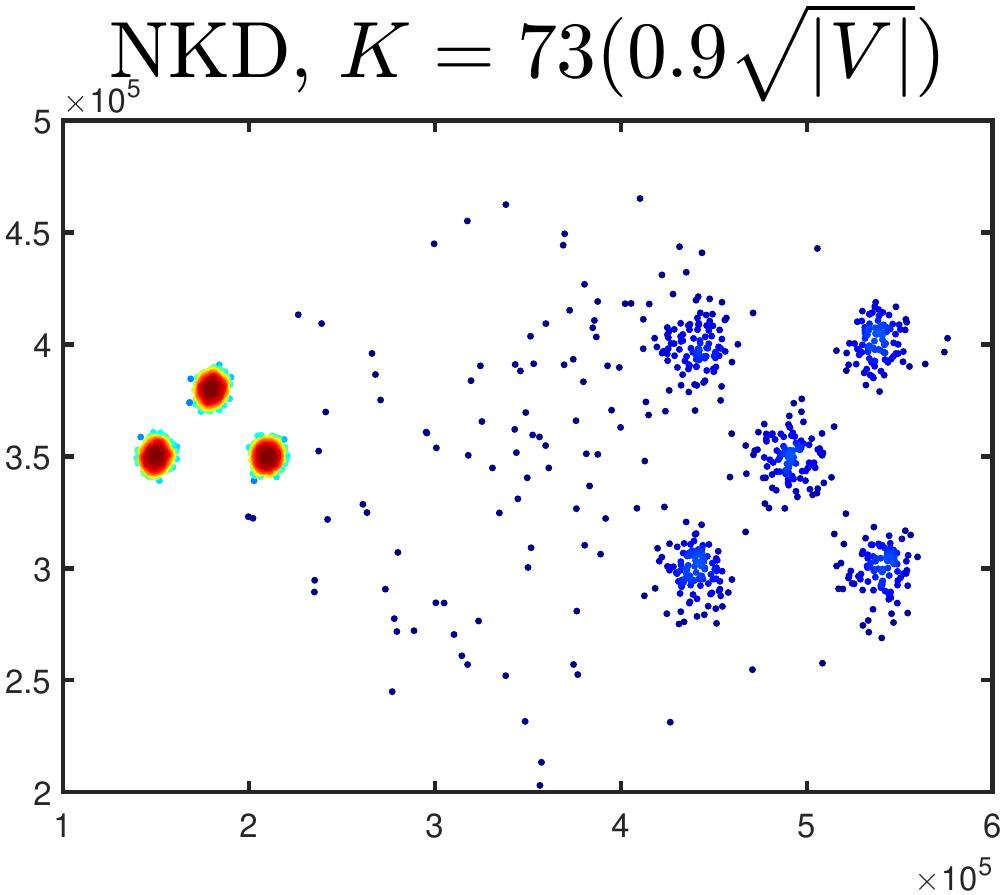}\
\includegraphics[width = 0.23\linewidth]{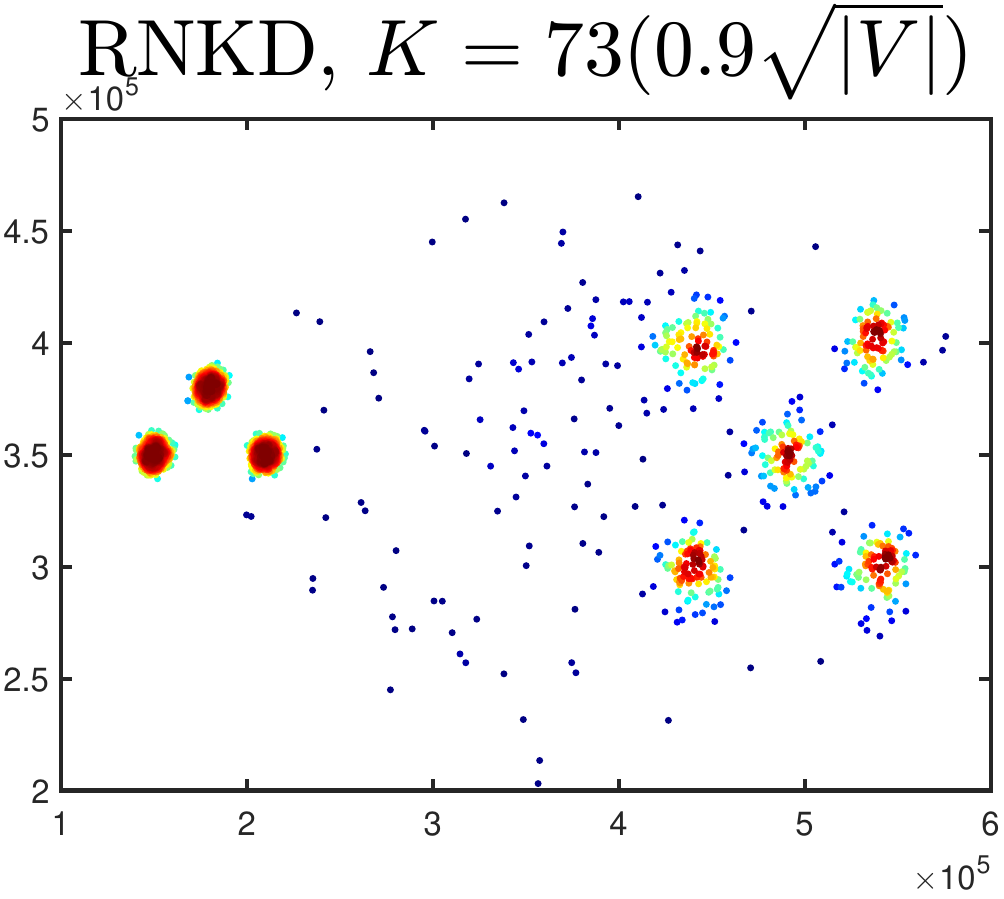}\
\includegraphics[width = 0.255\linewidth]{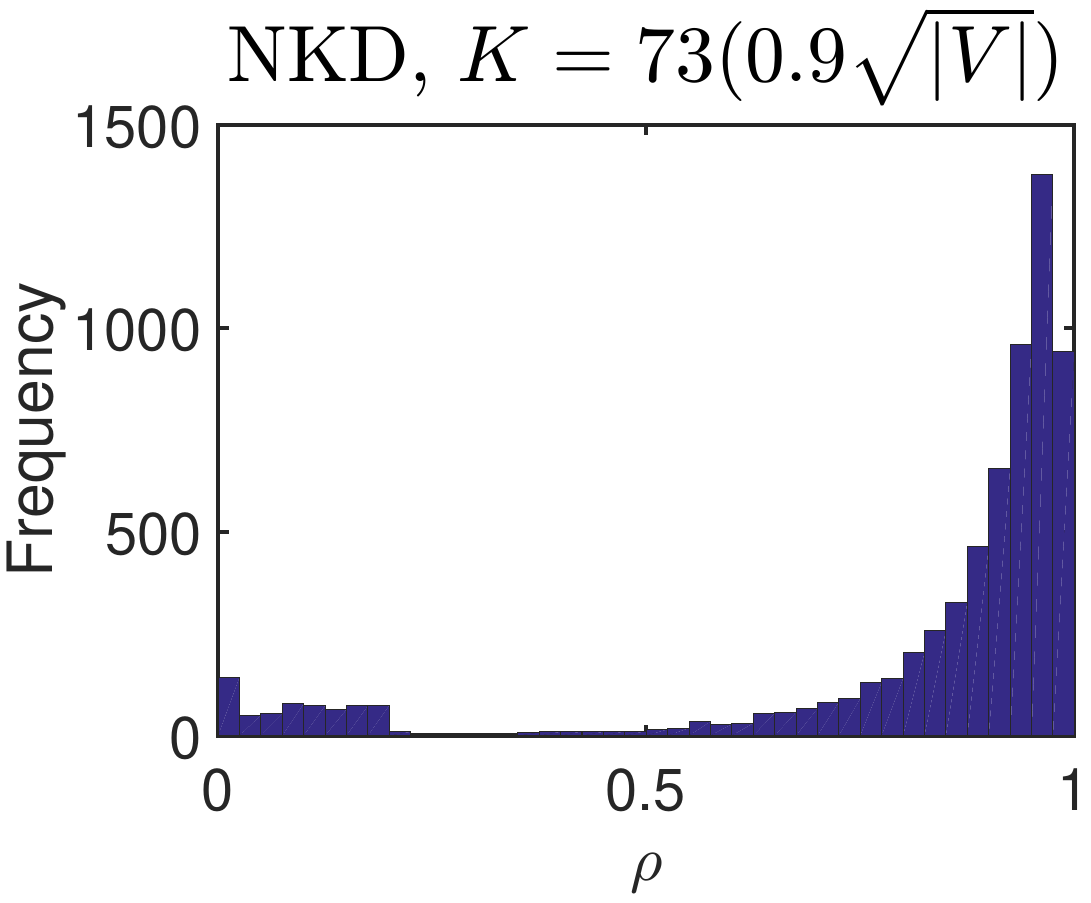}\
\includegraphics[width = 0.255\linewidth]{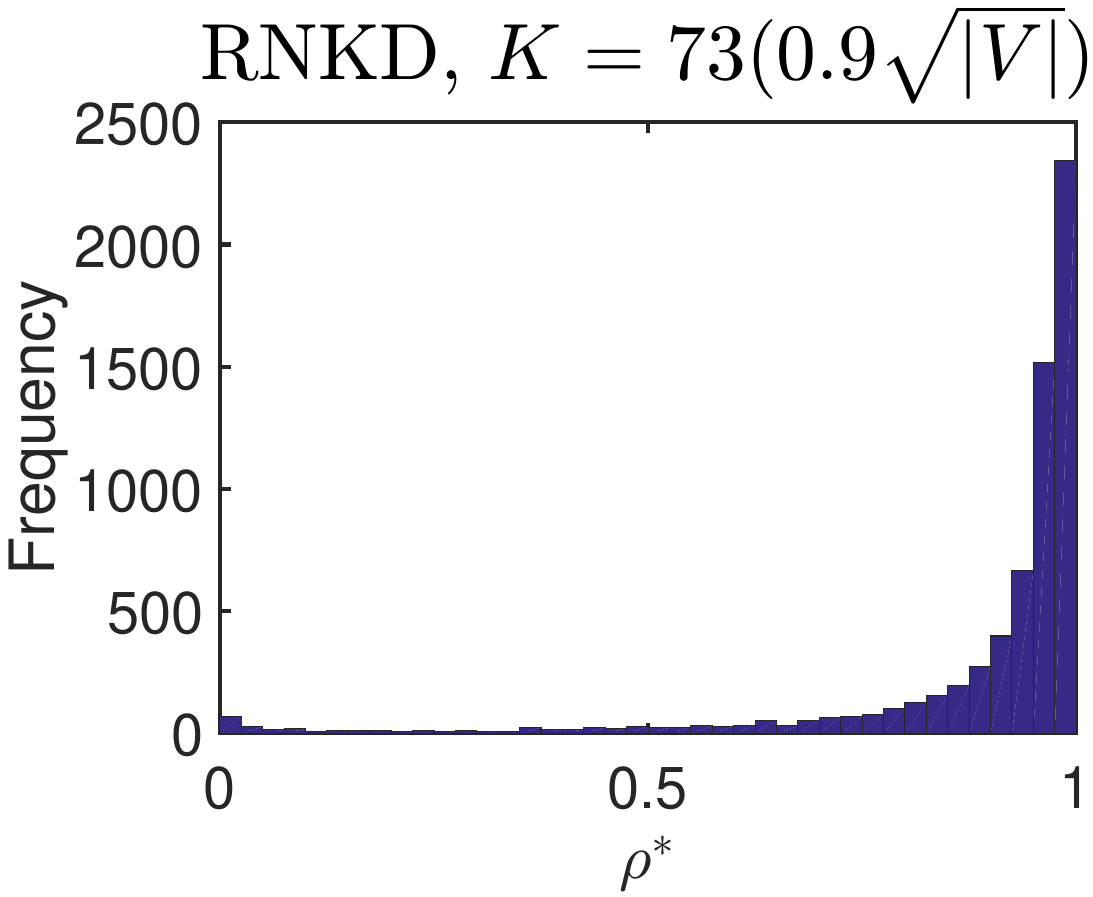}\\[1mm]
\includegraphics[width = 0.23\linewidth]{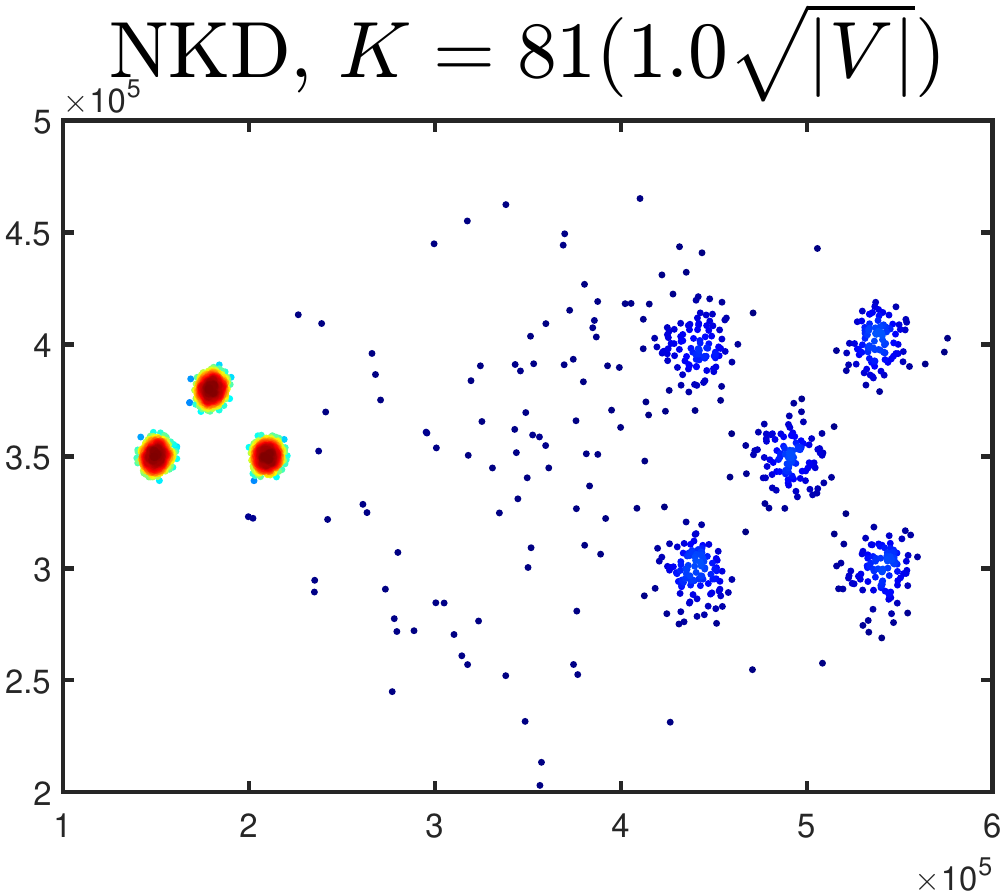}\
\includegraphics[width = 0.23\linewidth]{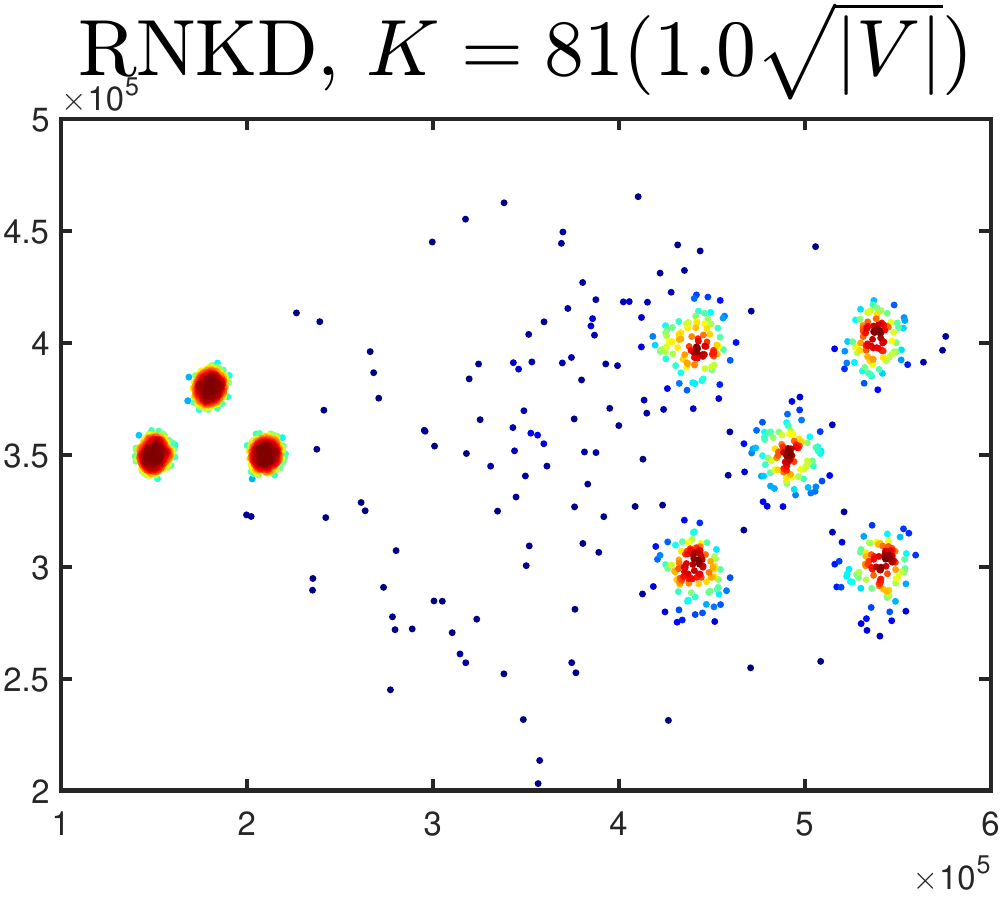}\
\includegraphics[width = 0.255\linewidth]{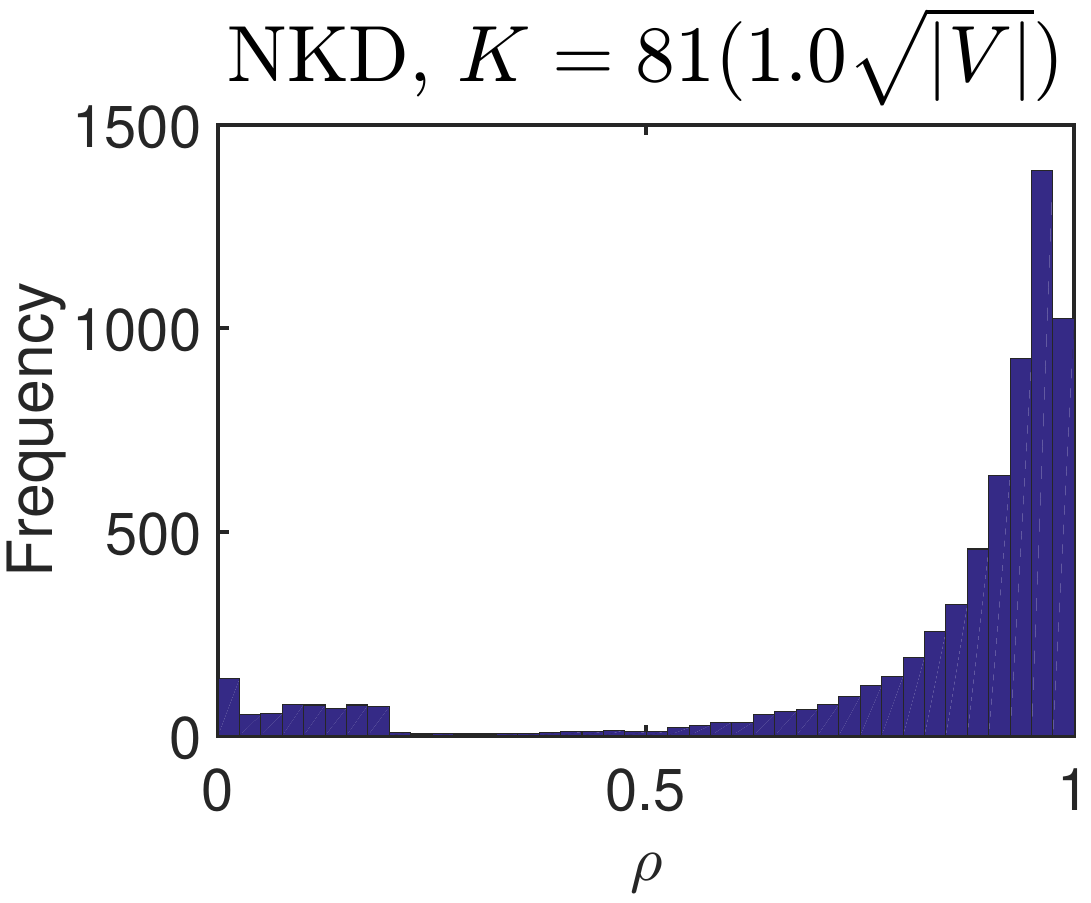}\
\includegraphics[width = 0.255\linewidth]{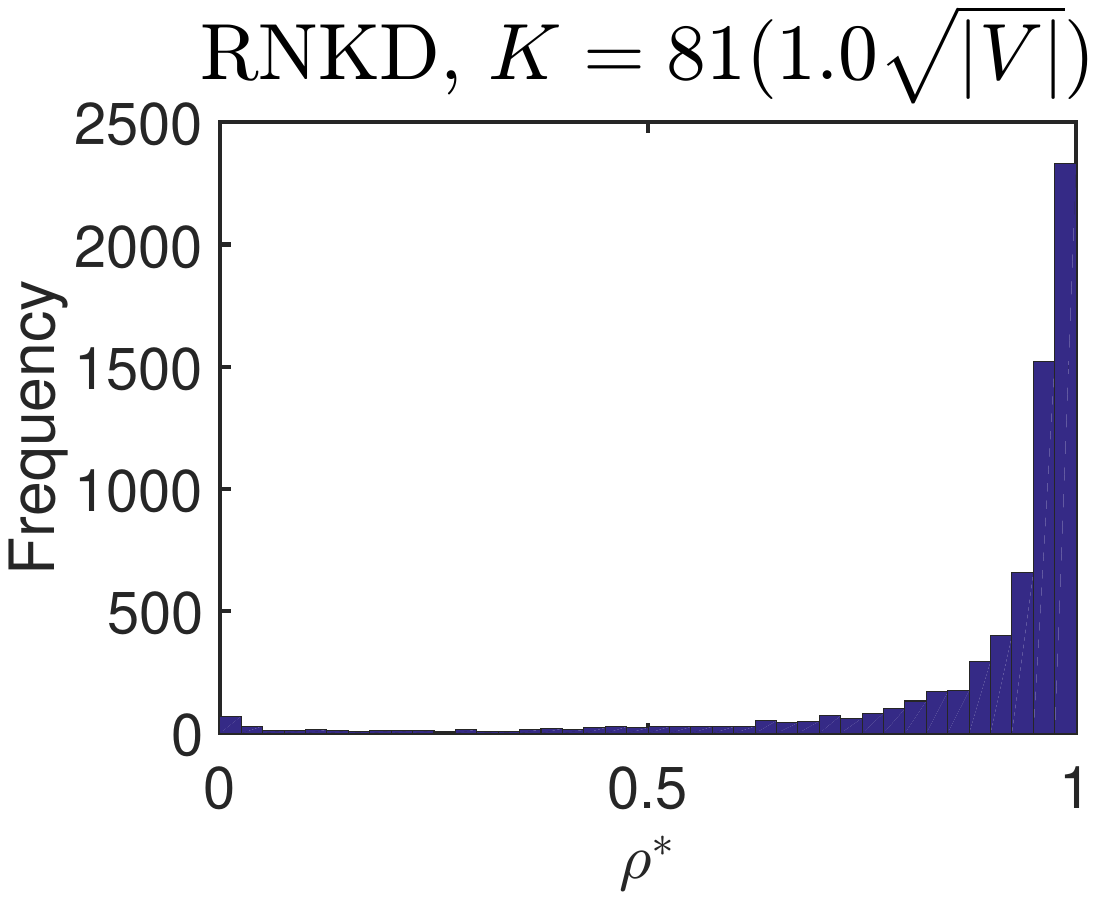}\\[1mm]
\caption{Columns from left to right are heat-map of NKD, heat-map of RNKD, distribution of NKD, and distribution of RNKD, respectively. Rows from top to bottom show the situations when $K$ takes the value from small to large.}
\label{fig:NKDAndRNKDToK}
\end{minipage}
\end{figure}

Another important issue is how sensitive the density measures are to the choice of parameter $K$. Figure~\ref{fig:NKDAndRNKDToK} shows the heat-map and the distribution of NKD and RNKD on the synthetic dataset in Figure~\ref{fig:density} when $K$ valued from $0.4\sqrt{|V|}$ to $\sqrt{|V|}$. From this figure, we can observe that, when $K\geq0.5\sqrt{|V|}$, both NKD and RNKD follow similar distributions despite varying of $K$. This indicates, for this representative dataset, NKD and RNKD are robust to the parameter $K$ in the range $[0.5\sqrt{|V|}, \sqrt{|V|}]$. In Section \ref{ssec:ParaAnal}, we will show that the proposed RECOME algorithm also achieves good performance when $K$ is in this range.

\section{RECOME Clustering Algorithm}
\label{sec:RECOME}
Now we are in the position to present RECOME based on NKD and RNKD. We first identify {\it core objects}
corresponding to data points of peak relative density. These core objects
serve as centers of sub-clusters, called {\it atom clusters}, which will be
further merged through connected paths on a KNN graph. Thus, RECOME is, in
essence, an agglomerative hierarchical clustering method.

\subsection{Finding core objects and atom clusters}
From the definition of RNKD, we know that there exist objects with unit
RNKD. Such objects are good candidates for cluster centers as they have
local maxima values in density. Formally,
\begin{definition} \label{defCoreObje}
An object $u$, $u\in V$, is called a \textbf{core object} if $\rho^*(u)=1$.
Denote the set of core objects by $O = \{u|u\in V,\rho^*(u)=1\}$.
\end{definition}

For a non-core object $u$ (namely, $\rho^*(u) < 1$), we define its {\it
Higher Density Nearest-neighbor} (HDN), $\pi(u)$ as
\begin{equation}
\pi(u)=\mathop{\arg\min}_{\rho(v)>\rho(u), v\in V}\{\text{d}(u, v)\}.
\label{equ:HDN}
\end{equation}
By the definition of $\rho^*(u)$, we can see that $\pi(u)$ exists in $N(u)$. In other words, the distance between an object and its HDN is small enough for most objects. So, HDN can be seen as a discrete approximation of gradients, which are hard to compute directly.


HDN allows us to construct a directed graph $\mathcal{G} = (V, A)$, where each
vertex is an object and a direct edge exists from a non-core object to its
higher density nearest neighbor, namely, $A = \{\langle u, \pi(u)\rangle|u\in V\backslash O\}$.

In $\mathcal{G}$, starting from any non-core object and following the directed
edges, we will eventually reach a core object. In other words, $\mathcal{G}$
can be partitioned into trees with disjoint vertices, where each tree is rooted
at a core object. A core object and its descendants thus form a cluster,
called an {\it atom cluster}. Due to the bijective relation between an atom
cluster and its core object, we use the two terms interchangeably. Formally, for a core object $o\in O$, the atom cluster rooted at $o$ is given by,

\begin{equation}
\label{equ:atomClus}
\{o\}\cup\{v\in V\backslash O | \mbox{$v$ is connected to $o$ in $\mathcal{G}$}\}.
\end{equation}

Atom clusters form the basis of final clusters, however, they themselves tend to be
too fine. A true cluster may consist of many atom clusters. This happens
when many local maximals exist in one true cluster. Thus, a merging step is needed to selectively combine atom clusters into desirable clusters.
%

\subsection{Merging Atom Clusters}

To merge the selected atom clusters to obtain a better clustering result, we treat each
core object as the representative object of the atom cluster that it belongs to.
The problem is thus transformed from merging atom clusters to merging core objects.
To do so, we define the KNN graph as follows,

\begin{definition} \label{def:KNNGraph}
A \textbf{KNN graph} $G_K=(V,E)$, is an undirected graph, where
$V$ consists of all objects in the given dataset and $E=\{\langle
u,v\rangle |u\in N(v)\land v \in N(u)\}$.
\end{definition}

Therefore, two objects $u$ and $v$ are directly connected in graph $G_K$ if they
are the top $K$ nearest neighbors of each other. Next, we define the notion of
$\alpha$-reachability between vertices in $G_K$.

\begin{definition}
\label{def:alphaReach}
Given $\alpha \in [0,1]$ and $u,v\in V$, if there exists a path $\langle u,w_1,\dots,w_s,v\rangle$ in $G_K$ that satisfies $\rho^*(w_i )>\alpha, i=1, 2,\ldots s$, then
\textbf{$v$ is $\alpha$-reachable from $u$}, denoted by $u\xrightarrow{\alpha}v$.
\end{definition}

Clearly, $\alpha$-reachability is reflexive, symmetric, and transitive for the core object set $O$ in $G_K$. Thus, it can be used as an equivalence relation among the core objects  to divide them into equivalence classes. All atom clusters associated
with core objects in the same equivalent class are merged into a single cluster.
Alternatively, we can view $\alpha$-reachability as a way to prune edges in
$G_K$. Core objects in the same equivalence class reside in the same connected component.

The choice of $\alpha$ is expected to affect the partition of equivalence classes. When $\alpha$ is small, say $\alpha = 0$, most core objects are merged together. On the other hand, when $\alpha$ is large, in the extreme case when $\alpha = 1$, every core object forms an equivalence class and atom clusters are the final clusters. The property and the selection of $\alpha$ will be analyzed in Section~\ref{ssec:ParaAnal}.
\subsection{Algorithm Description and Complexity Analysis}

\begin{algorithm}[!htb]
\caption{\small RECOME Clustering}
\label{alg:RECOME}
\KwIn{$V, K, \alpha$}
\KwOut{Clusters}
\For{$u\in V$}
{
	Compute the KNN set $N(u)$ of $u$\;
	$\rho(u)=\theta\sum\limits_{v\in N(u)}\text{exp}\left(-\frac{\text{d}(u, v)}{\sigma}\right)$\;
}
\For{$u\in V$}
{
	$\rho^*(u)=\frac{\rho(u)}{\max\limits_{v\in N(u)\cup\{u\}}\{\rho(v)\}}$\;
}
Let $O = \{u|u\in V,\rho^*(u)=1\}$\;
\For{$u\in V\backslash O$}
{
	$\pi(u)=\mathop{\arg\min}\limits_{\rho(v)>\rho(u), v\in V}\{\text{d}(u, v)\}$\;
}
Construct the directed graph $\mathcal{G}=(V,A)$, where $A = \{\langle u, \pi(u)\rangle|u\in V\backslash O\}$\;
\For{$o \in O$}
{
	Find the atom cluster $C_o$ w.r.t. $o$ by $C_o=\{o\}\cup\{u\in V\backslash O | \mbox{$u$ is connected to $o$ in $\mathcal{G}$}\}$\;
}
Construct the undirected KNN graph $G_K=(V,E)$, where $E=\{\langle u,v\rangle |u\in N(v)\land v \in N(u)\}$\;
Merge the core objects and their corresponding atom clusters by $\alpha$-reachable relation in $G_K$ (see Definition \ref{def:alphaReach}). Then get the final clustering result\;
\end{algorithm}

To this end, we have presented the four main steps of the proposed RECOME clustering algorithm, i.e., computing NKD, calculating RNKD, discovering atom clusters, and merging atom clusters. Algorithm~\ref{alg:RECOME} summarizes the details of RECOME.

Let the number of objects be $n = |V|$. Computing the KNN set is the most time-consuming step with computational complexity of $\mathcal{O}(n^2+Kn\log n)$. Note that this step can be performed in parallel and accelerated using indexing structures such as kd-tree \cite{Brown2014Building} and R*-tree \cite{beckmann1990r}. Both $\rho^*(u)$ and $\pi(u)$ can be obtained from its $K$ nearest neighbors in $\mathcal{O}(K)$ time. After obtaining the KNN set $N(u)$ and $\rho^* (u)$ for each $u$, the KNN graph $G_K$
can be built by a linear scan of $N(u)$.  Merging core objects can
be accomplished using depth-first-search on $G_K$. The computation complexity of Algorithm 1 is thus $\mathcal{O}(n^2+Kn\log n)$. Specifically, when $K$ is fixed, the KNN sets need to be computed only once and thus the time complexity will reduce to $\mathcal{O}(Kn)$ for a different $\alpha$. The space complexity of RECOME is $\mathcal{O}(Kn)$ because it needs not to store the distance matrix.

\section{Fast Jump Discontinuity Discovery for Parameter $\alpha$}
\label{sec:FJDD}

As evident from the description of RECOME, parameter $\alpha$ is crucial to the clustering result for a fixed $K$. Take the example in Figure~\ref{fig:alphIncr}. Starting from the same set of core objects and atom clusters, different values of $\alpha$ may result in different numbers of clusters. In particular, as $\alpha$ increases, cluster granularity (i.e., the volume of clusters) decreases and cluster purity increases. Thus, a pertinent question is how to select a proper $\alpha$. For ease of presentation, we suppose parameter $K$ is fixed in this section.


\subsection{Problem Formalization}

\begin{figure}[!htb]
\centering
\includegraphics[width=1\textwidth]{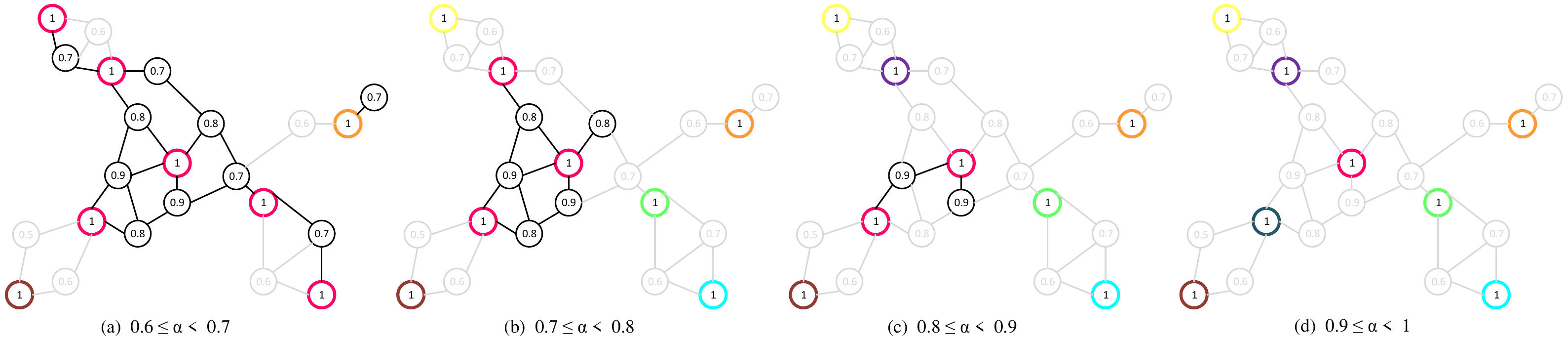}
\caption{Example of merging core objects results for different values of $\alpha$. KNN graph is shown and each object (in circle) is marked with its relative density. Core objects of the same color belong to the same final cluster.}
\label{fig:alphIncr}
\end{figure}

We observe that though $\alpha$ varies continuously in the interval
of $[0,1]$, the number of clusters computed by RECOME is a step function of
$\alpha$.  For instance, in Figure~\ref{fig:alphIncr}, only five clustering
outcomes are possible for the example dataset, corresponding to $\alpha$ in the ranges of
$[0, 0.6)$, $[0.6, 0.7)$, $[0.7, 0.8)$, $[0.8, 0.9)$, and $[0.9, 1)$. The numbers
of resulting clusters are 1, 3, 6, 7, and 8, respectively.  It is desirable to
have a small collection of $\alpha$ values (or ranges) that affect the clustering result as the processes of parameter tuning by developers or parameter selection by domain experts can be simplified.

We formalize the above intuition by first introducing the
notion of jump discontinuity set.

\begin{definition}
\label{defStepSet}
Given a data set $V$ and an input parameter $K$, an ascending list
$L = \{\alpha_1, \alpha_2,\ldots, \alpha_l\}$ is called a \textbf{jump
discontinuity} (JD) set if the number of resulting clusters from RECOME,
$\text{\#}(V,K, \alpha)$ is a step function of $\alpha$ with jump discontinuity
at $\alpha_1, \alpha_2,\ldots, \alpha_l \in [0,1]$ from left to right.
\end{definition}

Obviously, $\text{\#}(V,K, \alpha)$ is a non-decreasing function of
$\alpha$. By definition, each JD in $L$ yields a unique clustering
result. From all the JDs in $L$, we can produce all possible clusters using
RECOME. Recall that $O$ is the set of core objects and $|O|$ is the maximum number
of clusters attainable by RECOME. Trivially, $|L| \le |O|$.

\begin{figure}[!htb]
\centering
\subfloat[{\it Left-open capacity}]{
\label{fig:VertCap}
\includegraphics[width=0.45\textwidth]{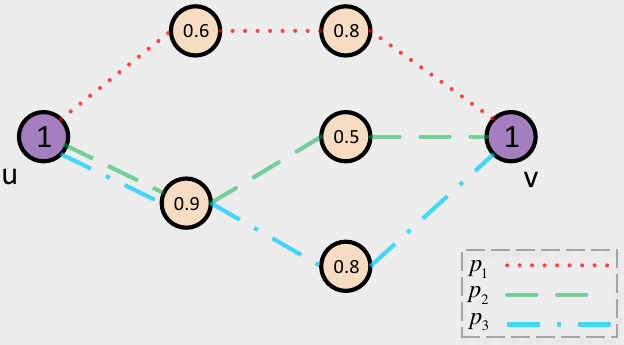}
}
\subfloat[{\it Edge capacity}]{
\label{fig:EdgeCap}
\includegraphics[width=0.45\textwidth]{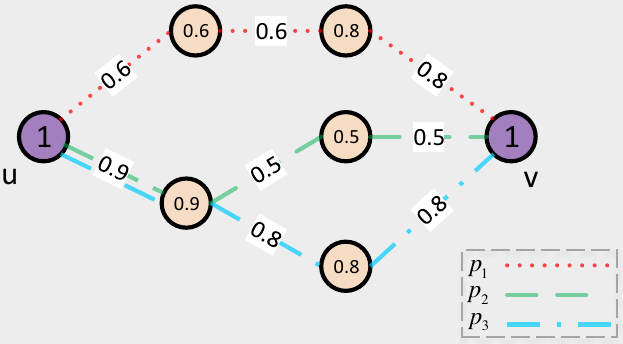}
}
\caption{(a) {\it Left-open capacity}. Each object (in circle) is marked with its relative density. There are two core objects (i.e., $u$ and $v$) and three paths (i.e., $p_1$, $p_2$, and $p_3$) between them. It can be seen that $c(p_1)=0.6$, $c(p_2)=0.5$, and $c(p_3)=0.8$, so we have $c(u,v)=\max\{0.6,0.5,0.8\}=0.8$. (b) {\it Edge capacity}. Each edge is marked with its weight. There are two core objects (i.e., $u$ and $v$) and three paths (i.e., $p_1$, $p_2$, and $p_3$) between them. It can be seen that $c^*(p_1)=0.6$, $c^*(p_2)=0.5$, and $c^*(p_3)=0.8$, so we have $c^*(u,v)=\max\{0.6,0.5,0.8\}=0.8$.}
\end{figure}


%

As described in Section~\ref{sec:RECOME}, the clustering number is only related to the merging of core objects, so we only need to discuss the effect of $\alpha$ on the merging process. Recall that $G_K$ is the KNN graph of dataset $V$. Without loss of generality, we suppose $G_K$ is connected. Consider a path $p = \langle u\equiv w_0, w_1, w_2, \ldots, w_s\equiv v\rangle$. Its {\it left-open path capacity} is defined as $c(p) =\min_{i>0}{\rho^*(w_i)}$. Suppose there are $l$ paths from $u$ to $v$, denoted by $p_1, p_2, \ldots, p_L$. The {\it left-open capacity} between $u$ and $v$ is defined as $c(u,v) = \max_{1\le l\le L}c(p_l)$. An example is shown in Figure~\ref{fig:VertCap}. In other words, $v$ is $\alpha$-reachable from $u$ iff $\alpha < c(u, v)$. Furthermore, we have the following proposition.


\begin{proposition}
\label{prop:JDset}
The JD set $L$ equals $\{0\}\cup\{c(u,v)|u,v\in O, u\not=v\}$.
\end{proposition}

\begin{proof}
See appendix.
\end{proof}

Considering that $0$ belongs to $L$ trivially, in the following section, we will focus on finding out $L^*\triangleq\{c(u,v)|u,v\in O, u\not=v\}$.


\subsection{Fast Jump Discontinuity Discovery}

Finding $L^*$ can be reduced to the problem {\it All-Pairs Bottleneck Paths} in vertex weighted graphs \cite{Shapira2011All}. Unfortunately, the state-of-the-art method for it has a time complexity of $\mathcal{O}(n^{2.575})$. To reduce the time complexity, we need to exploit the characteristics of our problem.

To transform our problem, we define the edge weight function $\text{w}(\langle u, v\rangle)=\min\{\rho^*(u),\rho^*(v)\}$ for $\langle u, v\rangle\in E(G_K)$. For a path $p = \langle u\equiv w_0, w_1, w_2, \ldots, w_s\equiv v\rangle$, its {\it path edge-capacity} is defined as $c^*(p) =\min_{i>0}{\text{w}(\langle w_{i-1}, w_i\rangle)}$. Suppose that there are $l$ paths from $u$ to $v$, denoted by $p_1, p_2, \ldots, p_L$. The {\it edge capacity} between $u$ and $v$ is defined as $c^*(u,v) = \max_{1\le l\le L}c^*(p_l)$. An example is shown in Figure~\ref{fig:EdgeCap}.

\begin{proposition}
\label{prop:tran}
$\forall u, v\in O, u\not=v$, the left-open capacity between $u$, $v$ equals the edge capacity $c^*(u,v)$ in the transformed graph, i.e., $c(u,v)=c^*(u,v)$.
\end{proposition}

%

Proposition \ref{prop:tran} implies that $L^*\triangleq\{c(u,v)|u,v\in O, u\not=v\}=\{c^*(u,v)|u,v\in O, u\not=v\}$. Therefore, it suffices to determine $\{c^*(u,v)|u,v\in O, u\not=v\}$. This problem can be reduced to the problem of {\it All-Pairs Bottleneck Paths} in edge weighted graphs
(edge-APBP) \cite{Shapira2011All}. For edge-APBP, the following property holds:
\begin{thm}
\label{thm:edgeAPBP}
Suppose $G$ is a connected edge weighted graph and $T$ is the \textbf{maximum spanning tree} of $G$. Then, $\forall u,v\in V(G)=V(T),u\not=v$, $c_G^*(u,v)=c_T^*(u,v)$.
\end{thm}
\begin{proof}
See \cite{Shapira2011All}.
\end{proof}
$c_G^*(u,v)$ and $c_T^*(u,v)$ denote the {\it edge capacity} between $u$ and $v$ in $G$ and $T$, respectively.  Suppose the maximum spanning tree of $G_K$ is $T_K$, and then according to Theorem~\ref{thm:edgeAPBP}, we can extract $L^*$ from $T_K$ using Algorithm~\ref{algo:extract}.

\begin{algorithm}[h]
\caption{Extracting-JD}
\label{algo:extract}
\KwIn{$T_K$}
\KwOut{$L^*$}
Sort $E(T_K)$ by edge weight decreasingly and suppose the result is $\{e_1,e_2,...,e_{n-1}\}$\;
Set $L^*\leftarrow\emptyset$, $T_K\leftarrow T_K-E(T_K)$ (remove all edges from $T_K$)\;
\For{$i$ from 1 to $n-1$}
{
	Suppose $u$ and $v$ are the two ends of $e_i$\;
	Suppose $C_u$ and $C_v$ are the two components in $T_K$ that contain $u$ and $v$, respectively\;
	\If{both $C_u$ and $C_v$ contain at least one core object}
	{
		$L^*\leftarrow L^* \cup\{\text{w}(e_i)$\}\;
	}
	$T_K\leftarrow T_K+e_i$\;
}
\end{algorithm}

\begin{algorithm}[h]
\caption{Fast Jump Discontinuity Discovery}
\label{algo:FJDD}
\KwIn{KNN graph $G_K$ and core object set $O$}
\KwOut{Sorted JD list $L$}
Define the edge weight function $\text{w}(\langle u, v\rangle)=\min\{\rho^*(u),\rho^*(v)\}$ for $\langle u, v\rangle\in E(G_K)$\;
Compute the maximum spanning tree of $G_K$ using Prim's algorithm and denote it as $T_K$\;
Set $L\leftarrow$Extracting-JD$(T_K)\cup\{0\}$\;
\end{algorithm}

In the Algorithm~\ref{algo:extract}, the sorting step is most time consuming with computational complexity $\mathcal{O}(n\log n)$. The Prim's maximum spanning tree algorithm has a time complexity of $\mathcal{O}(|E|+n\log n)$, where $|E|=\mathcal{O}(Kn)$. Therefore, the computation complexity of Algorithm~\ref{algo:FJDD} is $\mathcal{O}(Kn+n\log n)$. Finally, the Fast Jump Discontinuity Discovery algorithm is summarized in Algorithm~\ref{algo:FJDD}.

\section{Experiments}
\label{sec:Expe}
In this section, we evaluate RECOME over synthetic and real-world datasets, and compare it with six other representative algorithms. We introduce the experiment setup in Section \ref{ssec:ExpeSetu}, and then present the experimental results and analysis in Section \ref{ssec:ExpeResu}. In section \ref{ssec:ParaAnal}, we conduct the parameter analysis of RECOME. Our implementation uses Microsoft Visual C++ 2015 14.0.24720.00, and all experiments are conducted on a workstation (Windows 64 bit, 4 Intel 3.2 GHz processors, 4 GB of RAM). Our code has been released in: {\scriptsize\url{https://github.com/gyla1993/RECOME-A-new-density-based-clustering-algorithm-using-relative-KNN-kernel-density}}.

\subsection{Experiment Setup}
\label{ssec:ExpeSetu}
\subsubsection{Datasets}
\begin{figure}[!htb]
\centering
\includegraphics[width = \linewidth]{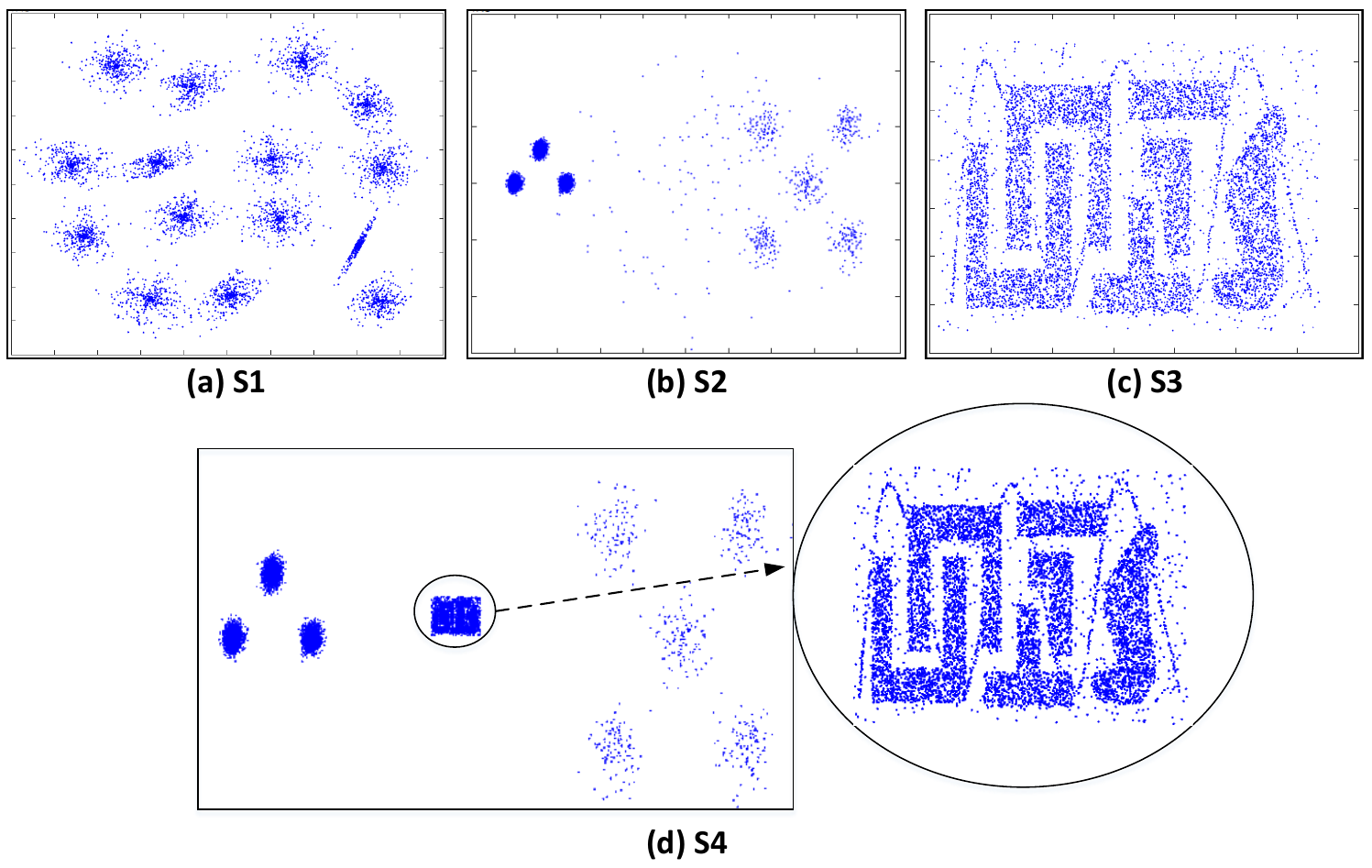}
\caption{Two-dimensional synthetic datasets.}
\label{fig:ExamTwoDim}
\end{figure}

\textit{Two-dimensional synthetic datasets}: Four representative datasets S1, S2, S3, and S4 are used. S1 comprises 5,000 objects and 15 Gaussian clusters. S2 is an unbalanced dataset, which contains 8 classes of different density and size, 6500 objects with 117 noisy objects injected. S3 is comprised of 6 classes of non-convex shape and 8000 objects; S4 is the mixture of S2 and S3, which contains 14 classes of different shapes, densities, and scales. See Figure \ref{fig:ExamTwoDim} for details.

\textit{MNIST training set} \cite{lecun1998gradient}: It is a real-world dataset containing 60,000 examples of handwritten digits from 0 to 9, which has been widely used in data mining and machine learning. We select the subsets of different digits for clustering tasks, including M367 (all examples of digits ``3, 6, 7''), M3467, M0:8 (all examples of digits from 0 to 8), M0:9, sM3467 (small edition of M3467 for the visualization, generated by randomly selecting 900 examples of each digit), and sM0:8.

\textit{Olivetti face database} \cite{samaria1994parameterisation}: It is a real-world dataset containing 40 subjects of human face and each contains 10 examples. This dataset poses a challenge to the density estimation since the real number of clusters is comparable with the number of objects in each cluster (10 different pictures for each people). Two subsets, face10 (the examples of first 10 subjects) and face40 (the examples of all 40 subjects), have been selected for the evaluation.

\subsubsection{Baseline Methods and Settings}

We select 6 representative density-based clustering algorithms as baseline methods, i.e., DBSCAN \cite{ester1996density}, SNN \cite{Levent2003Finding}, KNNC \cite{tran2006knn}, FDP \cite{rodriguez2014clustering}, 3DC \cite{Liang2016delta}, STClu \cite{Wang2016Automatic}. Since many methods share a common density measure, in the implementation, we consider the two density measures.

(i) $\epsilon$ cut-off density. The parameter $\epsilon$
is set to the value of the top $\beta$ percent distance
among all object pairs. We vary
$\beta$ value in $\{1, 2, 3, 4, 5, 6, 7, 8, 9, 10\}$ according to the recommendation in \cite{rodriguez2014clustering}.

(ii) Density based on $K$ nearest neighbors. The parameter $K$ is searched from $(0,\sqrt{|V|}]$ with a step size of 10\% of the range, where $V$ is the number of objects. This range covers the recommended parameters of different clustering methods and is shown to provide a good trade-off between evaluating clustering performance and computation time.

The detailed settings of all methods are listed in TABLE \ref{tab:methSett}.

\begin{table}[htb]
\caption{The detailed settings of all methods.}
\renewcommand\arraystretch{1.1}
\centering
\scalebox{0.77}{
\begin{tabular}{c|c|p{0.8\textwidth}}
\hline
Methods & Density measure & Other settings\\\hline
DBSCAN \cite{ester1996density}	&	(i)	& The parameter $MinPts$ is determined by $\lambda \cdot \bar{\rho}$, where $\lambda$ is searched in $\{ 0.05, 0.1, 0.2, 0.4, 0.8, 1.6\}$ and $\bar{\rho}$ is the mean density. Each outlier is assigned to its nearest cluster.\\\hline
SNN	\cite{Levent2003Finding}	&	(ii) & For remained parameters $\epsilon$ and $MinPts$, both of them are set to $\lambda K$,  where $\lambda$ is chosen from $\{0.2, 0.4, 0.5, 0.6, 0.8\}$, according to the recommendation in \cite{Levent2003Finding}.\\\hline
KNNC \cite{tran2006knn}	&	(ii)	& No other parameters.\\\hline
FDP	\cite{rodriguez2014clustering}	&	(i)	& The objects with top $C_t$ gamma values are chosen as cluster centers, where $C_t$ is the real class number.\\\hline
3DC	\cite{Liang2016delta}	&	(i)	& No other parameters.\\\hline
STClu \cite{Wang2016Automatic}	&	(ii)	& No other parameters.\\\hline
RECOME	&	(ii)	& For parameter $\alpha$, enumerate all JD values extracted by \textit{FJDD} algorithm. Euclidean distance is specified as the distance function.\\\hline
\end{tabular}}
\label{tab:methSett}
\end{table}

%
%
%
%
%
%

The best performance is reported for all methods by searching through the afore-mentioned parameter space.

\subsubsection{Performance Metrics}

For the two-dimensional synthetic datasets, due to the lack of ground truth labels, we compare clustering results visually. For real-world datasets, two metrics are calculated based on the ground truth of the datasets. One is the normalized mutual information (NMI) \cite{Strehl2002Cluster}, which is one of the most widely used measures of clustering quality. Given a dataset $V$ of size $n$, suppose there are $C$ clusters and $C_t$ actual classes. Let $n_i$, $n^{(j)}$ and $n_i^{(j)}$ denote the number of objects in cluster $i$, actual class $j$, and both cluster $i$ and actual class $j$, respectively. Then, NMI can be computed by
$$
\text{NMI}=\dfrac{\sum_{i=1}^{C}\sum_{j=1}^{C_t}{\frac{n_i^{(j)}}{n}\log{\frac{n n_i^{(j)}}{n_i n^{(j)}}}}}{\sum_{i=1}^{C}{\frac{n_i}{n}\log\frac{n_i}{n}}\sum_{j=1}^{C_t}{\frac{n^{(j)}}{n}\log\frac{n^{(j)}}{n}}}.
$$

The other metric is the F value. Given a dataset $V=\{v_1,v_2,\dots,v_n \}$ with  cluster labels $\{c_1, c_2, \dots, c_n\}$ and actual class labels $\{l_1, l_2, \dots, l_n\}$, define
$$
\text{Correctness}(v_i,v_j )=
\begin{cases}
1& \text{if $l_i=l_j\Leftrightarrow c_i=c_j$}\\
0& \text{otherwise}
\end{cases}.
$$
Then, define $PB$ and $RB$ as:
$$PB=\frac{1}{n}\sum_{i=1}^n\frac{ \sum\limits_{\scriptscriptstyle i\not=j, c_i=c_j}\text{Correctness}(v_i, v_j)}{|\{v_j|i\not=j, c_i=c_j\}|},$$
$$RB=\frac{1}{n}\sum_{i=1}^n\frac{ \sum\limits_{\scriptscriptstyle i\not=j, l_i=l_j}\text{Correctness}(v_i, v_j)}{|\{v_j|i\not=j, l_i=l_j\}|}.$$
$PB$ and $RB$ refer to precision b-cubed and recall b-cubed \cite{han2011data}, respectively. Then F value is computed by $F = \frac{2\times PB\times RB}{PB\times RB}$. We use this measure because $PB$ and $RB$ have been shown superior than other indices \cite{Amig2009A}.

Both NMI and F fall in $[0, 1]$, and a higher value denotes better clustering performance.

%
%

\subsection{Experimental Results}
\label{ssec:ExpeResu}
\subsubsection{Results on Synthetic Dataset}

\begin{figure}[!thb]
\centering
\begin{minipage}{0.75\linewidth}
\subfloat[DBSCAN]{
  \label{fig:DBSCANsyn}
\includegraphics[width = 0.225\linewidth]{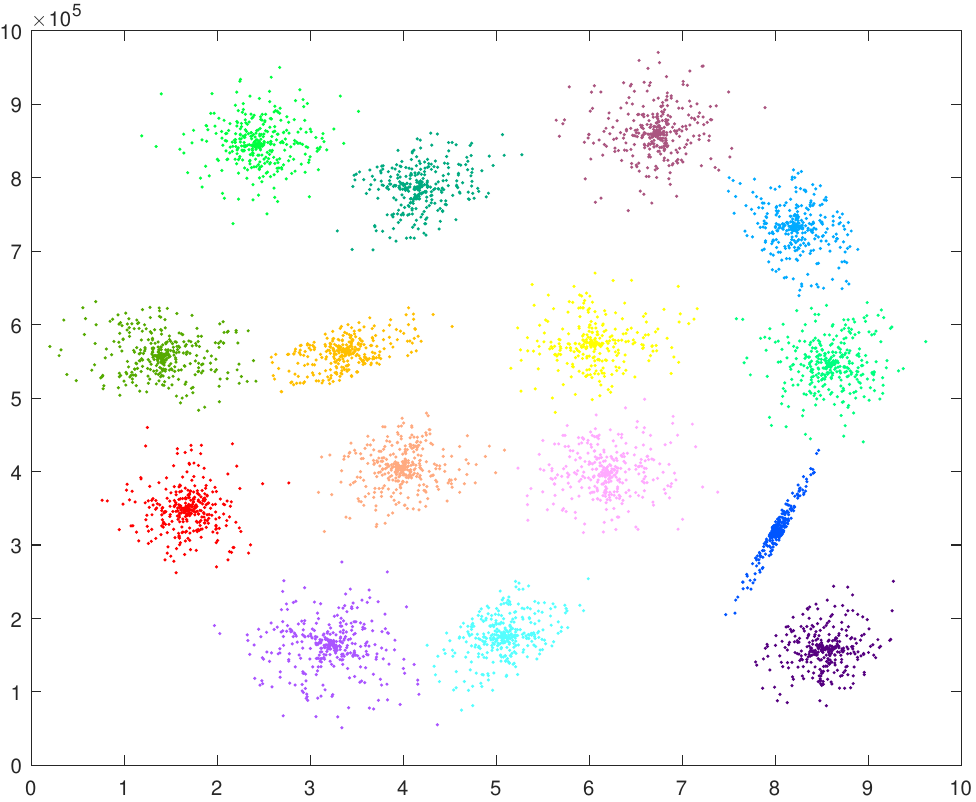}\
\includegraphics[width = 0.225\linewidth]{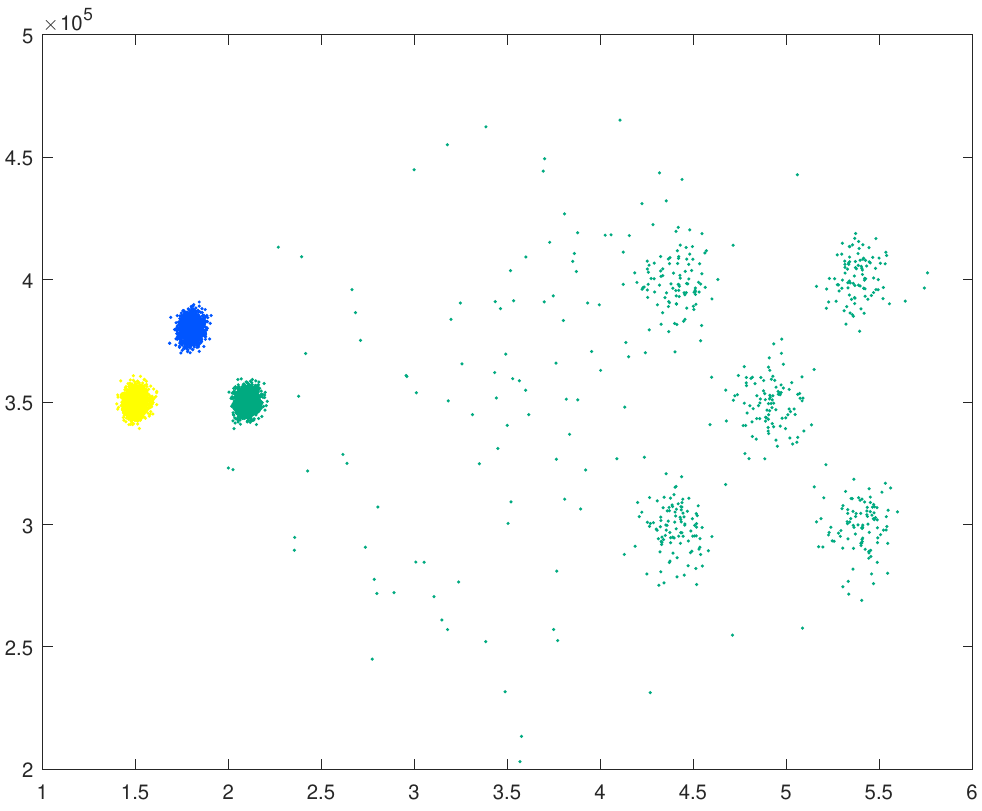}\
\includegraphics[width = 0.23\linewidth]{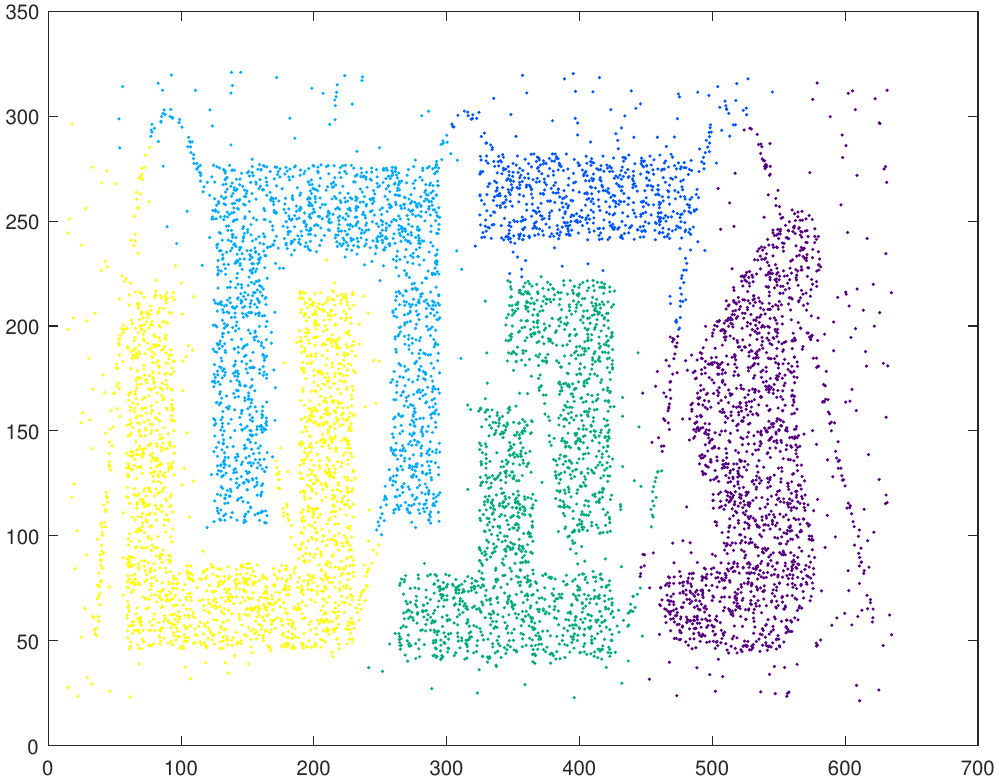}\
\includegraphics[width = 0.225\linewidth]{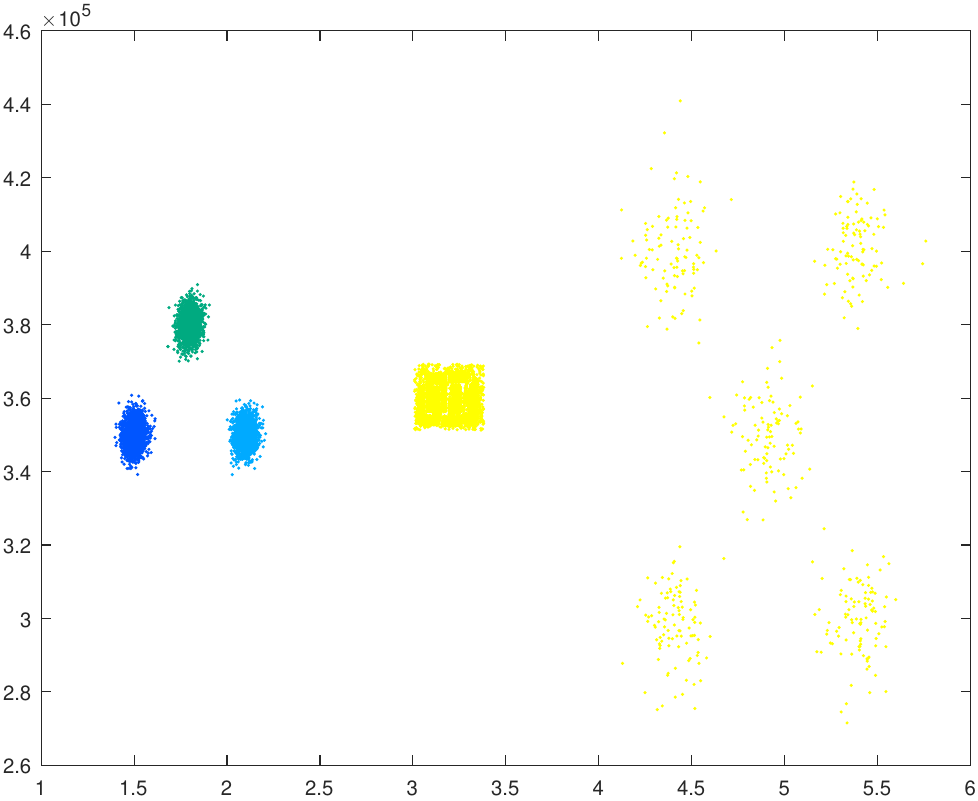}
}\
\subfloat[SNN]{
  \label{fig:SNNsyn}
\includegraphics[width = 0.225\linewidth]{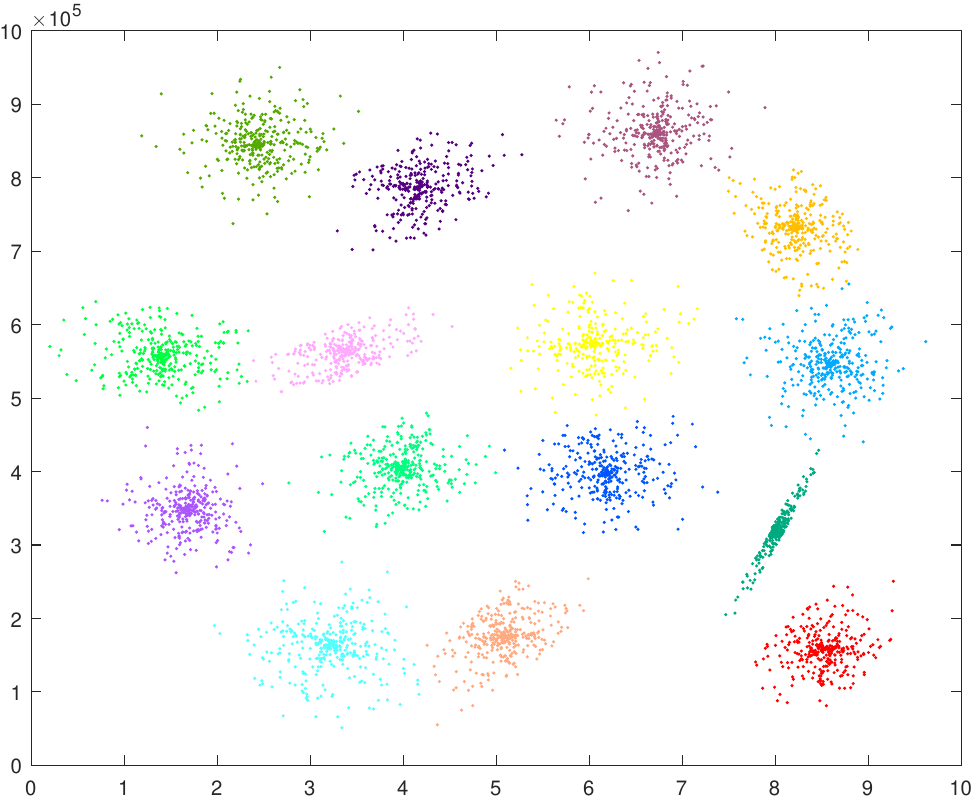}\
\includegraphics[width = 0.225\linewidth]{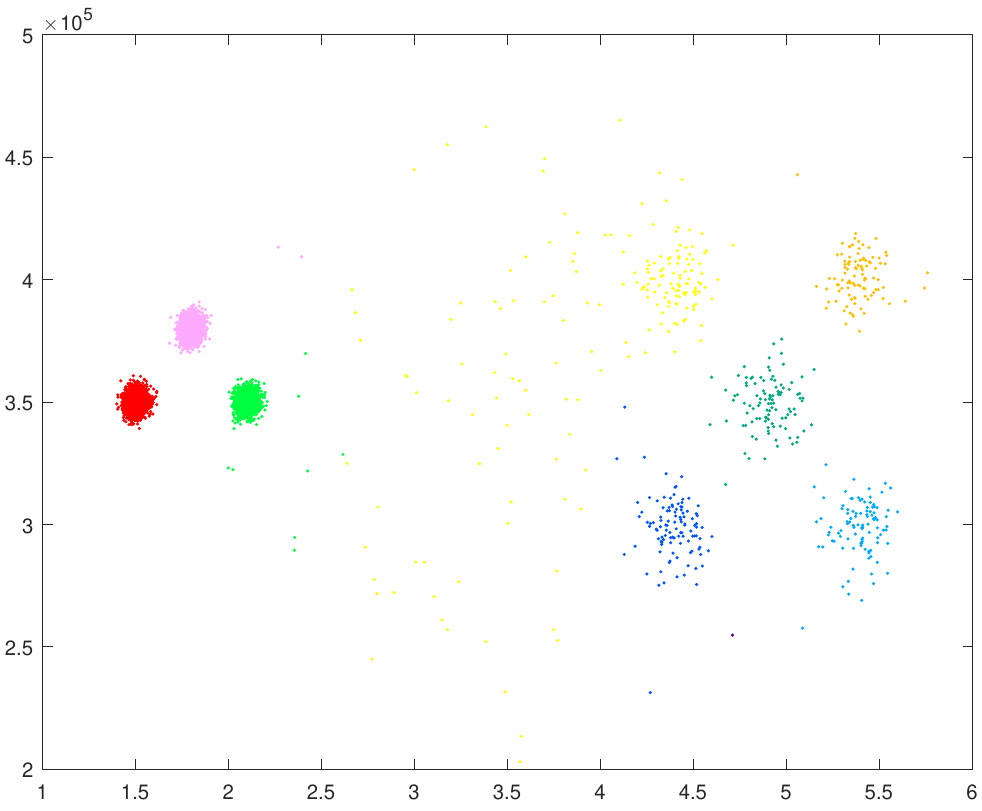}\
\includegraphics[width = 0.23\linewidth]{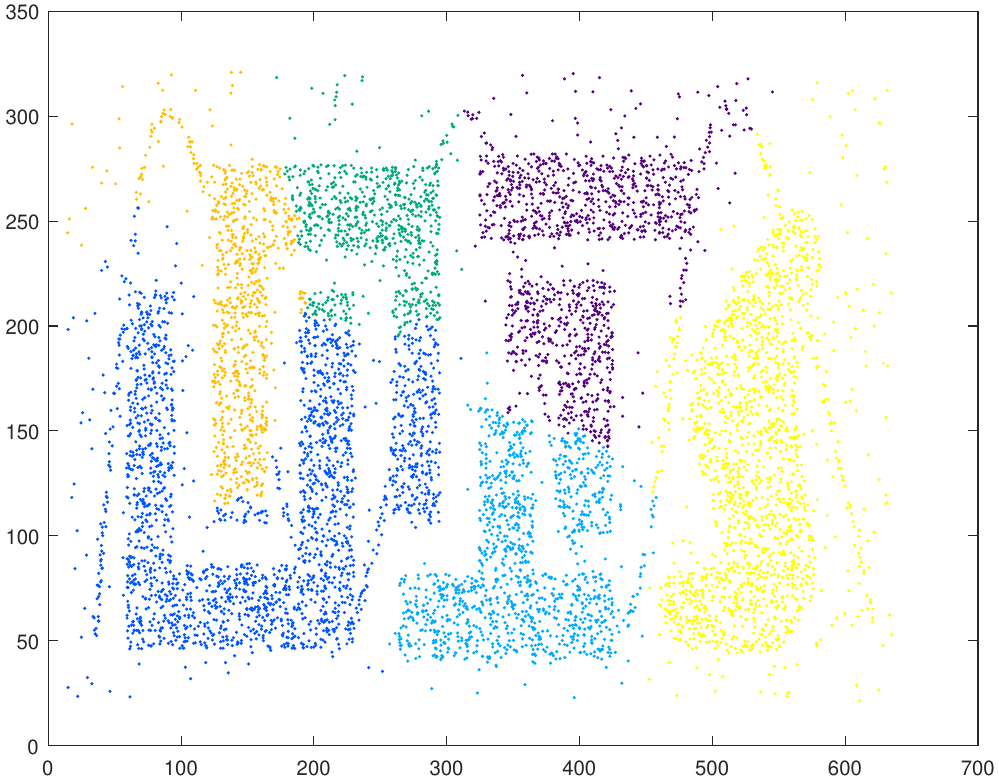}\
\includegraphics[width = 0.225\linewidth]{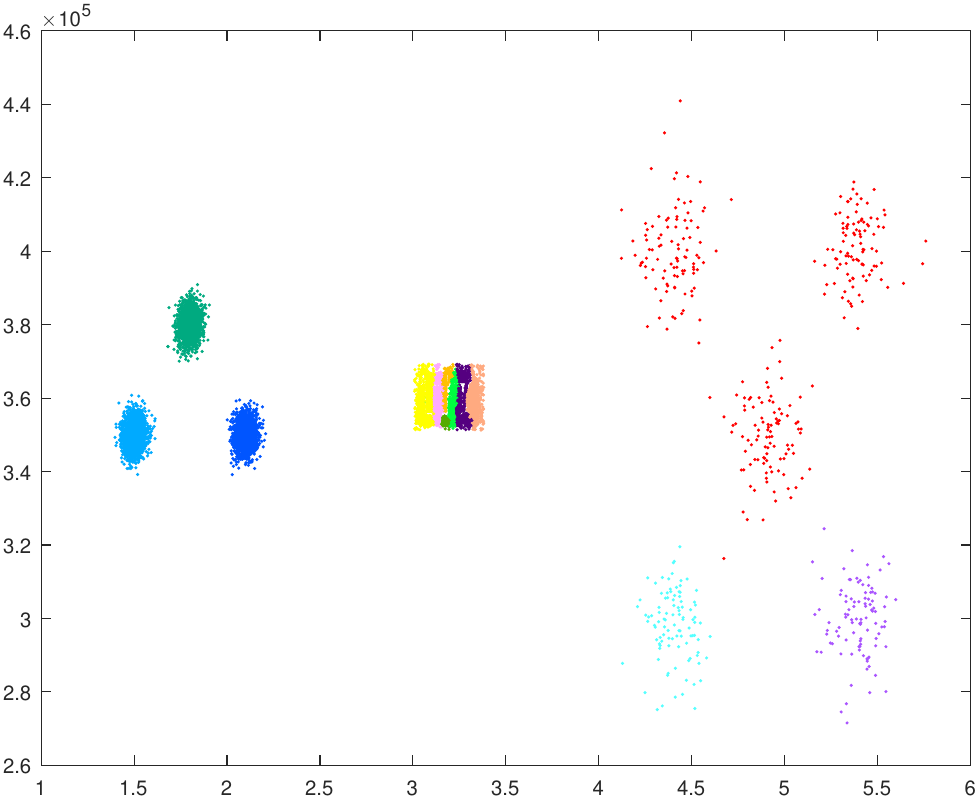}
}\
\subfloat[KNNC]{
  \label{fig:KNNCsyn}
\includegraphics[width = 0.225\linewidth]{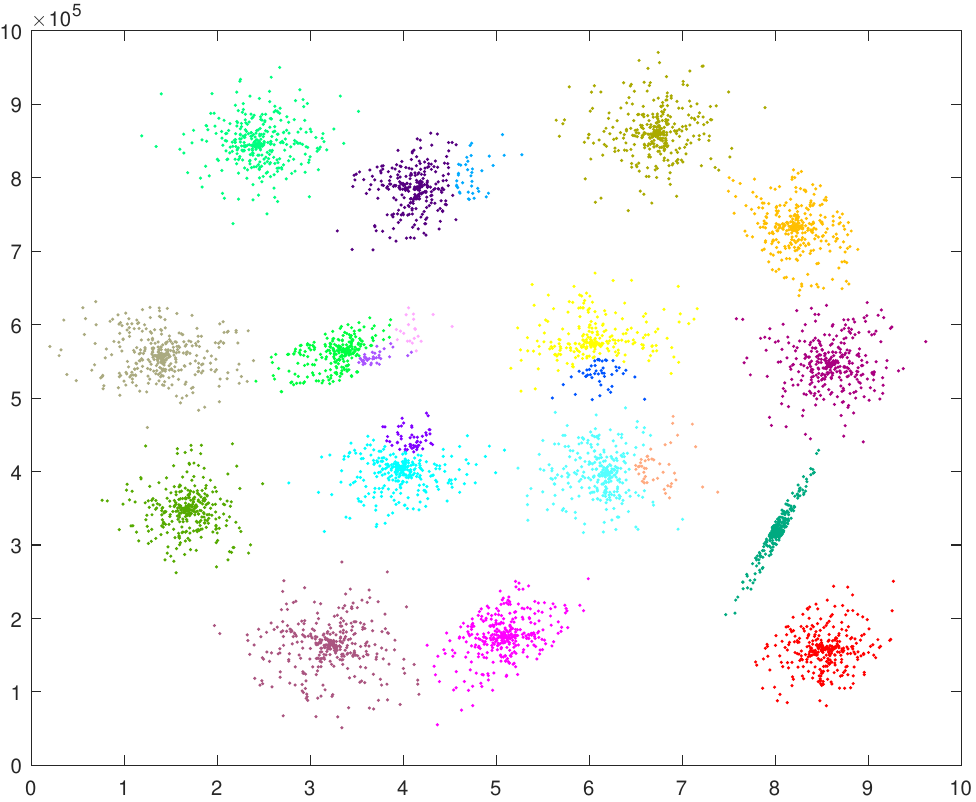}\
\includegraphics[width = 0.225\linewidth]{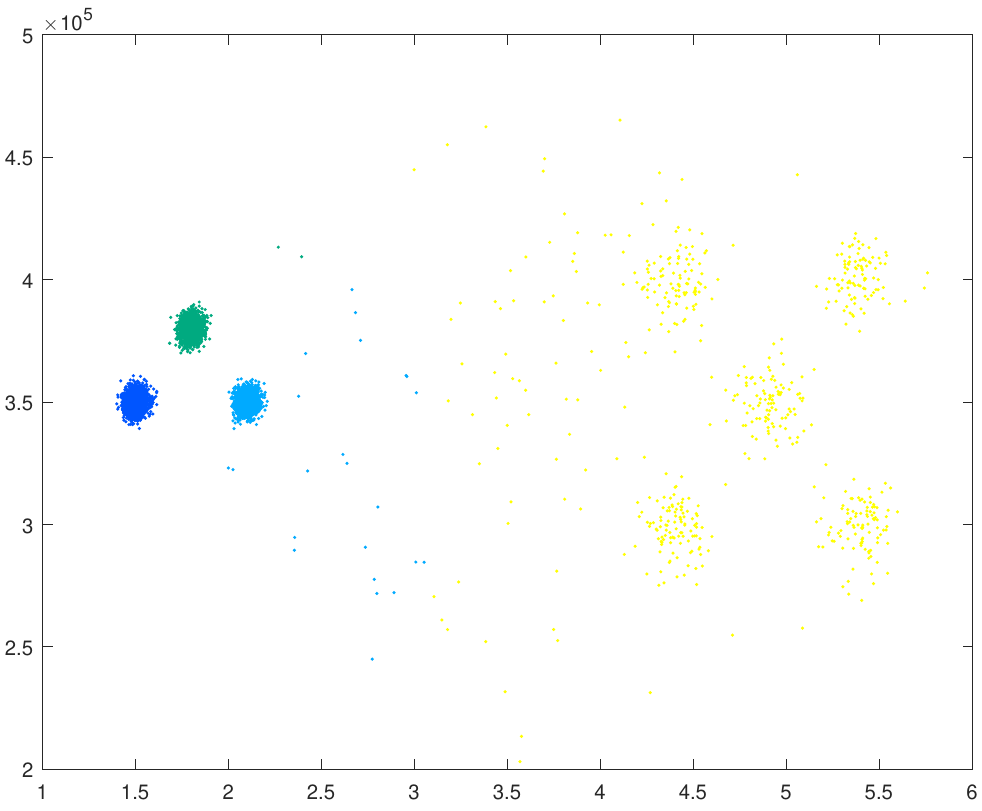}\
\includegraphics[width = 0.23\linewidth]{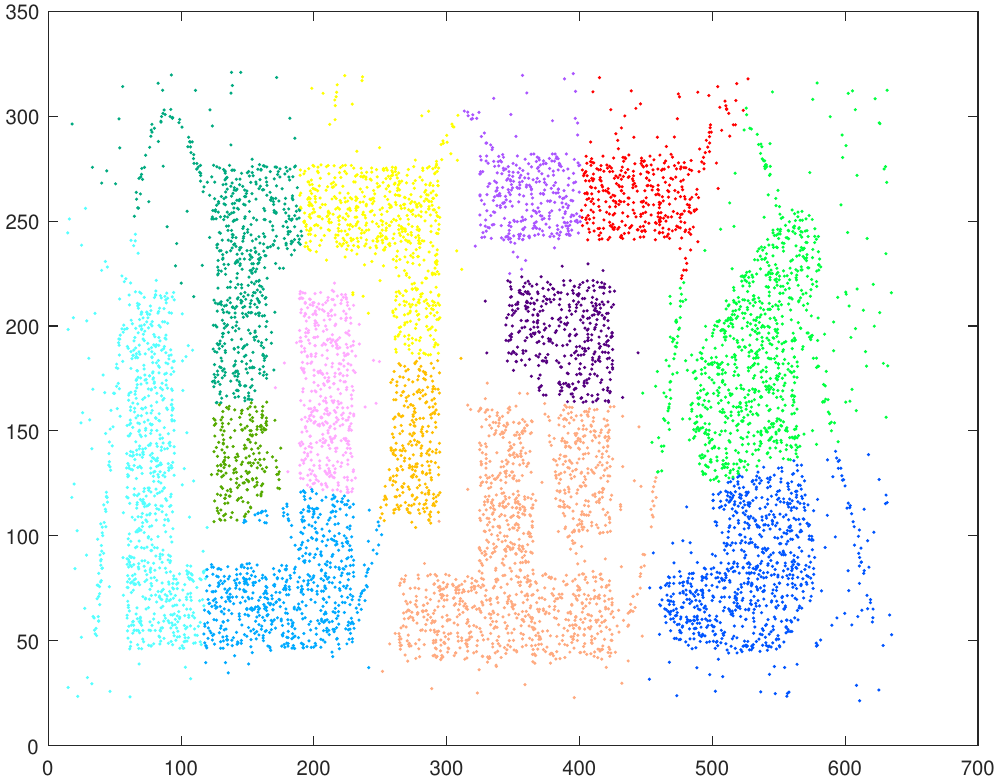}\
\includegraphics[width = 0.225\linewidth]{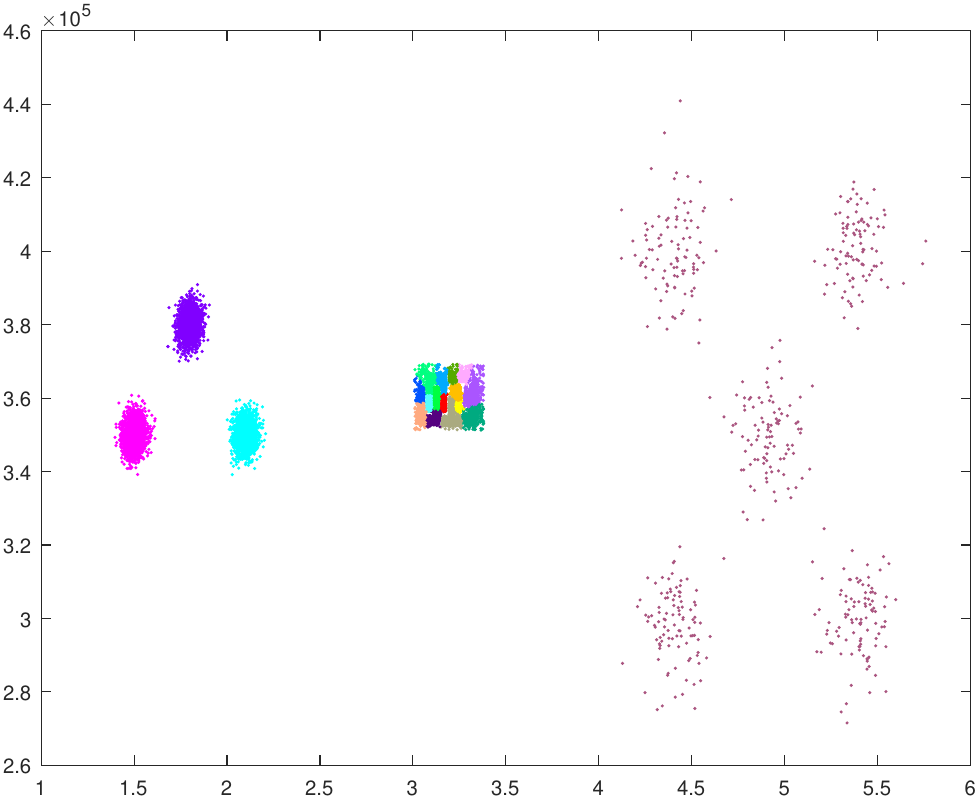}
}\
\subfloat[FDP]{
  \label{fig:FDPsyn}
\includegraphics[width = 0.225\linewidth]{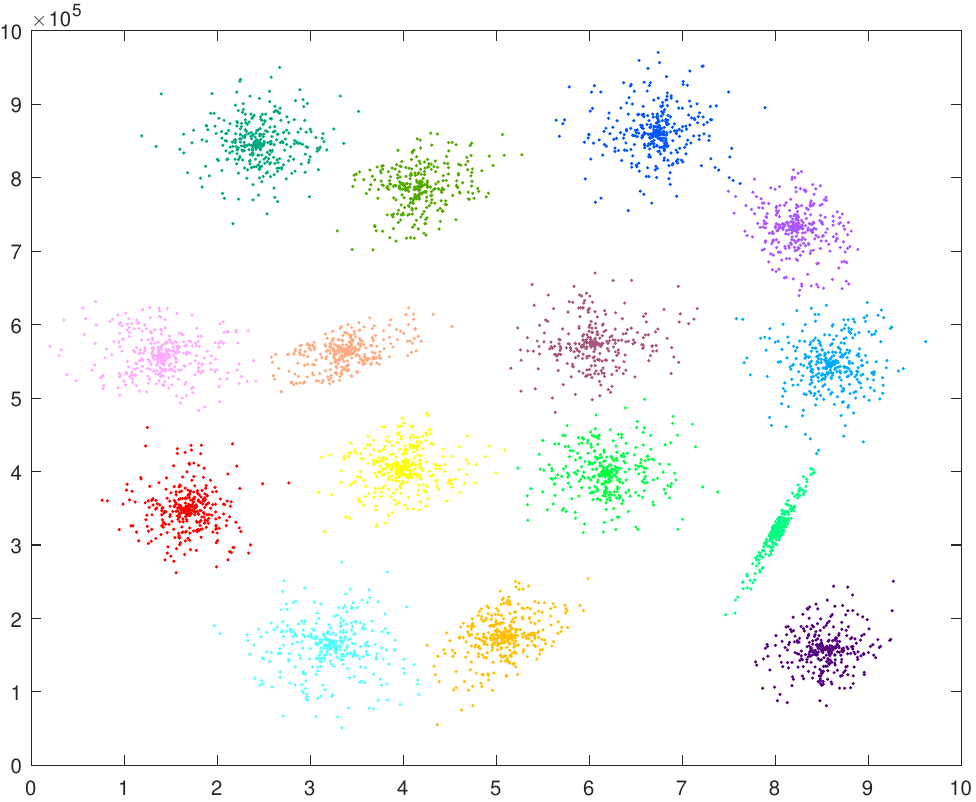}\
\includegraphics[width = 0.225\linewidth]{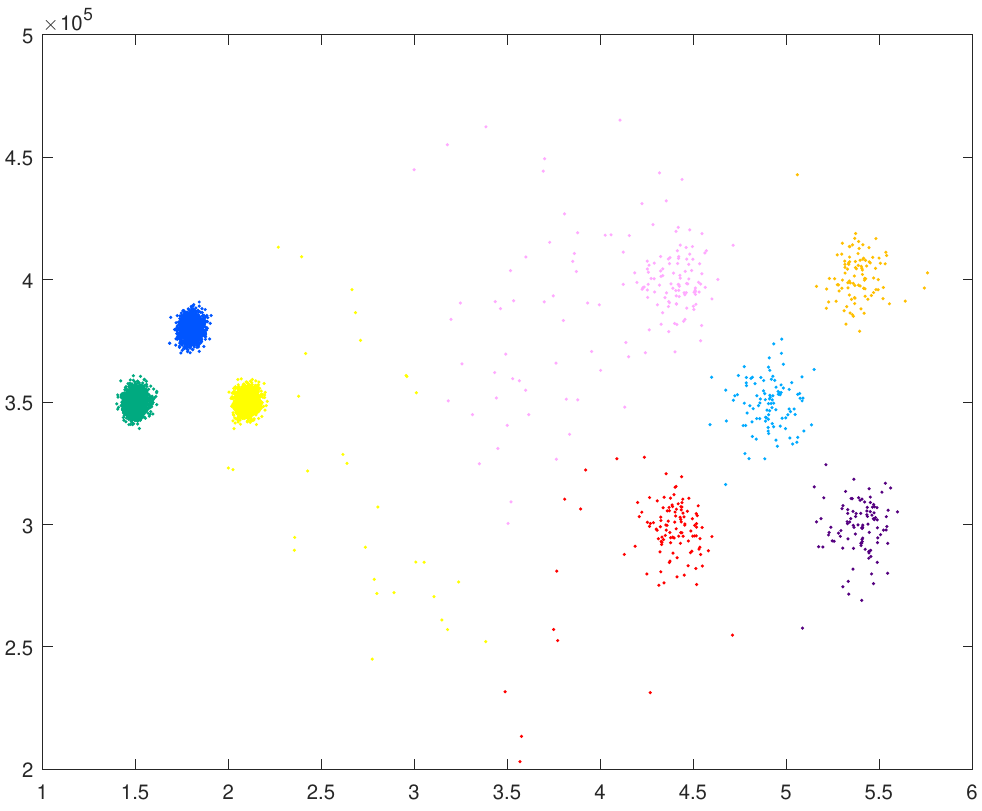}\
\includegraphics[width = 0.23\linewidth]{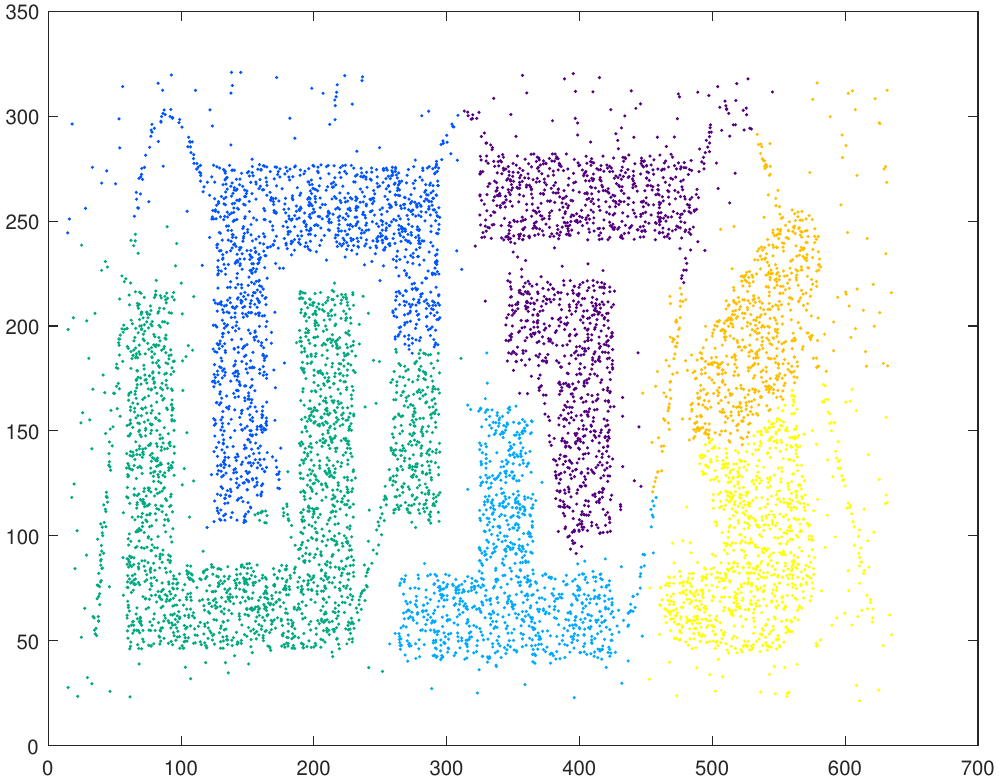}\
\includegraphics[width = 0.225\linewidth]{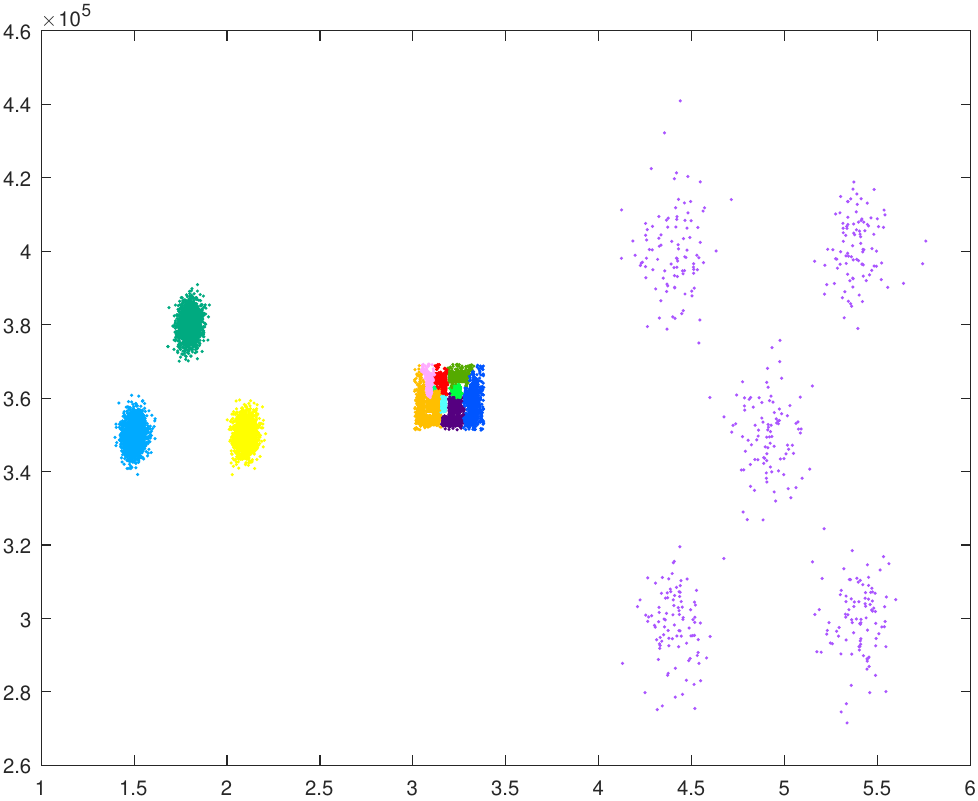}
}\
\subfloat[3DC]{
  \label{fig:3DCsyn}
\includegraphics[width = 0.225\linewidth]{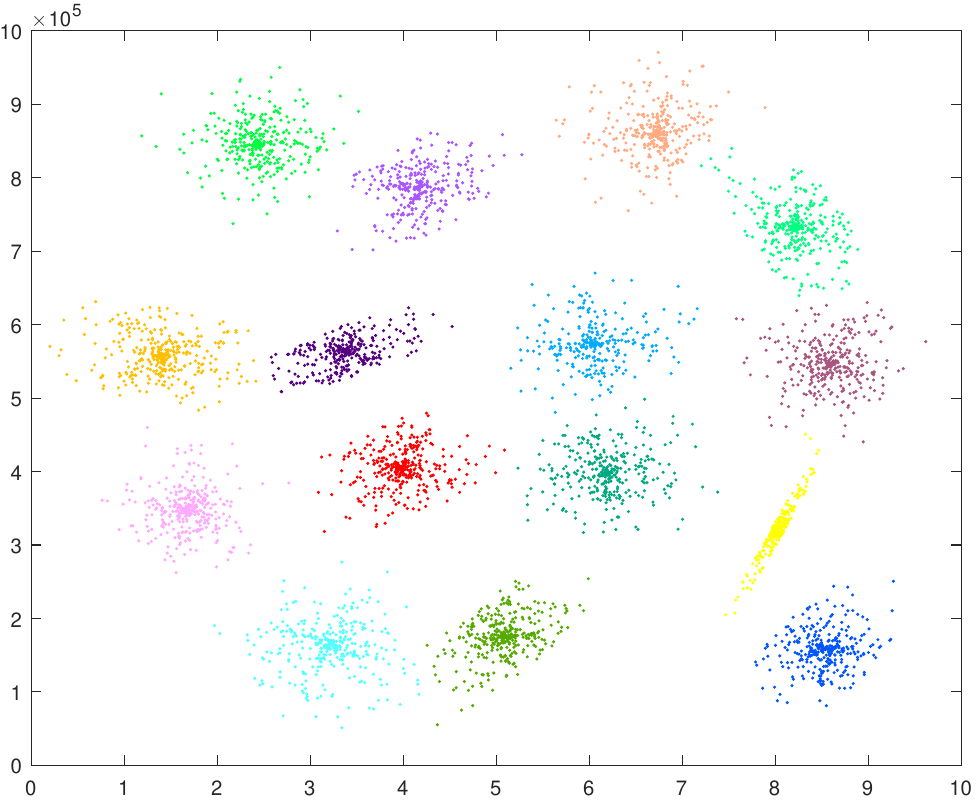}\
\includegraphics[width = 0.225\linewidth]{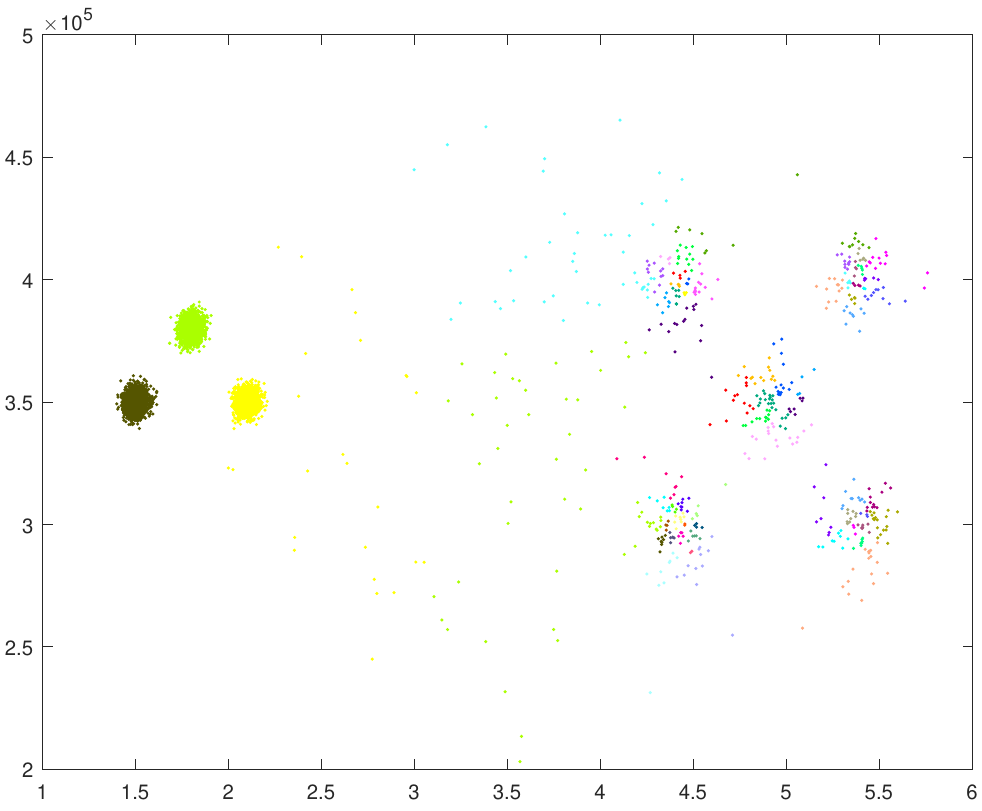}\
\includegraphics[width = 0.23\linewidth]{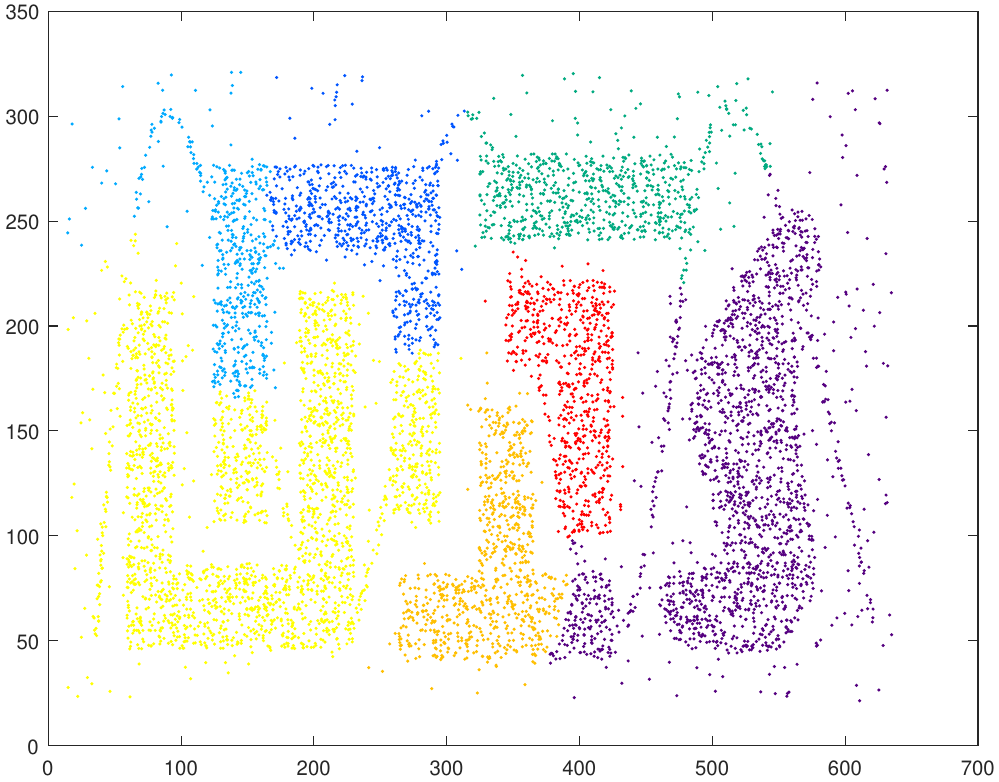}\
\includegraphics[width = 0.225\linewidth]{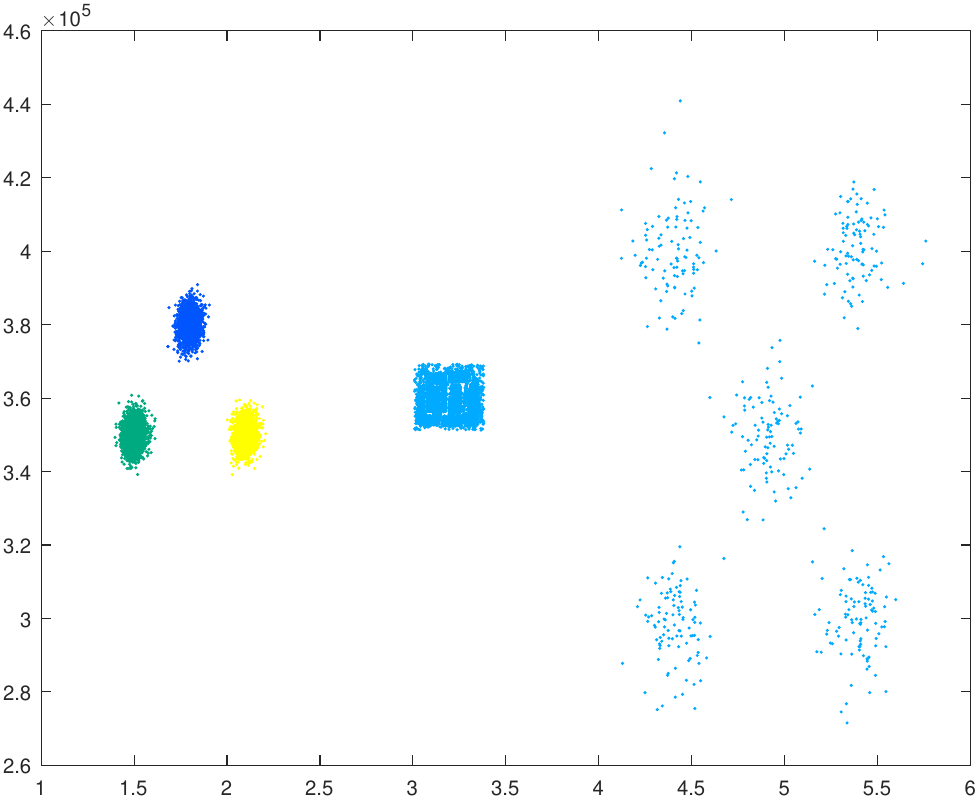}
}\
\subfloat[STClu]{
  \label{fig:STClusyn}
\includegraphics[width = 0.225\linewidth]{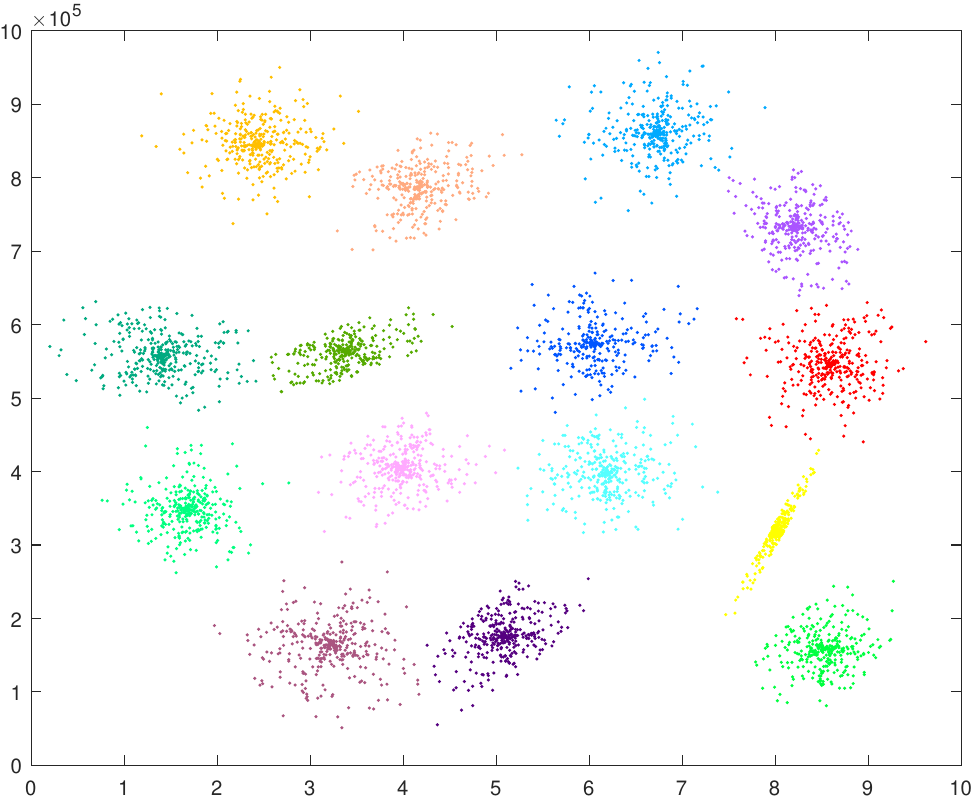}\
\includegraphics[width = 0.225\linewidth]{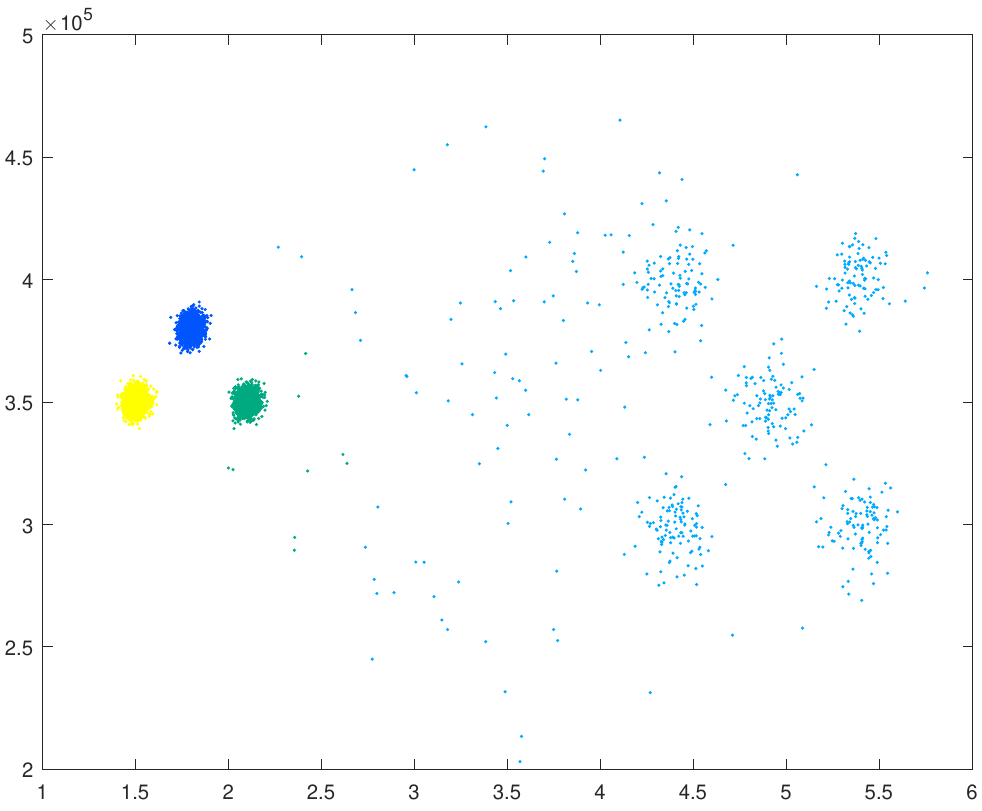}\
\includegraphics[width = 0.23\linewidth]{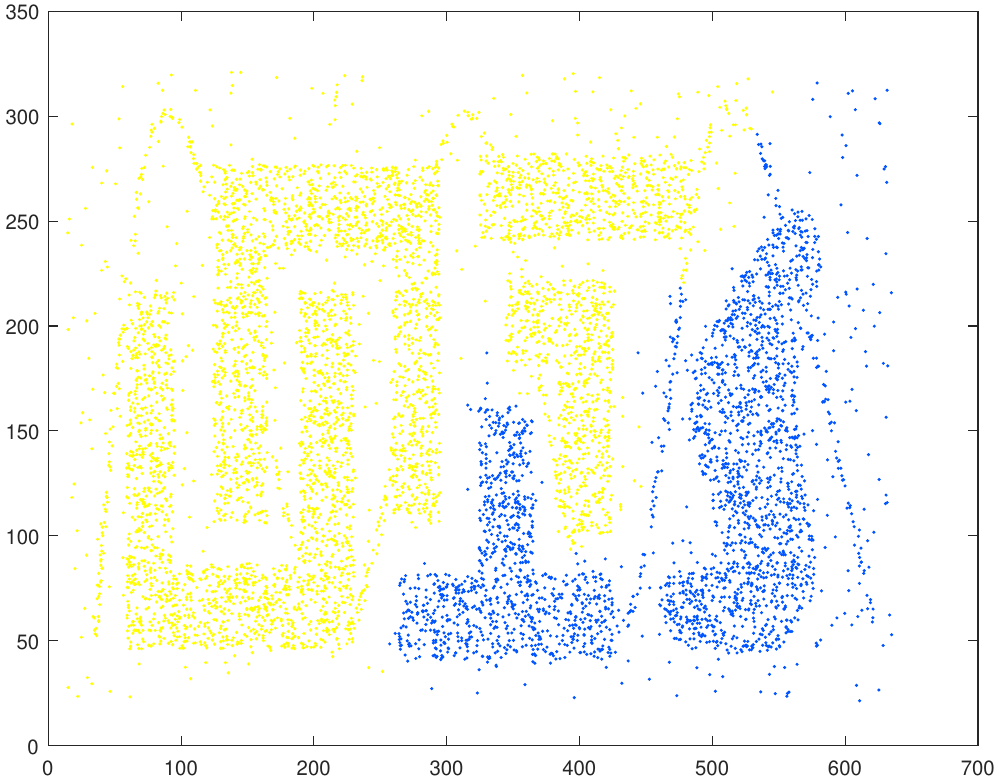}\
\includegraphics[width = 0.225\linewidth]{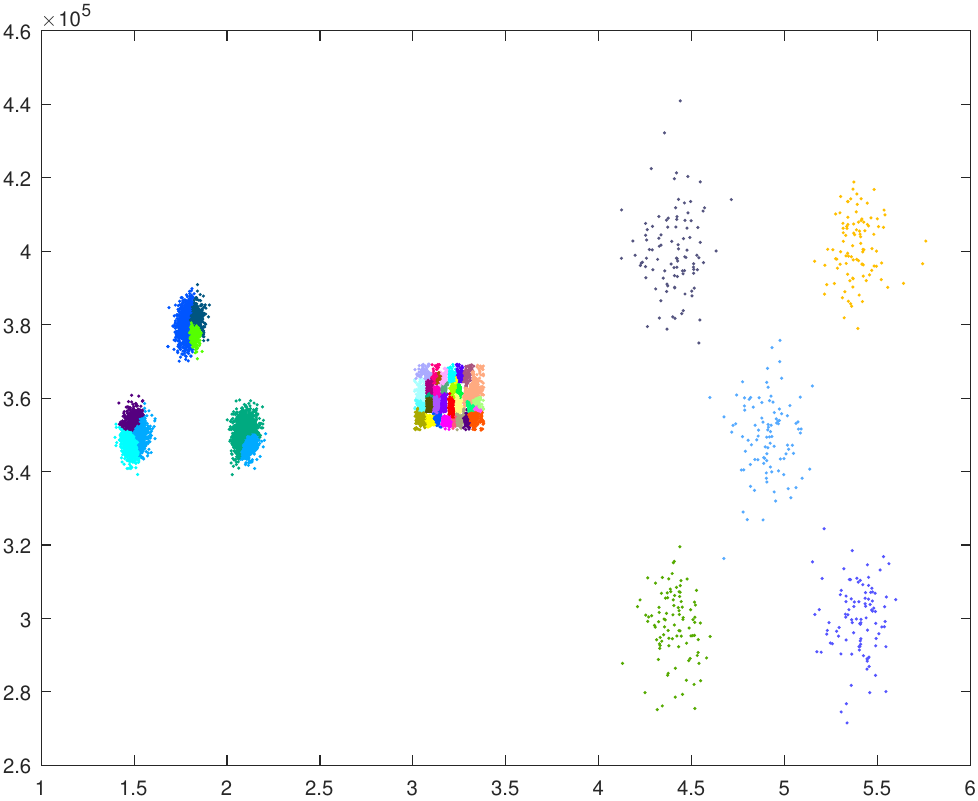}
}\
\subfloat[RECOME]{
  \label{fig:RECOMEsyn}
\includegraphics[width = 0.225\linewidth]{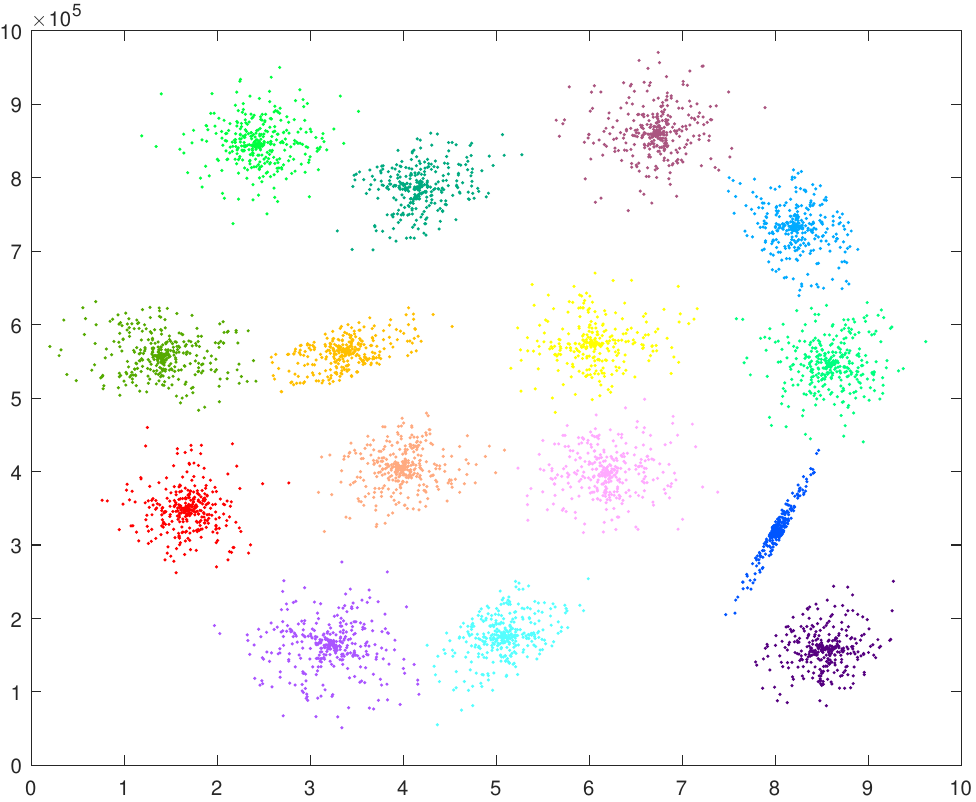}\
\includegraphics[width = 0.225\linewidth]{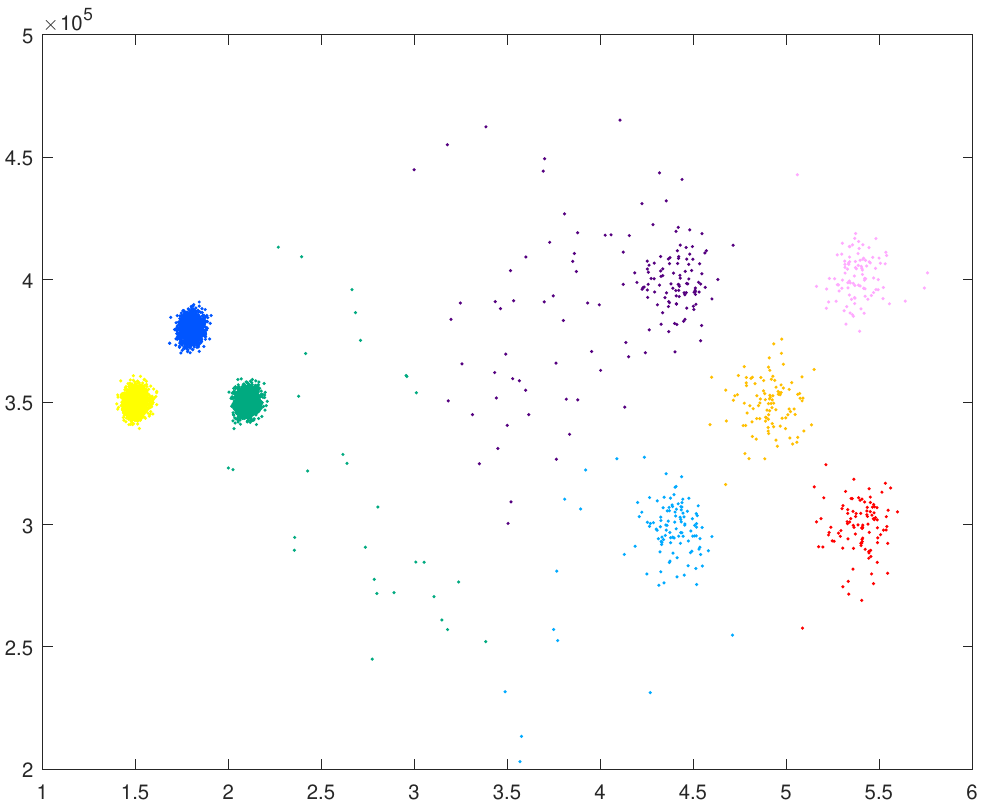}\
\includegraphics[width = 0.23\linewidth]{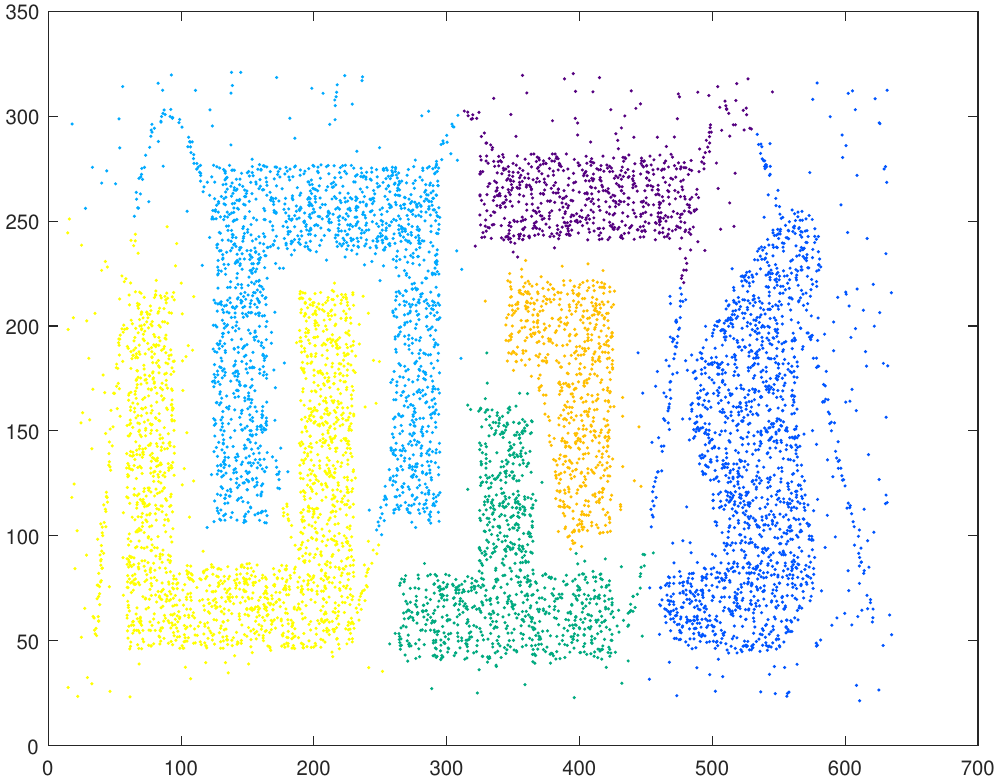}\
\includegraphics[width = 0.225\linewidth]{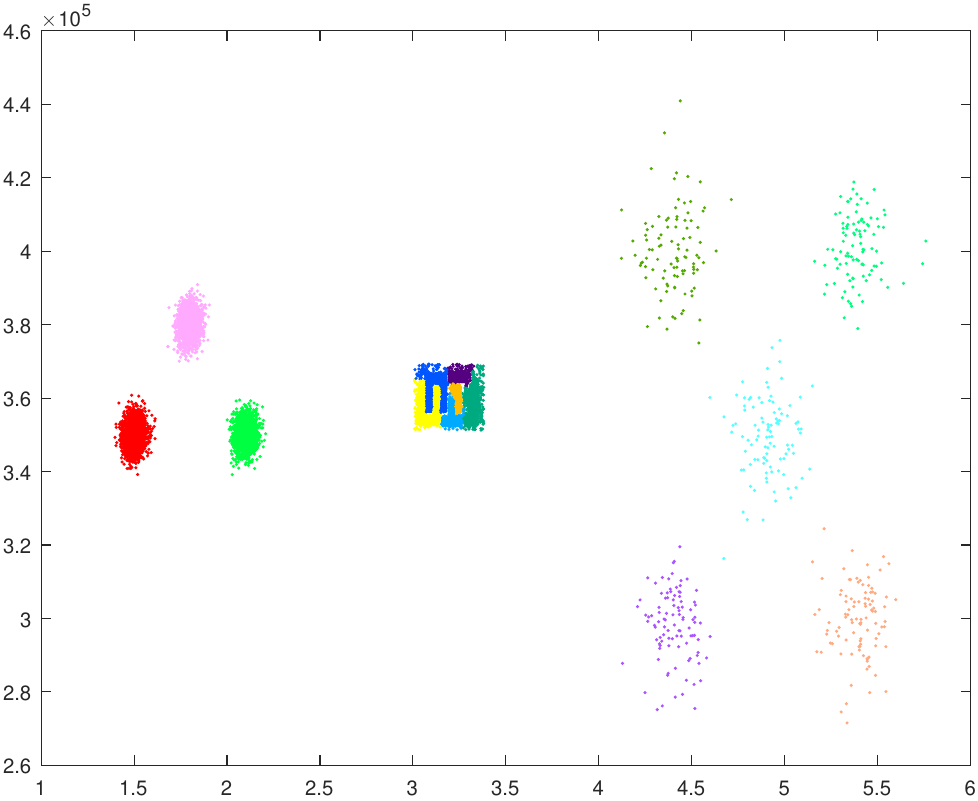}
}\
\caption{Clustering results on  two-dimensional synthetic datasets. Different colors are used to represent different output clusters.}
\label{fig:synResu}
\end{minipage}
\end{figure}

Figure \ref{fig:synResu} shows the clustering results on the two-dimensional datasets. From the last row of this figure, we can observe that, RECOME achieves desirable results for all the four datasets.

For dataset S1 (the first column) which contains only trivial Gaussian clusters, desirable results are achieved by all methods except for KNNC. For dataset S2 (the second column), which comprises unbalanced convex clusters and sparse noises, SNN, FDP and RECOME detect correct clusters. DBSCAN, KNNC and STClu overlook the clusters with little size. For dataset S3 (the third column), which is comprised of 6 clusters of nonconvex shape, a desirable result is output by RECOME. DBSCAN achieve a comparable and slightly poor performance. However, other methods rarely produce satisfactory results, mis-merge objects from different true clusters or subdivide a true cluster into different parts. For dataset S4 (the last column), which contains more complex clusters and proposes a big challenge to all baseline methods, only RECOME correctly detects the 14 clusters in it. These results indicate that RECOME is robust to the structure of cluster and own the potential to simultaneously discover clusters with different shapes, density, and scales.

\subsubsection{Results on MNIST Datasets}

\begin{figure}[!htb]
\centering
\begin{minipage}{0.85\linewidth}
\subfloat[DBSCAN]{
  \label{fig:DBSCANreal}
\includegraphics[width = 0.16\linewidth]{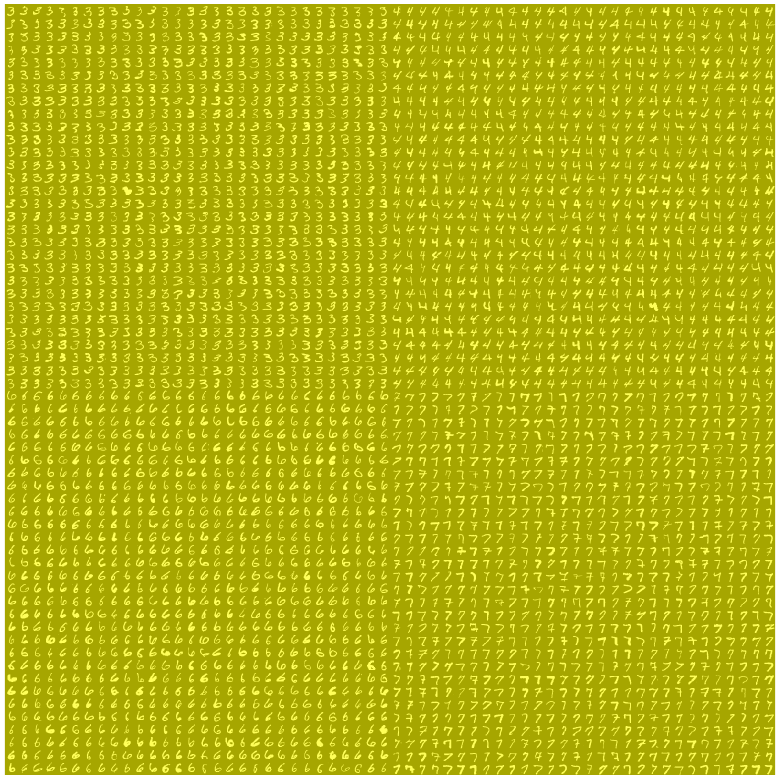}\
\includegraphics[width = 0.16\linewidth]{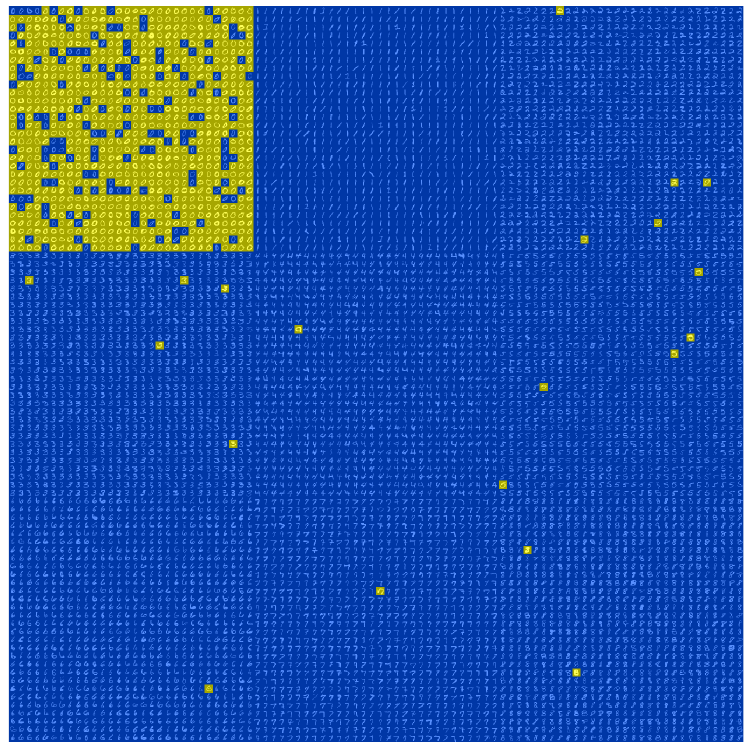}\
\includegraphics[width = 0.5\linewidth]{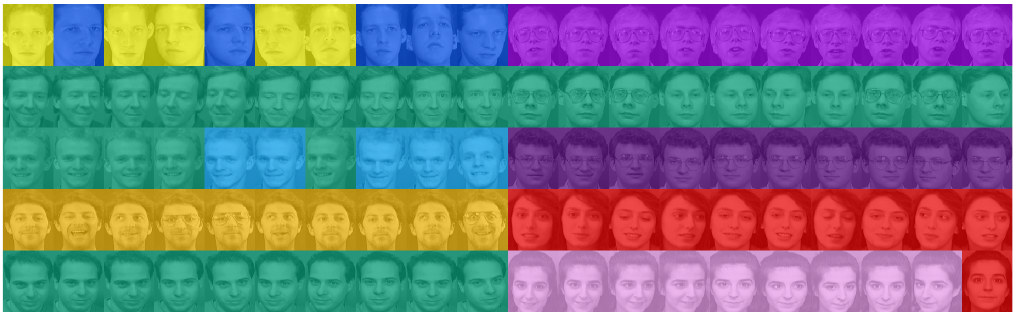}\
}\
\subfloat[SNN]{
  \label{fig:SNNreal}
\includegraphics[width = 0.16\linewidth]{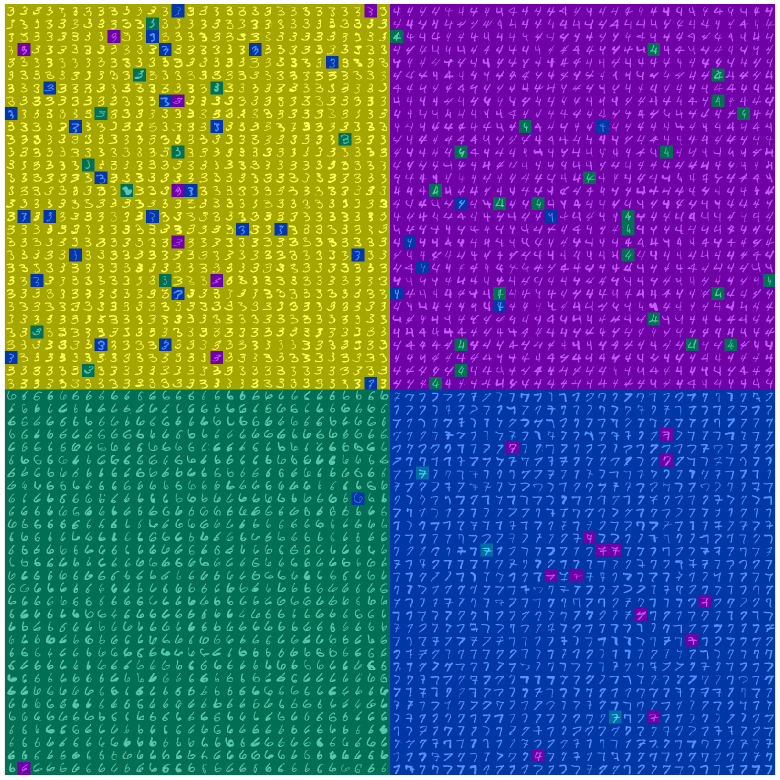}\
\includegraphics[width = 0.16\linewidth]{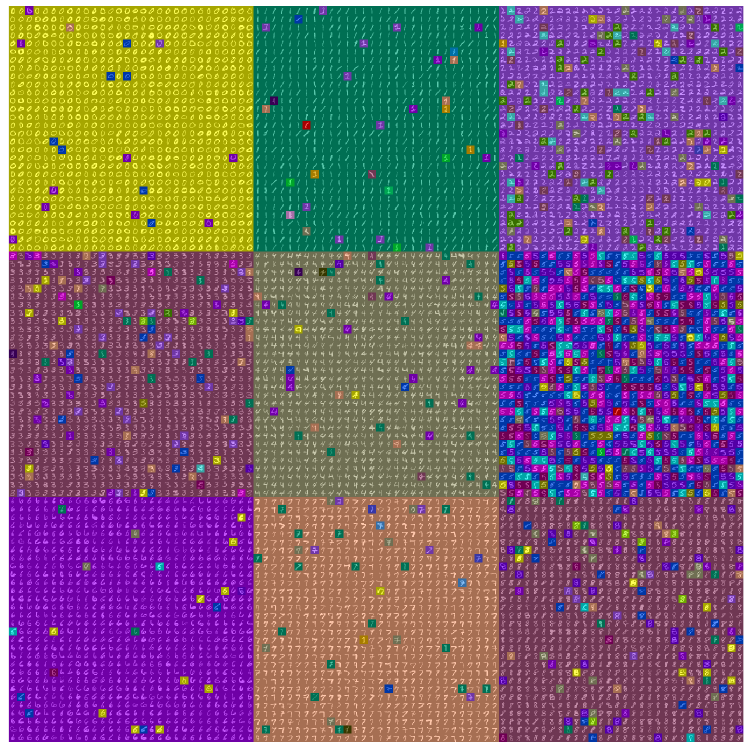}\
\includegraphics[width = 0.5\linewidth]{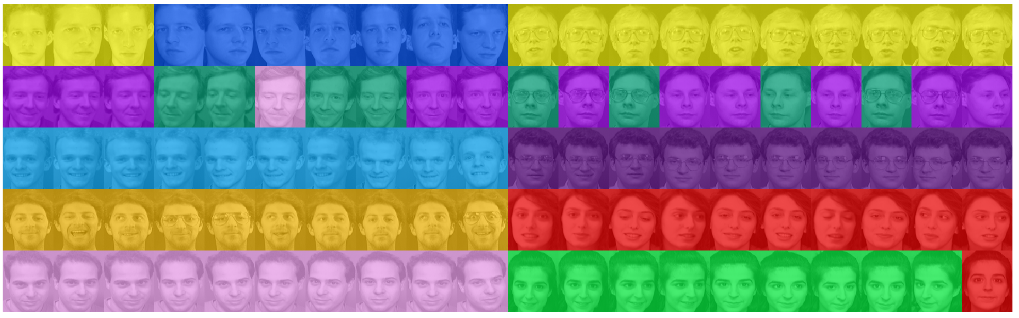}\
}\
\subfloat[KNNC]{
  \label{fig:KNNCreal}
\includegraphics[width = 0.16\linewidth]{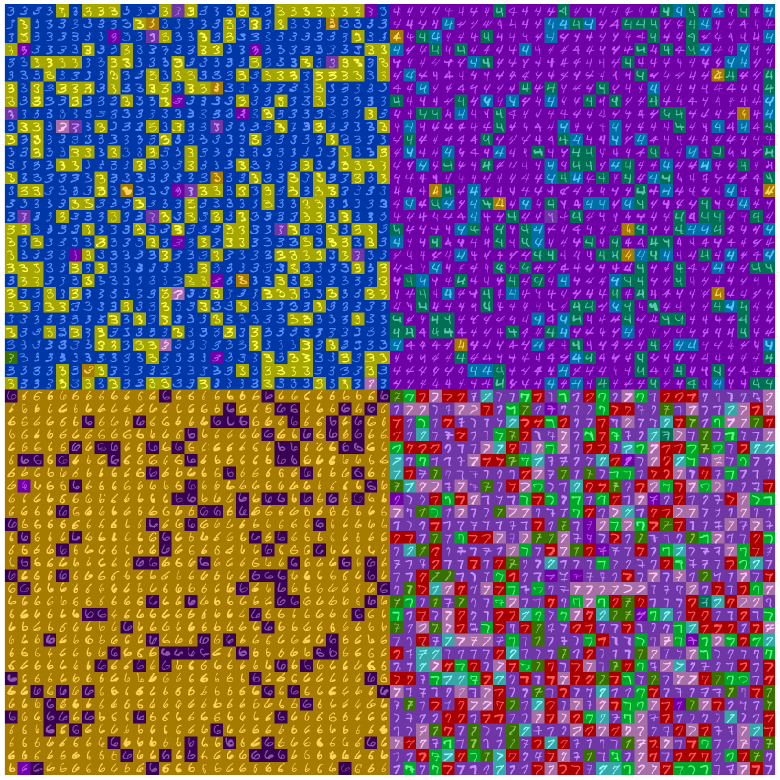}\
\includegraphics[width = 0.16\linewidth]{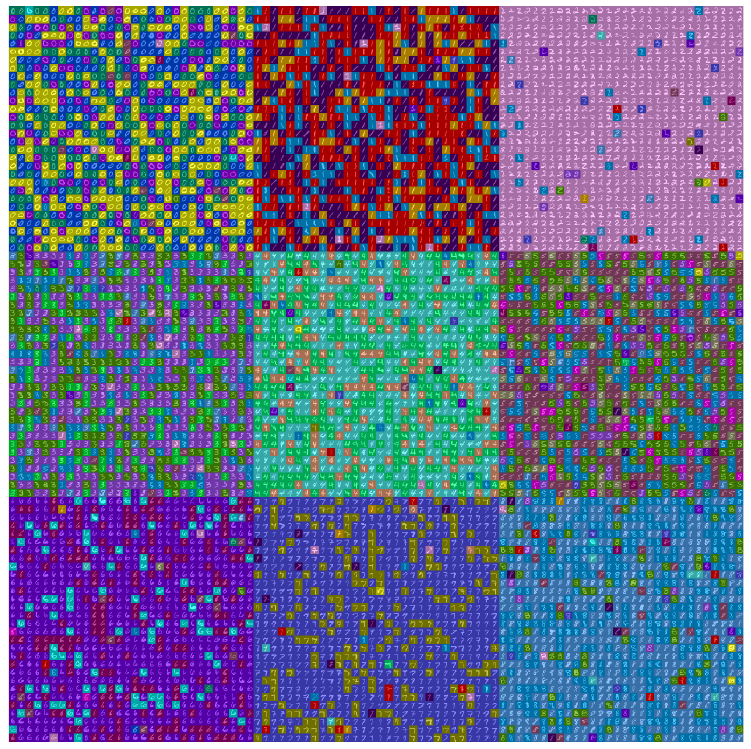}\
\includegraphics[width = 0.5\linewidth]{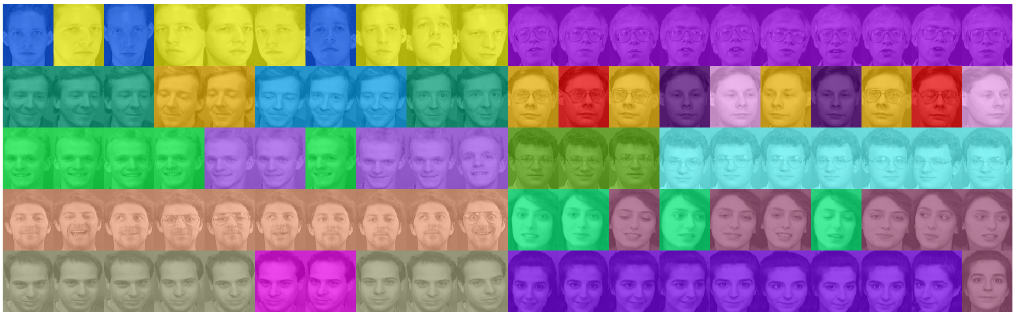}\
}\
\subfloat[FDP]{
  \label{fig:FDPreal}
\includegraphics[width = 0.16\linewidth]{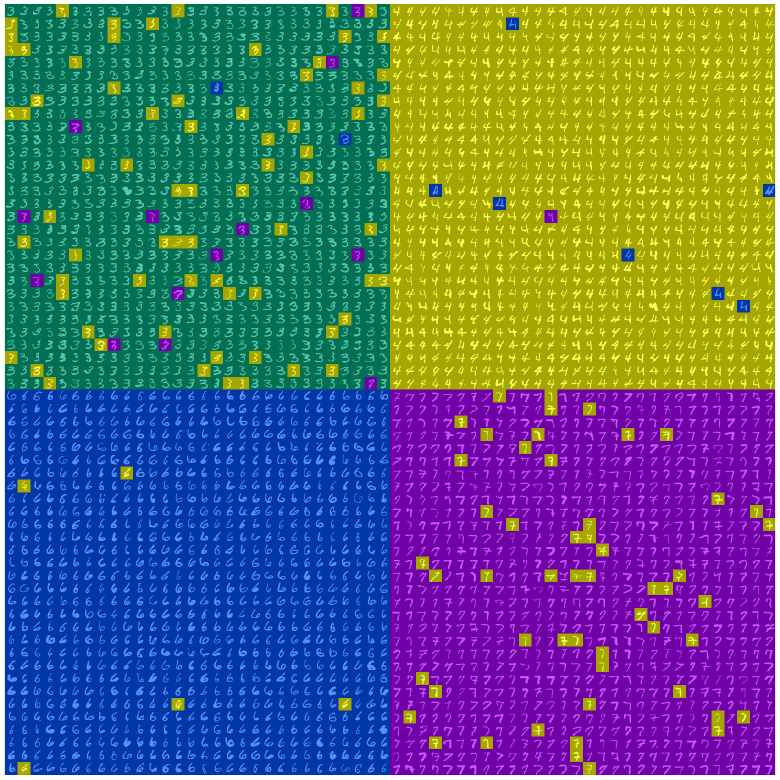}\
\includegraphics[width = 0.16\linewidth]{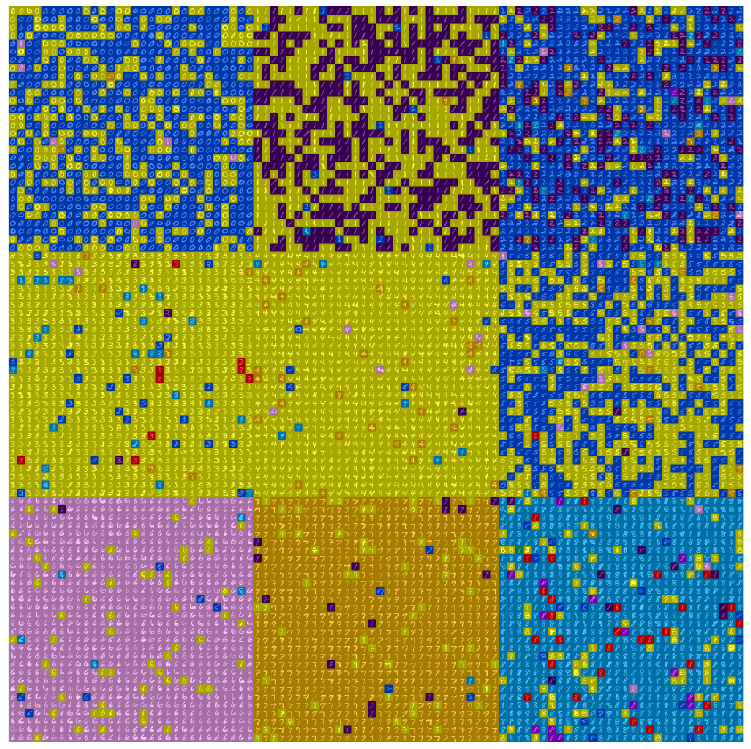}\
\includegraphics[width = 0.5\linewidth]{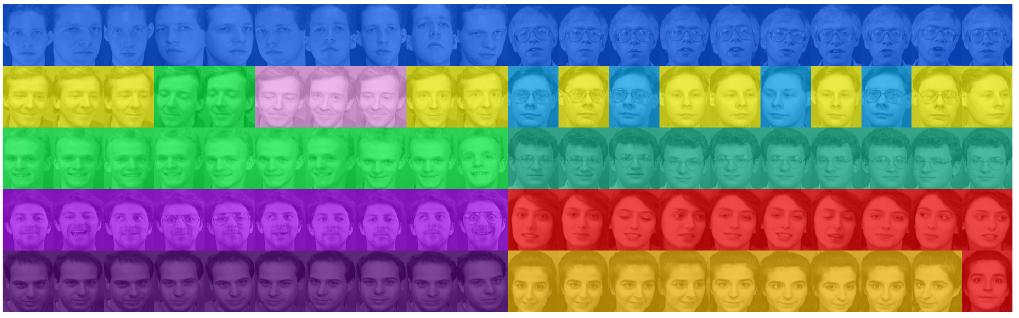}\
}\
\subfloat[3DC]{
  \label{fig:3DCreal}
\includegraphics[width = 0.16\linewidth]{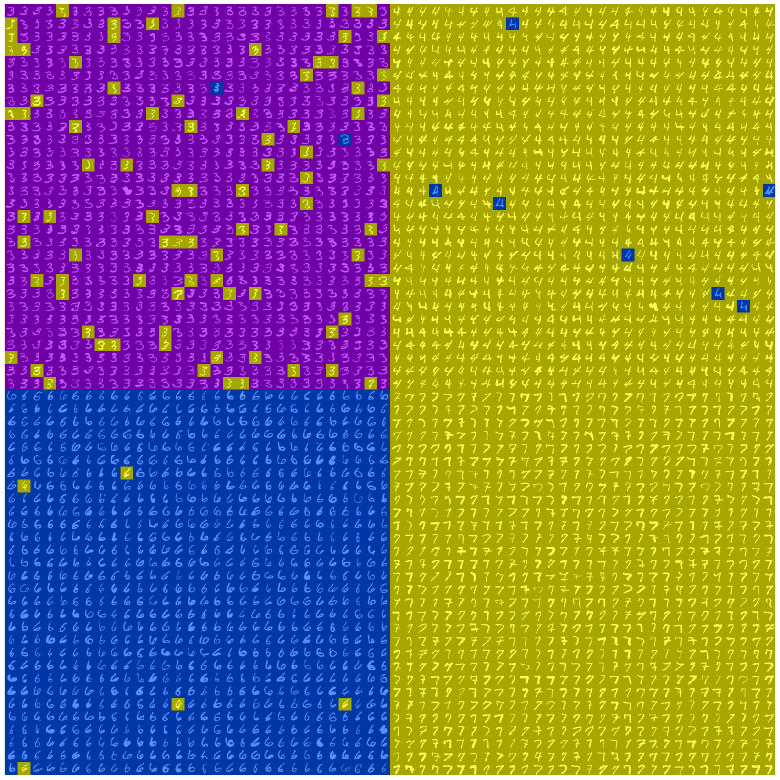}\
\includegraphics[width = 0.16\linewidth]{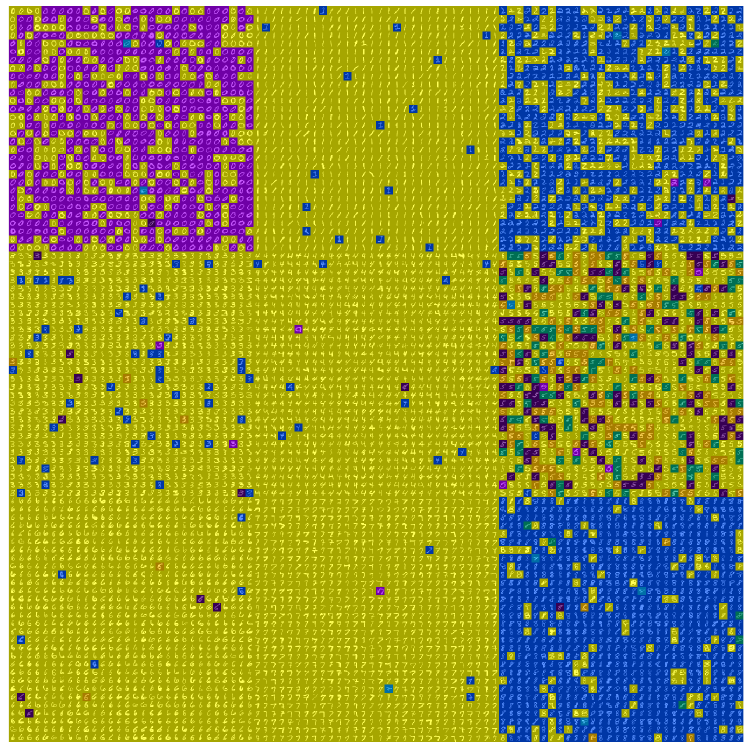}\
\includegraphics[width = 0.5\linewidth]{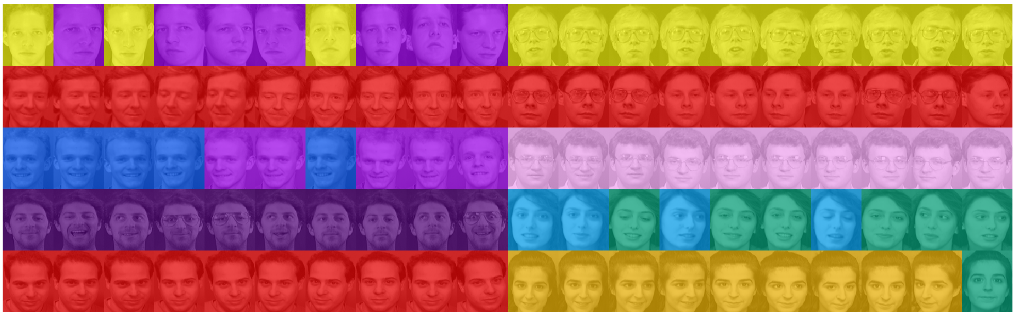}\
}\
\subfloat[STClu]{
  \label{fig:STClureal}
\includegraphics[width = 0.16\linewidth]{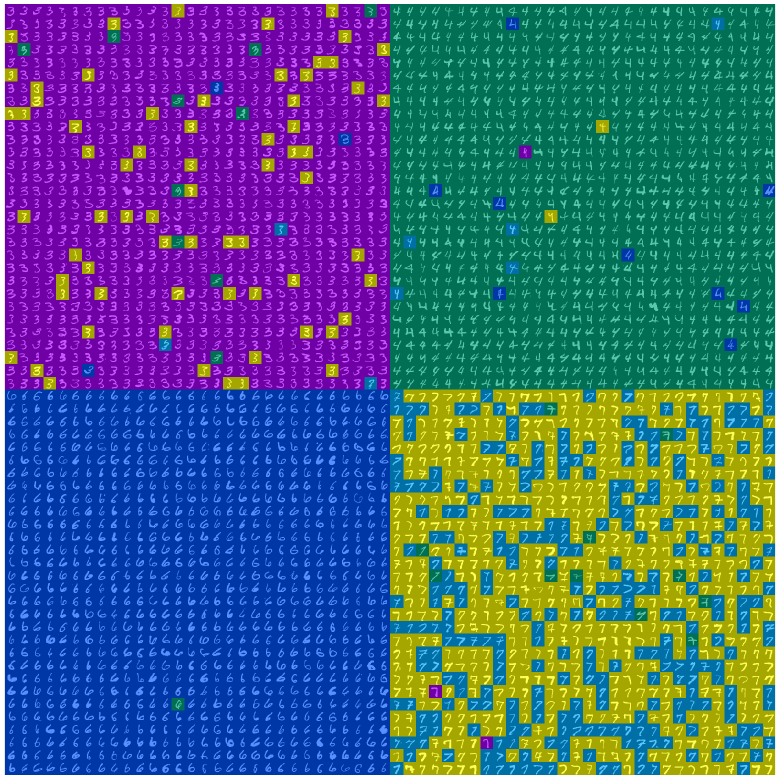}\
\includegraphics[width = 0.16\linewidth]{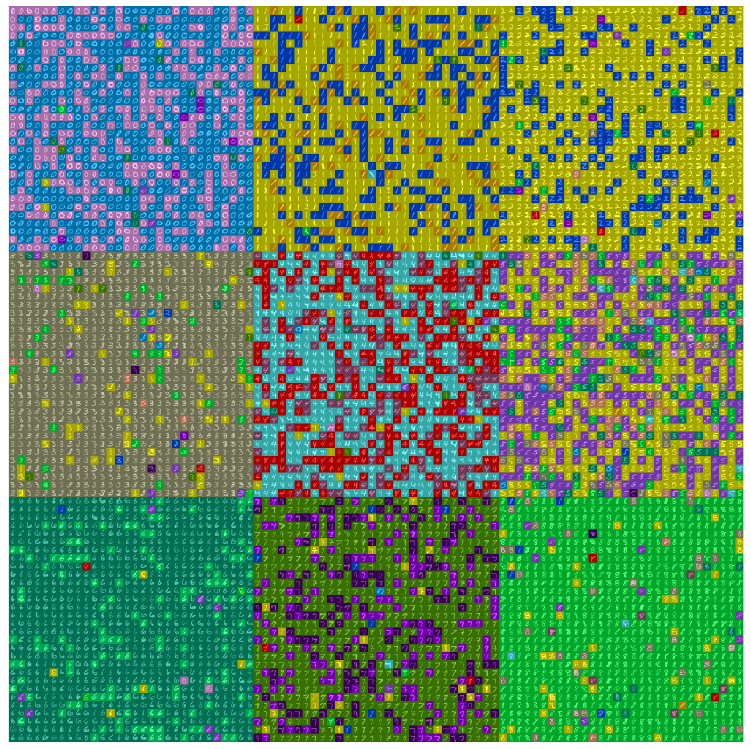}\
\includegraphics[width = 0.5\linewidth]{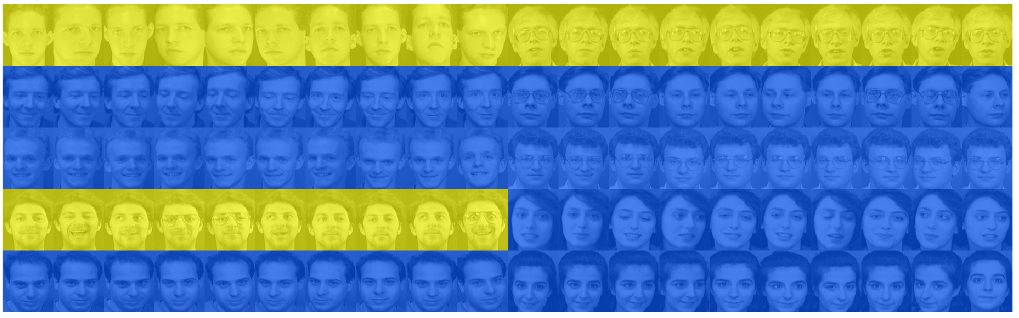}\
}\
\subfloat[RECOME]{
  \label{fig:RECOMEreal}
\includegraphics[width = 0.16\linewidth]{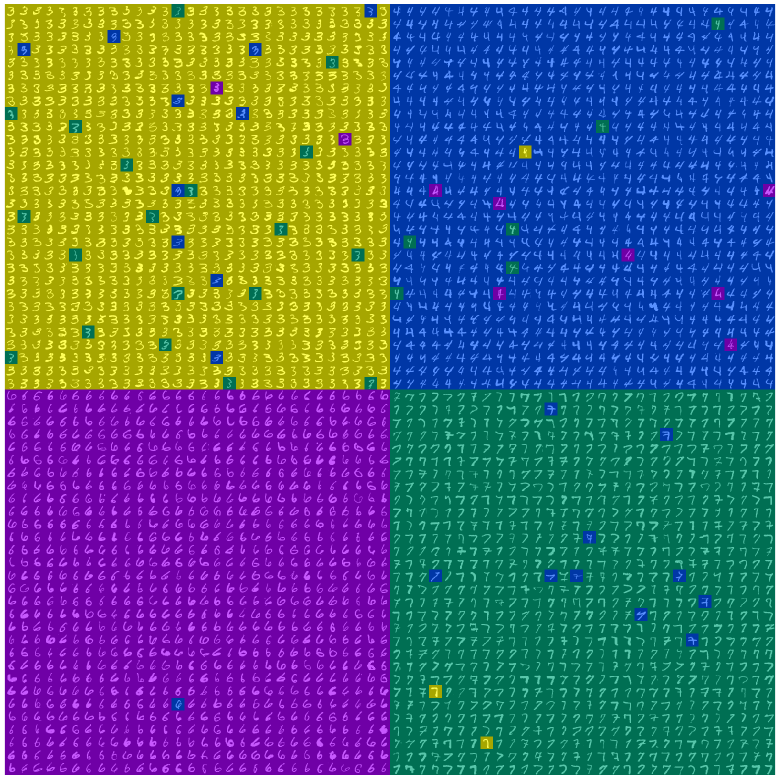}\
\includegraphics[width = 0.16\linewidth]{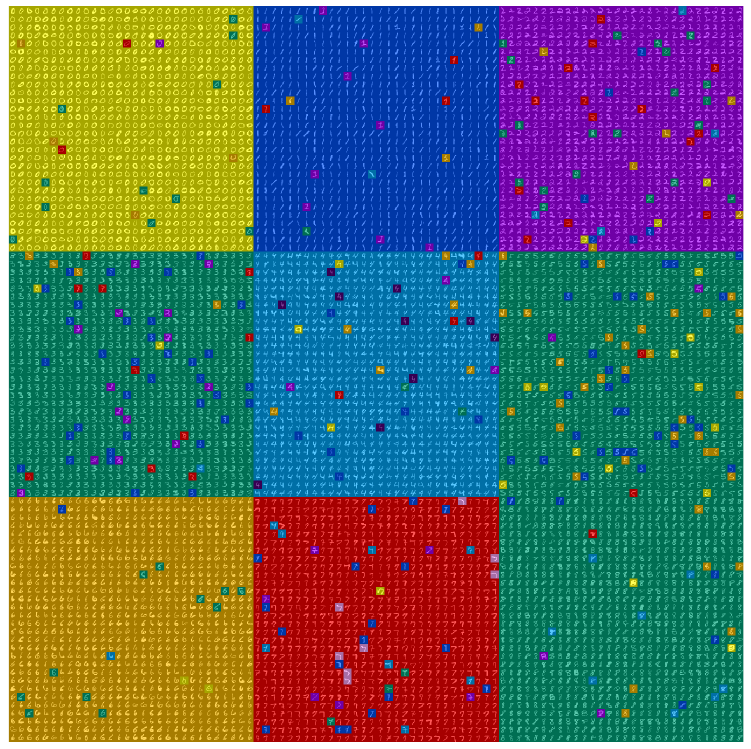}\
\includegraphics[width = 0.5\linewidth]{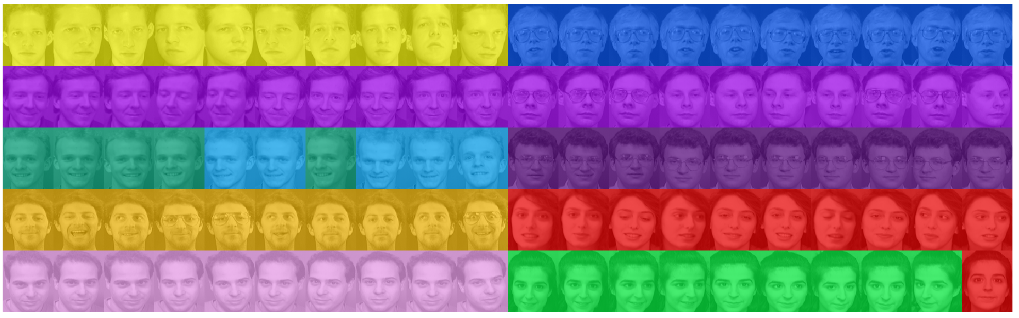}\
}\
\caption{Visualized clustering results on real datasets. Columns from left to right are visualizations for sM3467, sM0:8, and face10, respectively. Different clusters are marked by different colors.}
\label{fig:realResu}
\end{minipage}
\end{figure}

The clustering results on the MNIST datasets are presented in TABLE \ref{tab:resuOnMnist} and the visualization for sM3467 and sM0:8 are shown in Figure \ref{fig:realResu}.

\begin{table}[htb]
\caption{Performance comparison of the seven methods on the MNIST datasets. $C$ is the cluster number identified by the algorithms (or utilized as prior knowledge for FDP) and $C_t$ gives the true cluster number.}
\renewcommand\arraystretch{1.1}
\centering
\scalebox{0.85}{
\begin{tabular}{ccccccccc}
\hline
 & 			     & DBSCAN & SNN    &KNNC    &FDP 	&3DC 	&STClu	&RECOME\\\hline
 M367 & $C$      & 3      & 3      & 6 		& (3) 	& 	3	& 	3	& 3\\
($C_t=3$) & NMI  & .89    & .70    & .77 	& .93 	& 	.93	& 	.96	&\textbf{.98}\\
 	      & F    & .95    & .79    & .78 	& .97 	& 	.97	&\textbf{.98}& .96\\
 	  & Time(s)  & 545    & 587    & 26 	& 96 	& 	124	& 	57	& 64\\\hline
 sM3467 & $C$    & 1      & 5      & 13 	& (4) 	& 	3	& 	5	&4\\
($C_t=4$)&NMI    & .00    & .91    & .72 	& .87 	& 	.76	& 	.85	&\textbf{.94}\\
 	  & F        & .40    & .95    & .67 	& .92	& 	.81	&	.88	&\textbf{.97}\\
 	  & Time(s)  & 21     & 18     & 1  	& 3 	& 	4	& 	2	&2\\\hline
 M3467 & $C$     & 1      & 4      & 10 	& (4) 	& 	3	& 	5	&4\\
($C_t=4$)& NMI   & .00    & .90    & .76 	& .91 	& 	.80	& 	.90	&\textbf{.94}\\
 	  & F        & .40    & .95    & .74 	& .95 	& 	.83	& 	.92	&\textbf{.97}\\
 	  & Time(s)  & 956    & 1006   & 46 	& 170 	& 	235	& 	99	&116\\\hline
 sM0:8 & $C$     & 2      & 26     & 24 	& (9) 	& 	7	& 	17	&9\\
($C_t=9$) & NMI  & .20    & .76    & .68 	& .53 	& 	.35	& 	.68	&\textbf{.80}\\
 	  & F        & .33 	  & .77    & .54 	& .54 	& 	.42	& 	.62	&\textbf{.80}\\
 	  & Time(s)  & 106 	  & 95     & 5   	& 16 	& 	25	& 	10	&14\\\hline
 M0:8 & $C$      & 2 	  & 29     & 15 	& (9) 	& 	7	& 	19	&14\\
($C_t=9$) & NMI  & .14 	  & .78    & .75 	& .47 	& 	.04	& 	.73	&\textbf{.82}\\
 	  & F        & .30 	  & .80    & .69 	& .47 	& 	.22	& 	.68	&\textbf{.84}\\
 	  & Time(s)  & 5236   & 5409   & 190 	& 1172 	& 1886	& 1098	&762\\\hline
 M0:9 & $C$      & 2 	  & 40     & 20 	& (10) 	& 	8	& 	2	&12\\
($C_t=10$) & NMI & .09    & .72    & .74 	& .43 	& 	.29	& 	.26	&\textbf{.80}\\
 	  & F        & .24    & .71    & .70 	& .41 	& 	.37	& 	.19	&\textbf{.80}\\
 	  & Time(s)  & 6528   & 6162   & 198 	& 1721 	& 2321	& 1123	&918\\\hline
\end{tabular}}
\label{tab:resuOnMnist}
\end{table}

From TABLE \ref{tab:resuOnMnist}, we can see that RECOME has the best performance for all datasets except M367, on which a comparable result is produced by STClu. SNN achieves slightly lower scores for almost all datasets than RECOME at much longer running time. Among all the algorithms, KNNC has the shortest running time for all datasets, but its performance is not steady---it achieves passable performance on M0:8 and M0:9 but poor performance on the remaining datasets. For FDP, we error on the conservative side by feeding it the true cluster number. Even so, RECOME achieves higher NMI and F values than FDP and the gap becomes obvious as the increasing of class number. Similar to FDP, the results of 3DC are not desirable for the datasets with many classes. The performance of STClu is unstable---it achieves comparable results on M367 to M3467 and moderate results on sM0:8 and M0:8, but fails to work well on M0:9. This may be due to it based on statistical testing, which is largely depended on data quality. Despite our best efforts in parameter tuning, DBSCAN fail to perform well over all datasets but M367.

As shown in Figure \ref{fig:realResu}, for dataset sM3467, SNN and RECOME output almost correct results. FDP and STClu detect four main classes, but cluster excessive objects wrong classes. DBSCAN and 3DC fail to find the true cluster number for this dataset. For sM0:8, we can see that SNN and RECOME achieve the best performance but they both threat the two clusters of the digits ``3'' and ``8'' (the first column of the second row and the last column of the third row) as one. Besides, for the cluster of the digit ``5'', SNN divides it into several sub-clusters. In contrast, RECOME detects this cluster but merge it into other clusters. Unfortunately, all other methods fail to output satisfactory results for this dataset, detecting few meaningful clusters or attributing many objects into wrong clusters.

\subsubsection{Results on Olivetti Face Database}

\begin{table}[htb]
\caption{Performance comparison of the seven methods on the Olivetti face database, where $C$ means the cluster number identified by algorithm (or utilized as prior knowledge, for FDP) and $C_t$ means the true cluster number.}
\renewcommand\arraystretch{1.1}
\centering
\scalebox{0.77}{
\begin{tabular}{ccccccccc}
\hline
 & 			     & DBSCAN & SNN    &KNNC    &FDP 	&3DC 	&STClu	&RECOME\\\hline
 face10& $C$      & 11     & 10     & 19     & (10) 	& 9		& 	2	& 10\\
($C_t=10$)&NMI   & .87    & .88    & .87 	& .89 	& .83	& 	.42	&\textbf{.94}\\
 	        &F   & .81    & .81    & .73 	& .82 	& .73	& 	.32	&\textbf{.90}\\\hline
 face40 & $C$    & 42     & 50 	   & 66		& (40) 	& 	12	& 	2	& 37\\
($C_t=40$)& NMI  & .80    & .87    &\textbf{.89}& .83& 	.54	& 	.27	& .88\\
 	  & F 		 & .60    & .64    &\textbf{.67}& .60& 	.29	&	.09	& .66\\\hline
\end{tabular}}
\label{tab:resuOnFace}
\end{table}

The clustering results on face10 and face40 are presented in TABLE\ref{tab:resuOnFace}. The visualized results for face10 are shown in the rightmost column of Figure \ref{fig:realResu} (The visualization for face40 can be found in the supplementary material).

For face10, we can see that RECOME gets the highest score on both NMI and F metrics from TABLE \ref{tab:resuOnFace}. The other methods with the exception of STClu achieve similar scores, which are slightly lower than RECOME. Though the correct cluster number is found by RECOME, as shown in Figure \ref{fig:realResu}, it mistakenly merges two clusters (the second row) and subdivides a true cluster (the left subject of the third row) into two classes.

For dataset face40, as shown in TABLE \ref{tab:resuOnFace}, KNNC achieves the best result in NMI and F measures, but this comes at the cost of an abnormally large cluster number. RECOME and SNN get comparable scores lower than KNNC by one percent and two percent in the NMI value, respectively. Besides, DBSCAN and FDP achieve reasonable performance and have a gap of 0.06 in the F metrice compared with RECOME. Despite our best efforts in parameter tuning, 3DC and STClu fail to perform well for this dataset. This may be explained by statistical errors since there are only 10 pictures for each people as discussed in \cite{Liang2016delta, Wang2016Automatic}.

\subsection{Parameter Analysis}
\label{ssec:ParaAnal}

As discussed previously, parameters $K$ and $\alpha$ play fundamental roles in RECOME. Specifically, $K$ determines the density estimation and the structure of the KNN graph. The number of the final clusters are largely determined by $\alpha$. In this section, we present quantitative results on the impact of parameters $K$ and $\alpha$ on the performance of RECOME.

For a given dataset with $|V|$ objects, the possible value of the number of $K$ can vary from 1 to $|V|-1$. However, we find that when $K$ is large enough (up to $\sqrt{|V|}$), increasing $K$ has little or negative impacts on the performance of RECOME. Thus, in this set of experiments, $K$ is taken values from $(0, \sqrt{|V|}]$ with a step size of $\sqrt{|V|} / 40$.

\begin{figure}[]
\begin{minipage}{0.9\linewidth}
\centering
\subfloat[S1]{
 \label{fig:paraS1}
\includegraphics[width = 0.25\linewidth]{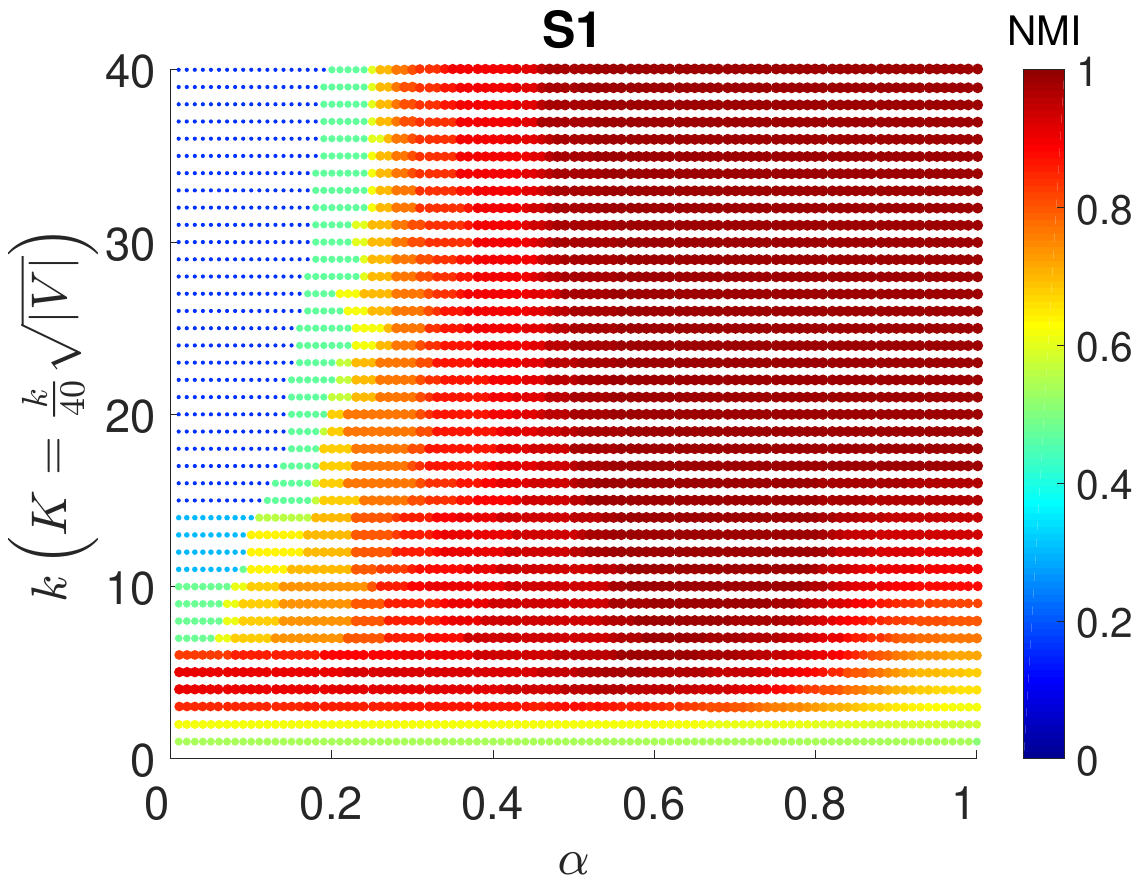}\
\includegraphics[width = 0.25\linewidth]{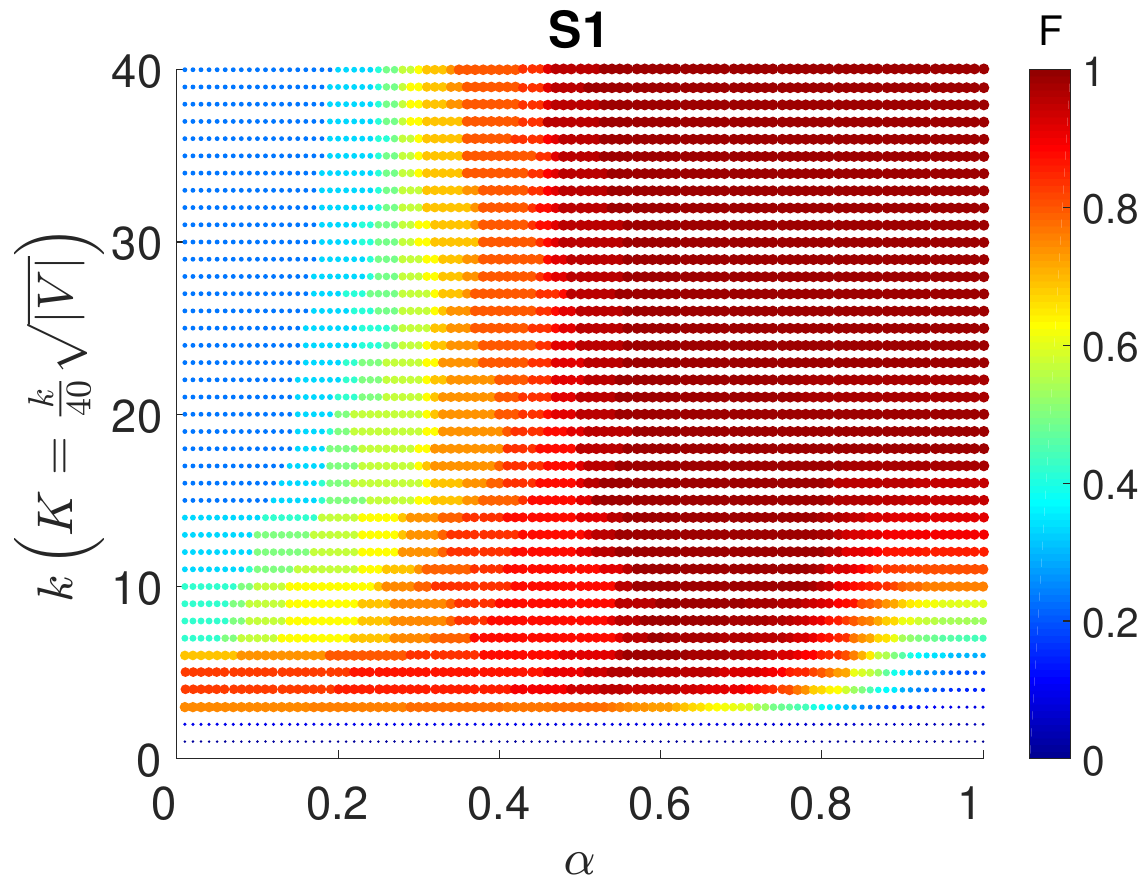}\
}
\subfloat[S2]{
  \label{fig:paraS2}
\includegraphics[width = 0.25\linewidth]{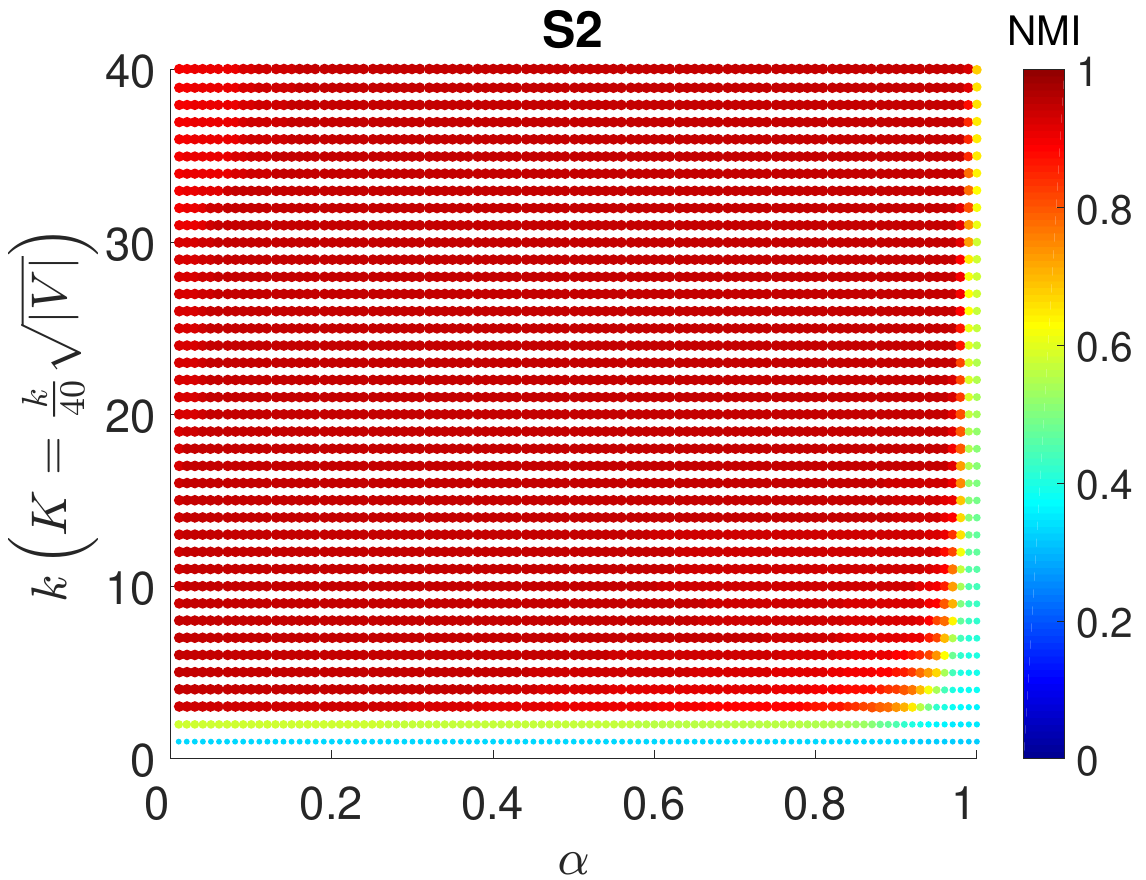}\
\includegraphics[width = 0.25\linewidth]{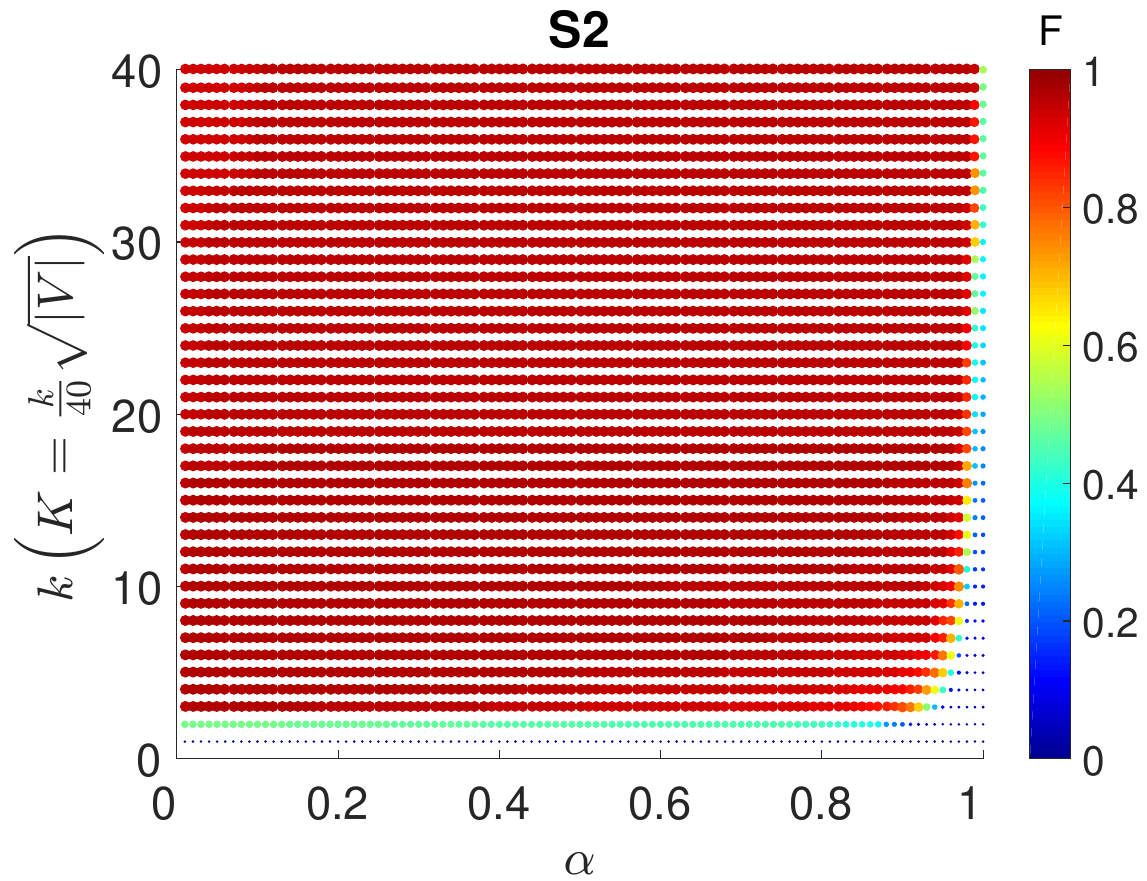}\
}\
\subfloat[S3]{
 \label{fig:paraS3}
\includegraphics[width = 0.25\linewidth]{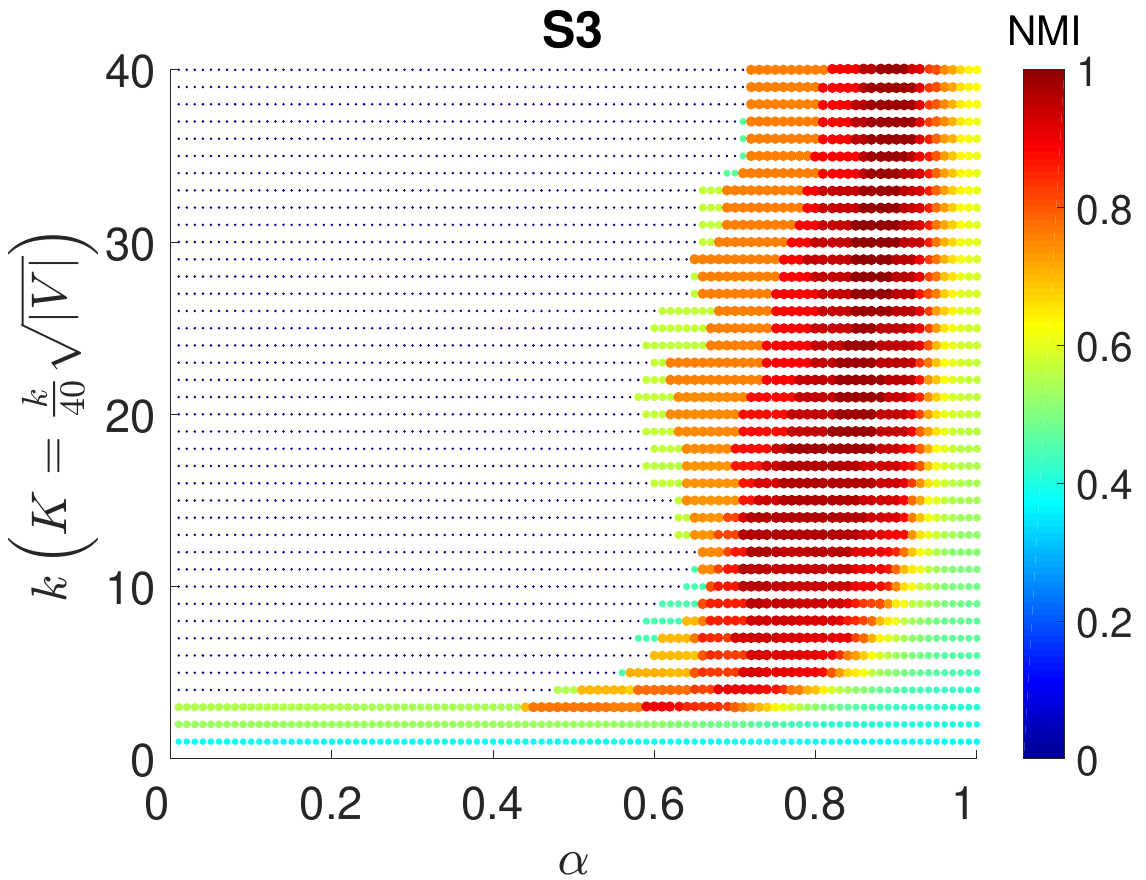}\
\includegraphics[width = 0.25\linewidth]{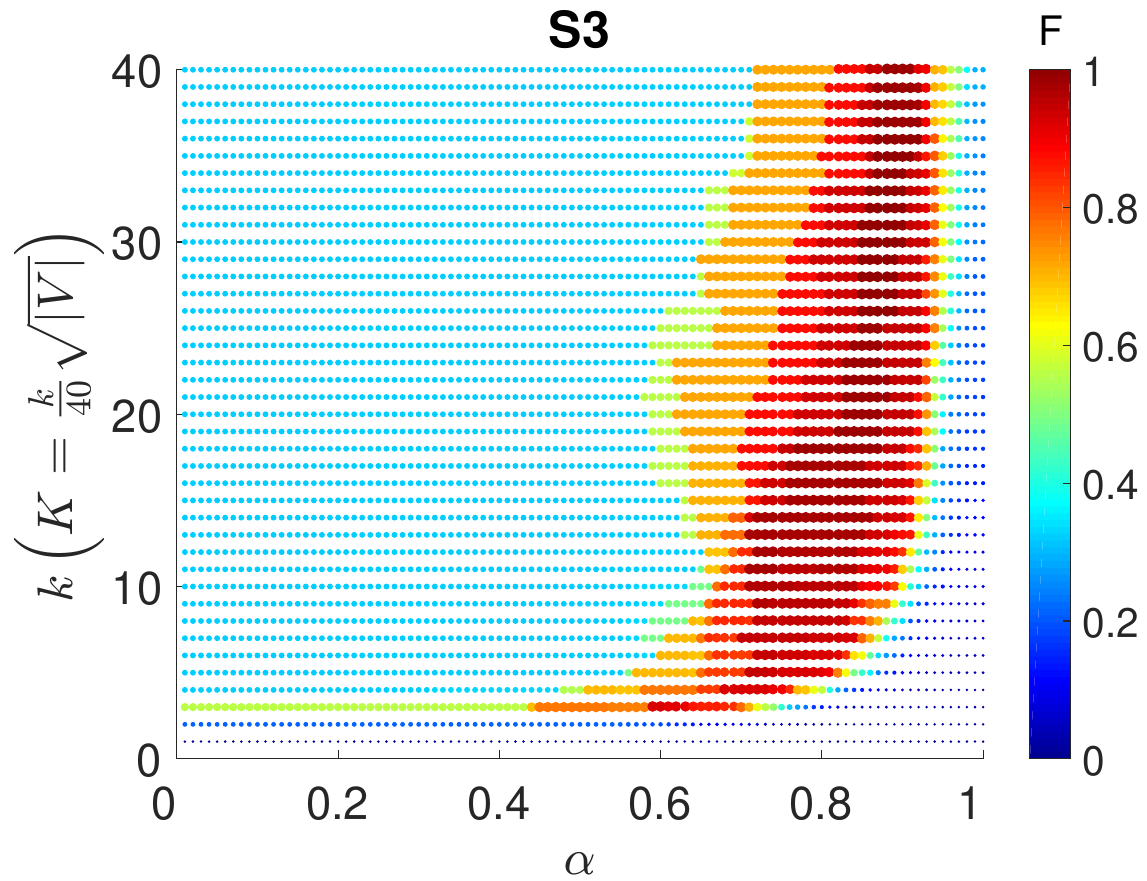}\
}
\subfloat[S4]{
  \label{fig:paraS4}
\includegraphics[width = 0.25\linewidth]{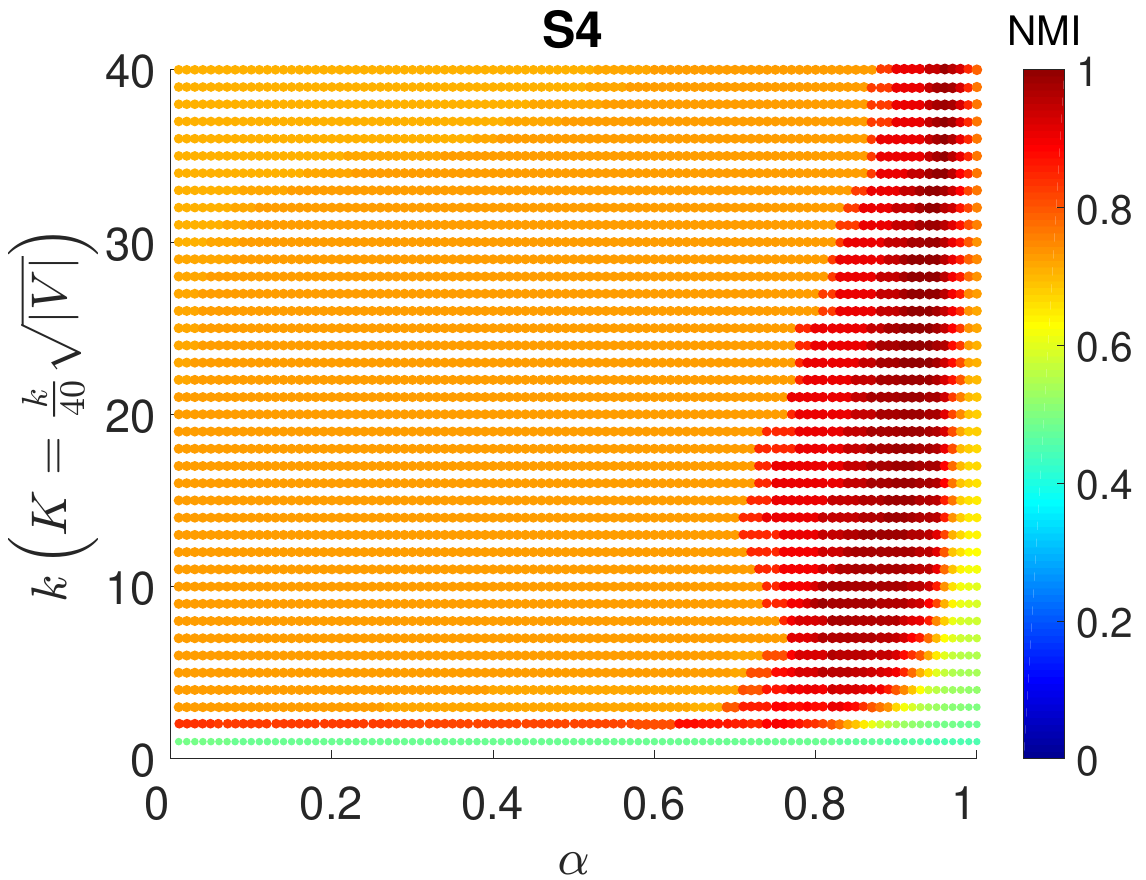}\
\includegraphics[width = 0.25\linewidth]{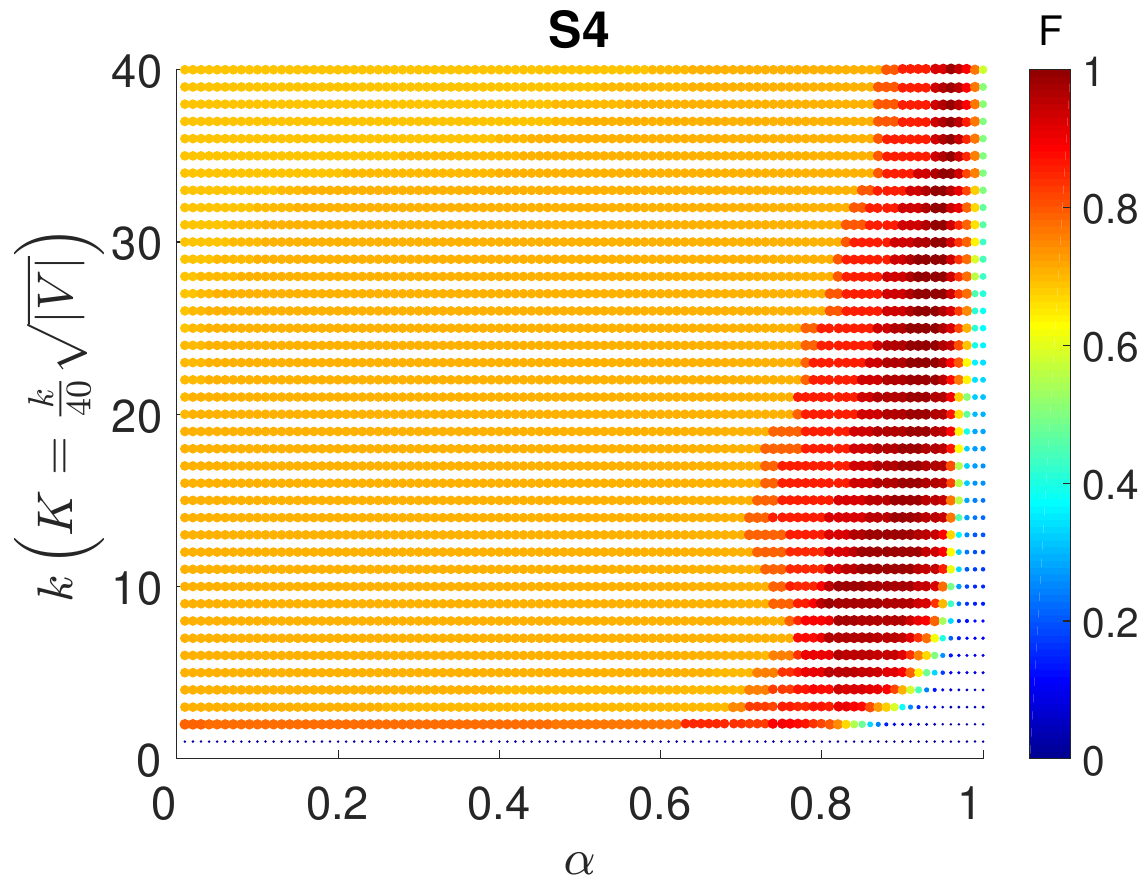}\
}\
\subfloat[M367]{
 \label{fig:paraM367}
\includegraphics[width = 0.25\linewidth]{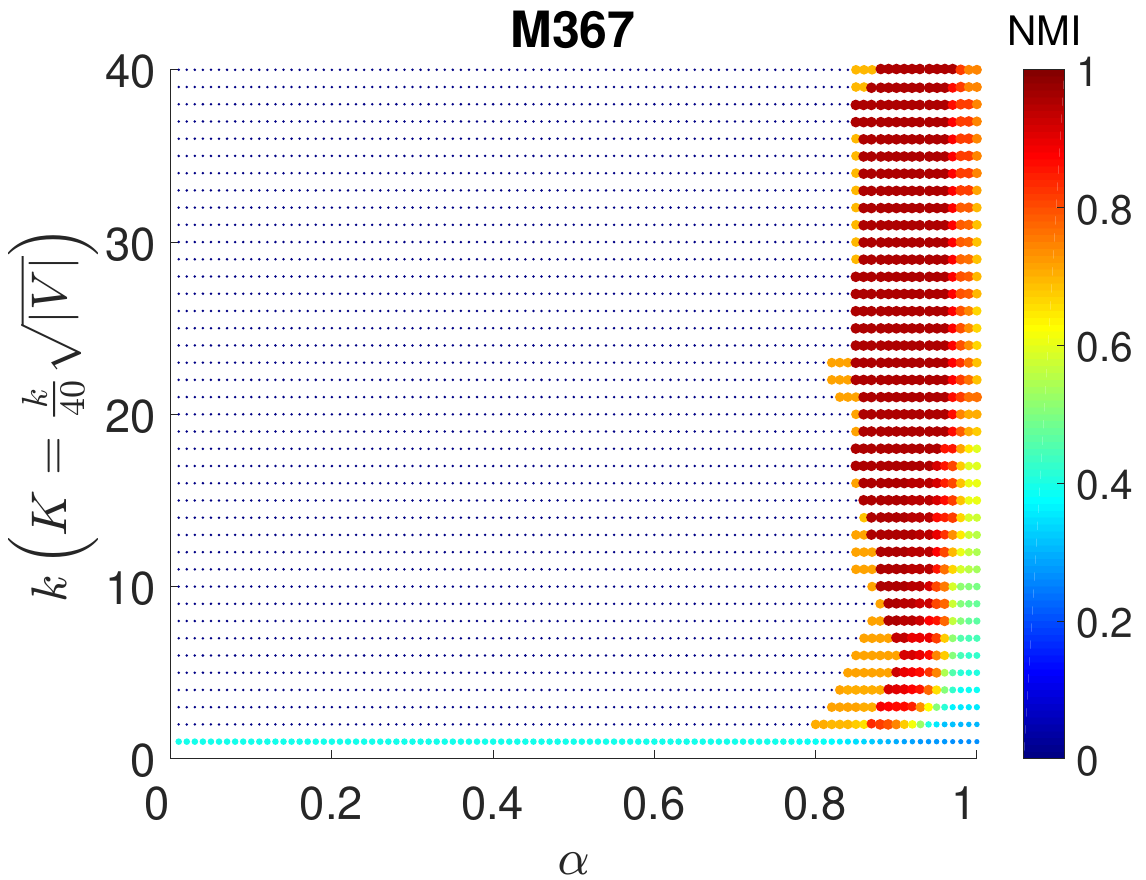}\
\includegraphics[width = 0.25\linewidth]{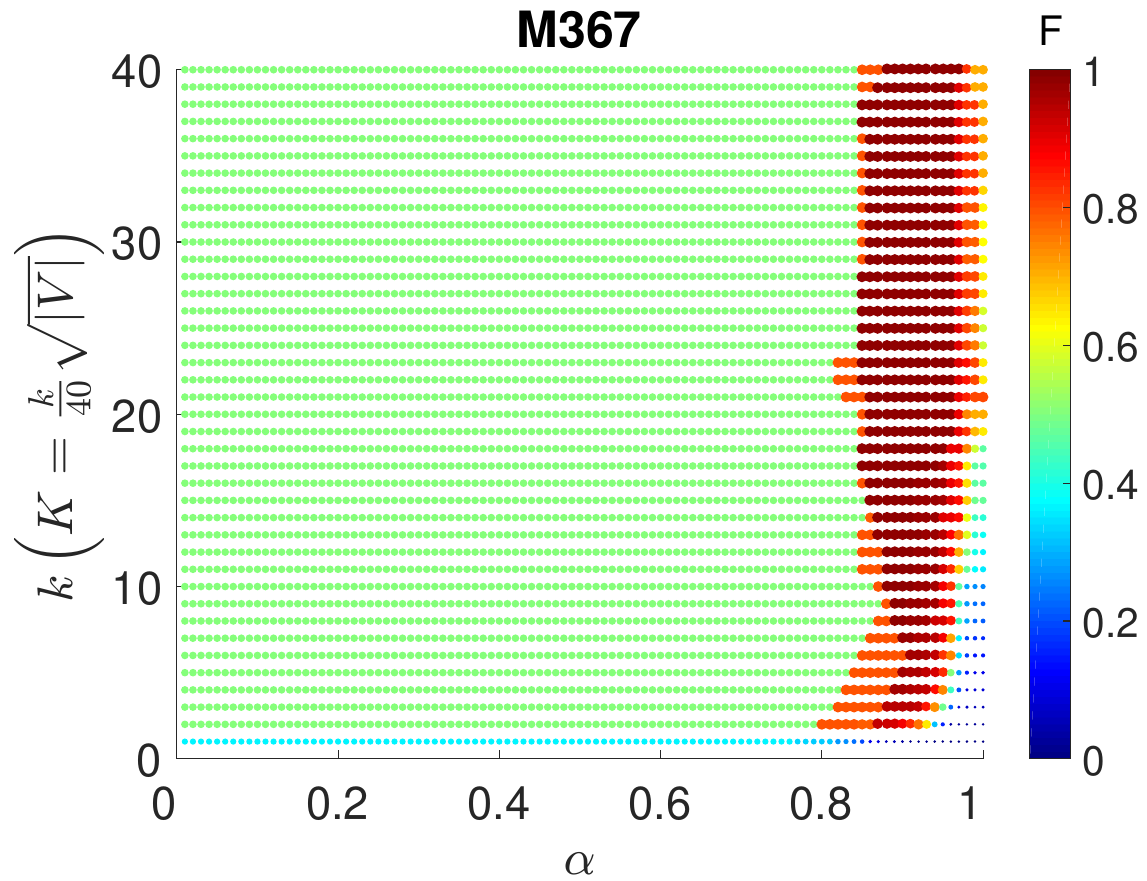}\
}
\subfloat[sM3467]{
 \label{fig:parasM3467}
\includegraphics[width = 0.25\linewidth]{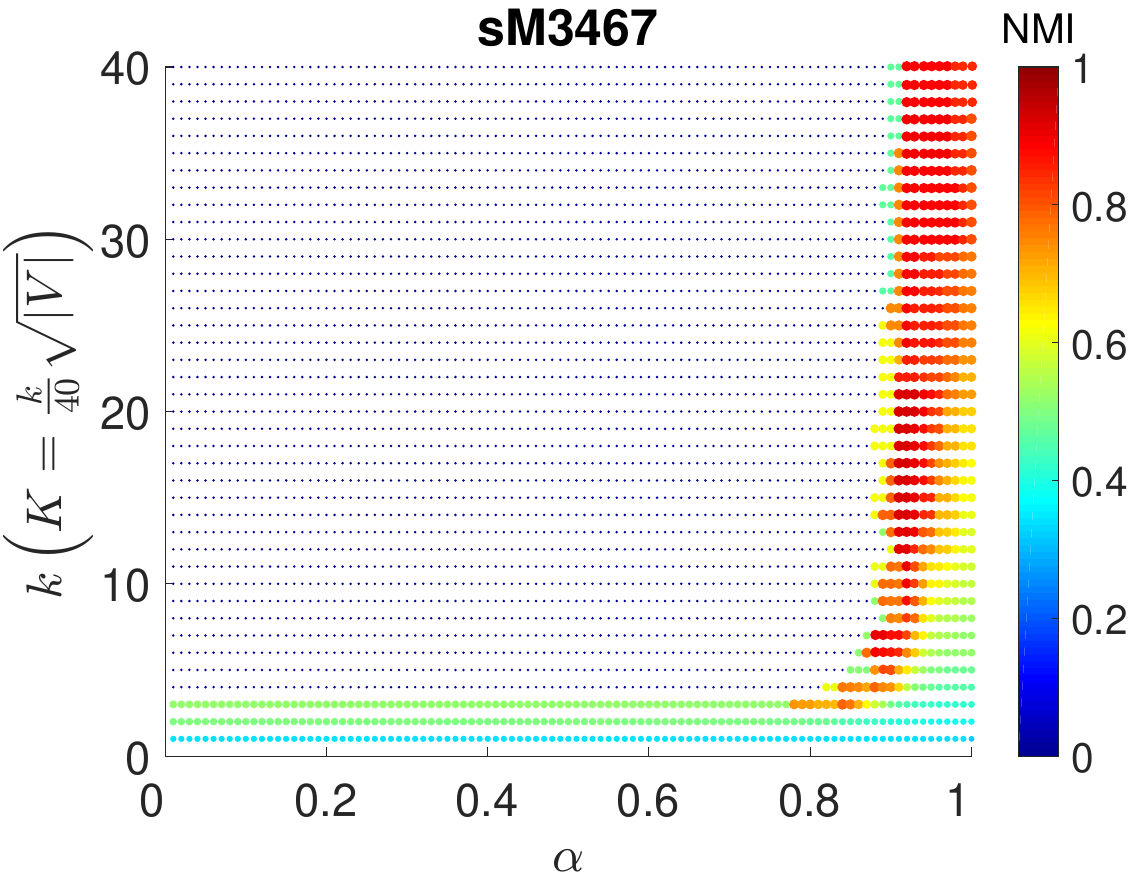}\
\includegraphics[width = 0.25\linewidth]{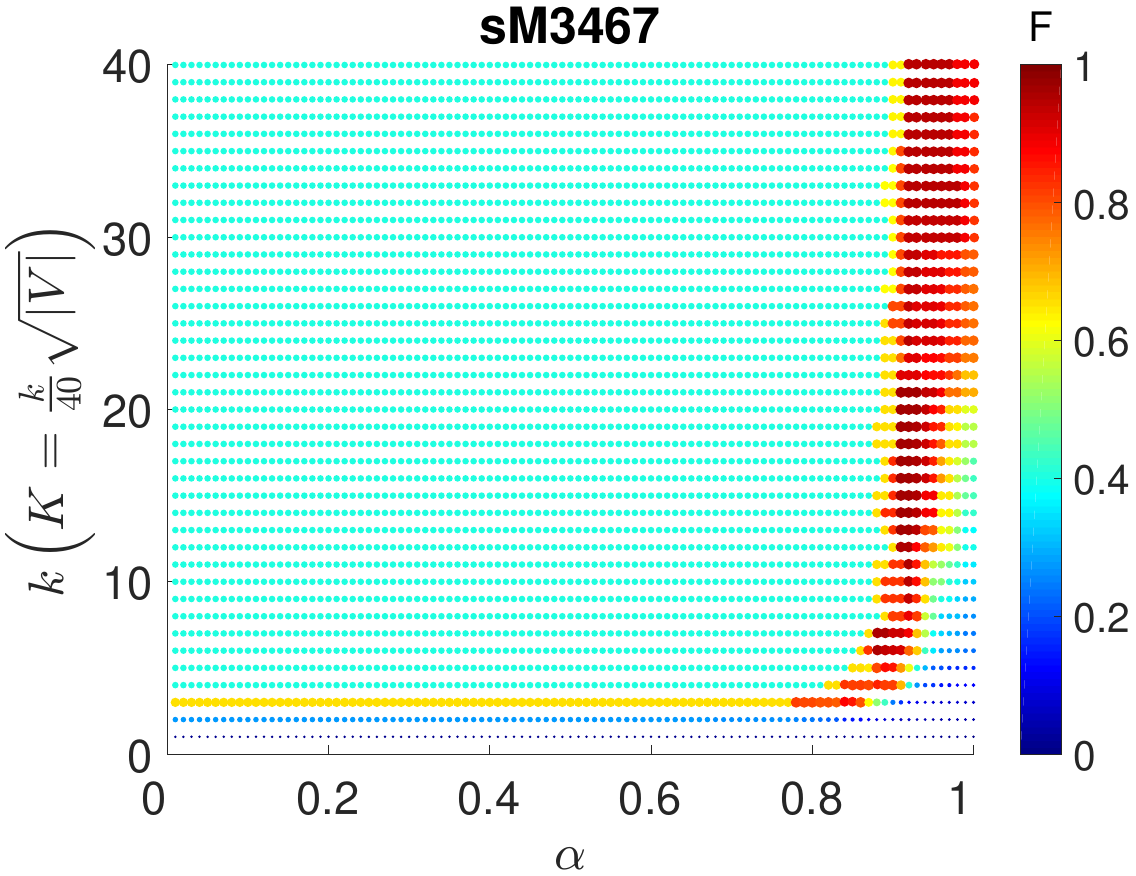}\
}\
\subfloat[M3467]{
 \label{fig:paraM3467}
\includegraphics[width = 0.25\linewidth]{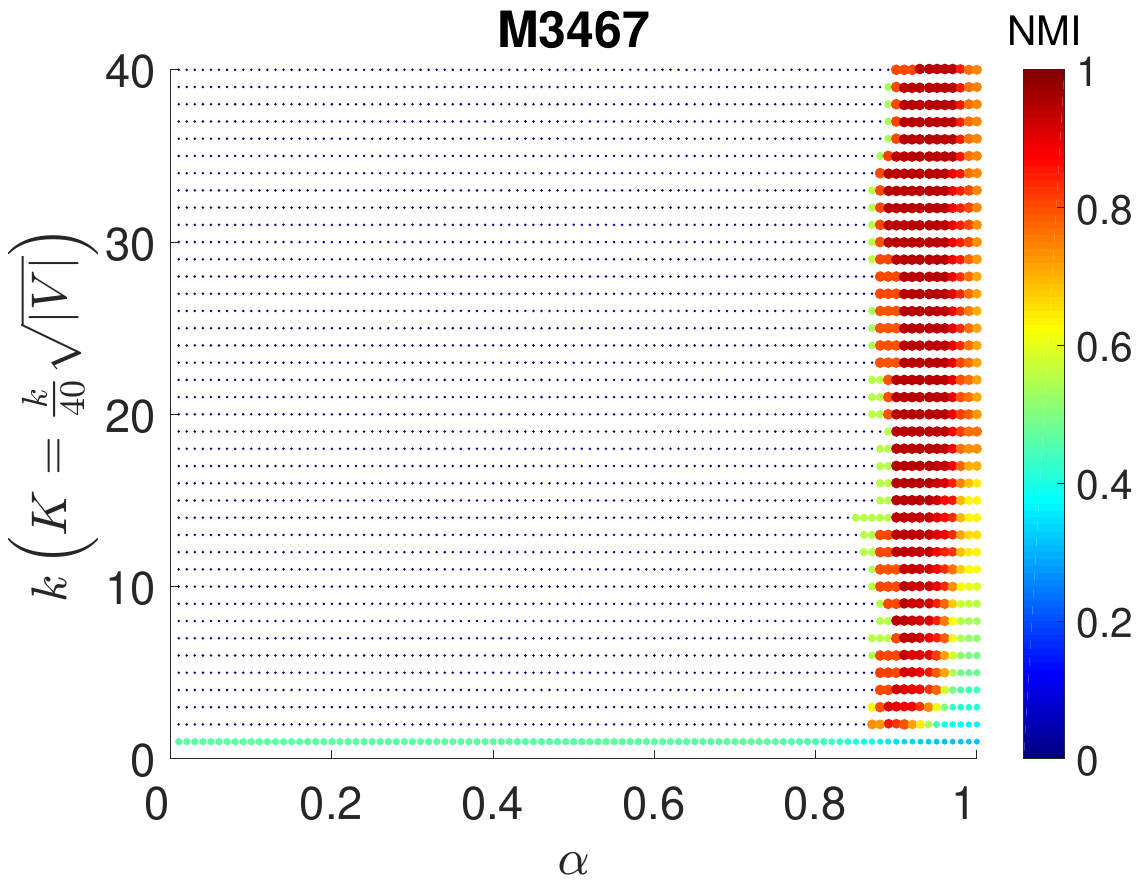}\
\includegraphics[width = 0.25\linewidth]{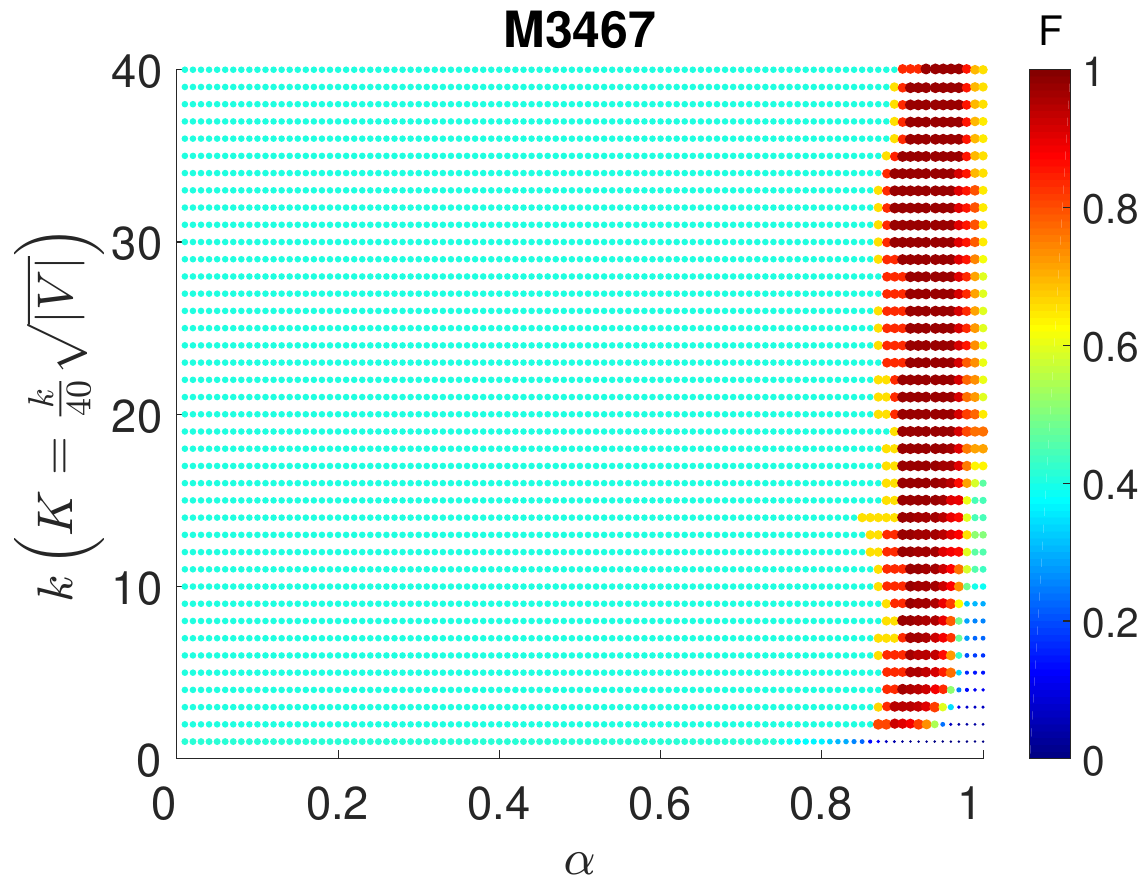}\
}
\subfloat[sM0:8]{
  \label{fig:parasM0-8}
\includegraphics[width = 0.25\linewidth]{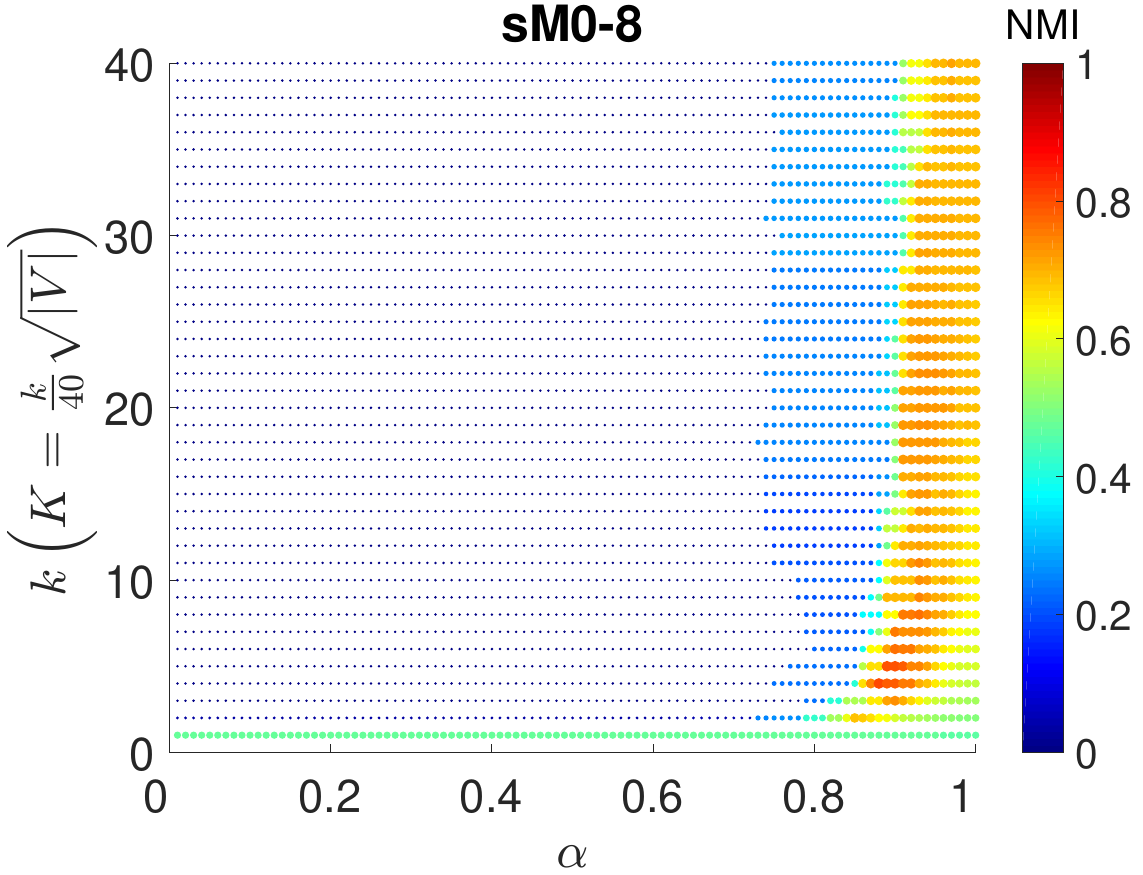}\
\includegraphics[width = 0.25\linewidth]{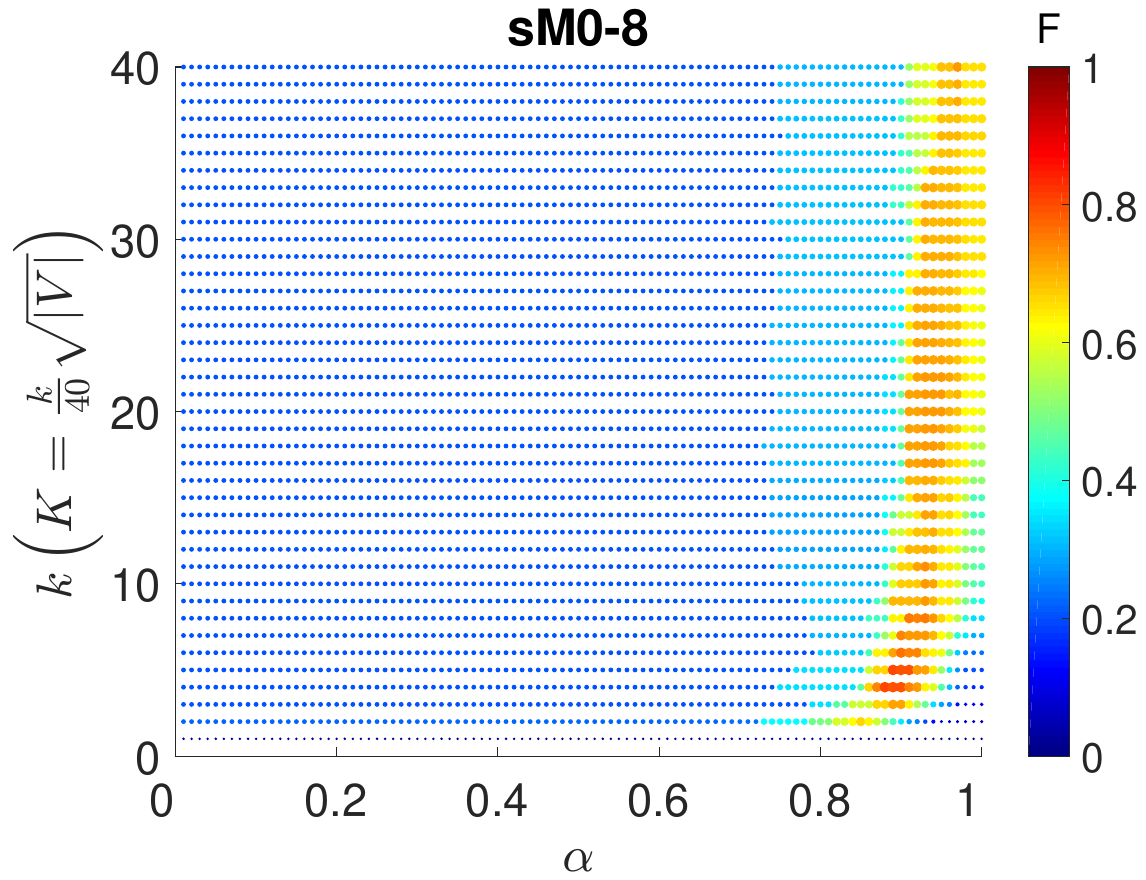}\
}\
\subfloat[M0:8]{
  \label{fig:paraM0-8}
\includegraphics[width = 0.25\linewidth]{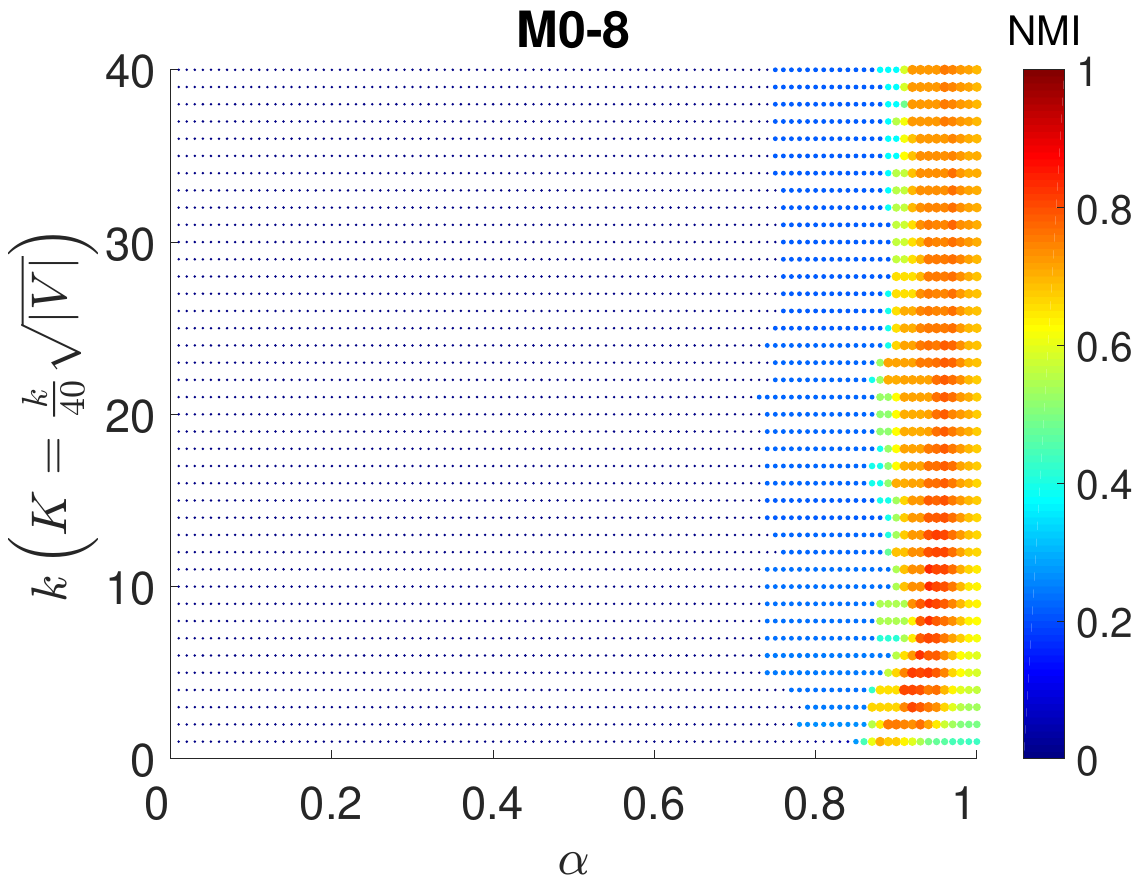}\
\includegraphics[width = 0.25\linewidth]{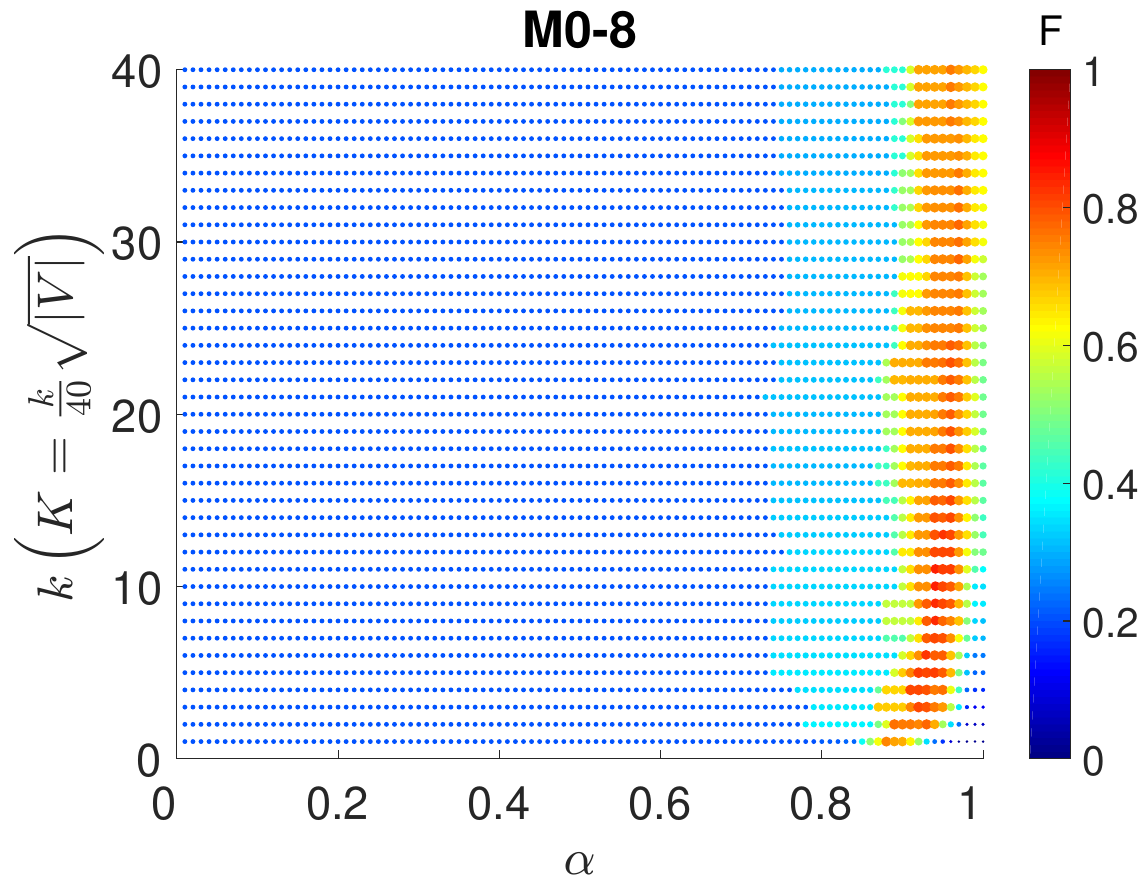}\
}
\subfloat[M0:9]{
  \label{fig:paraM0-9}
\includegraphics[width = 0.25\linewidth]{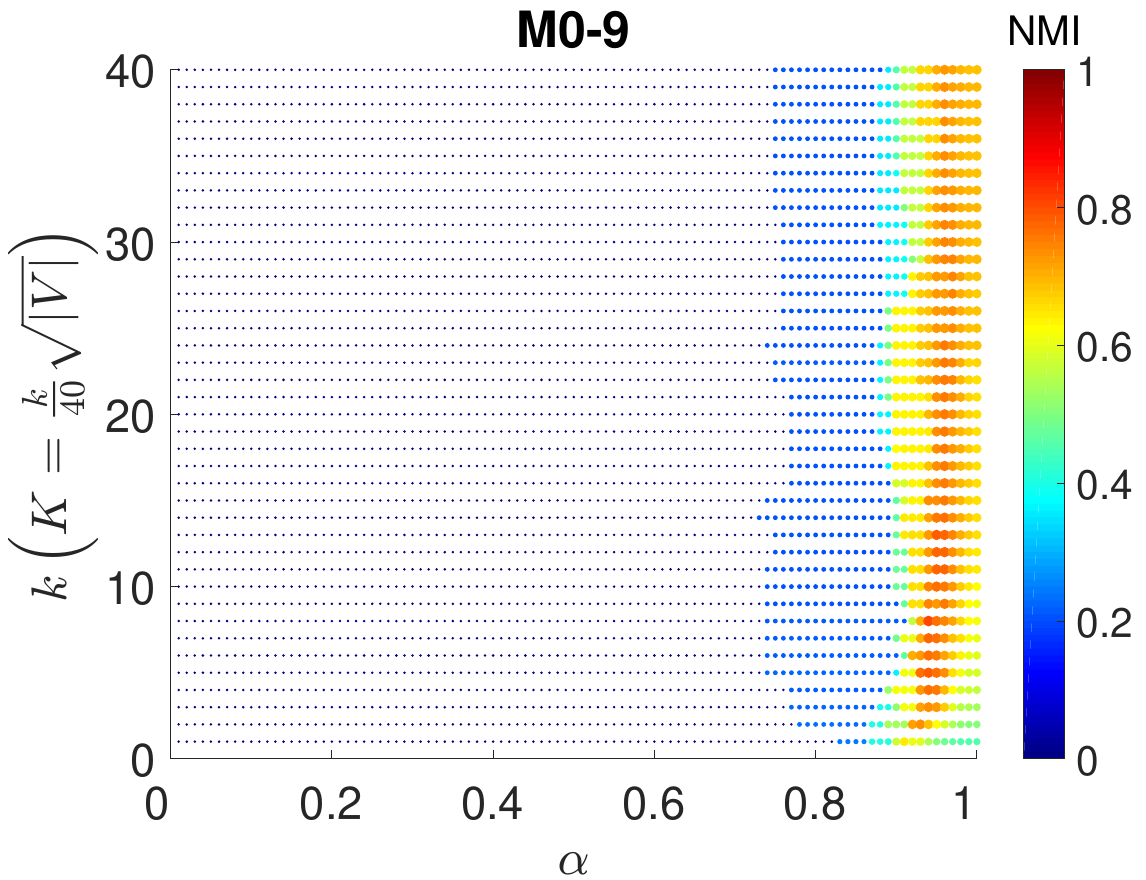}\
\includegraphics[width = 0.25\linewidth]{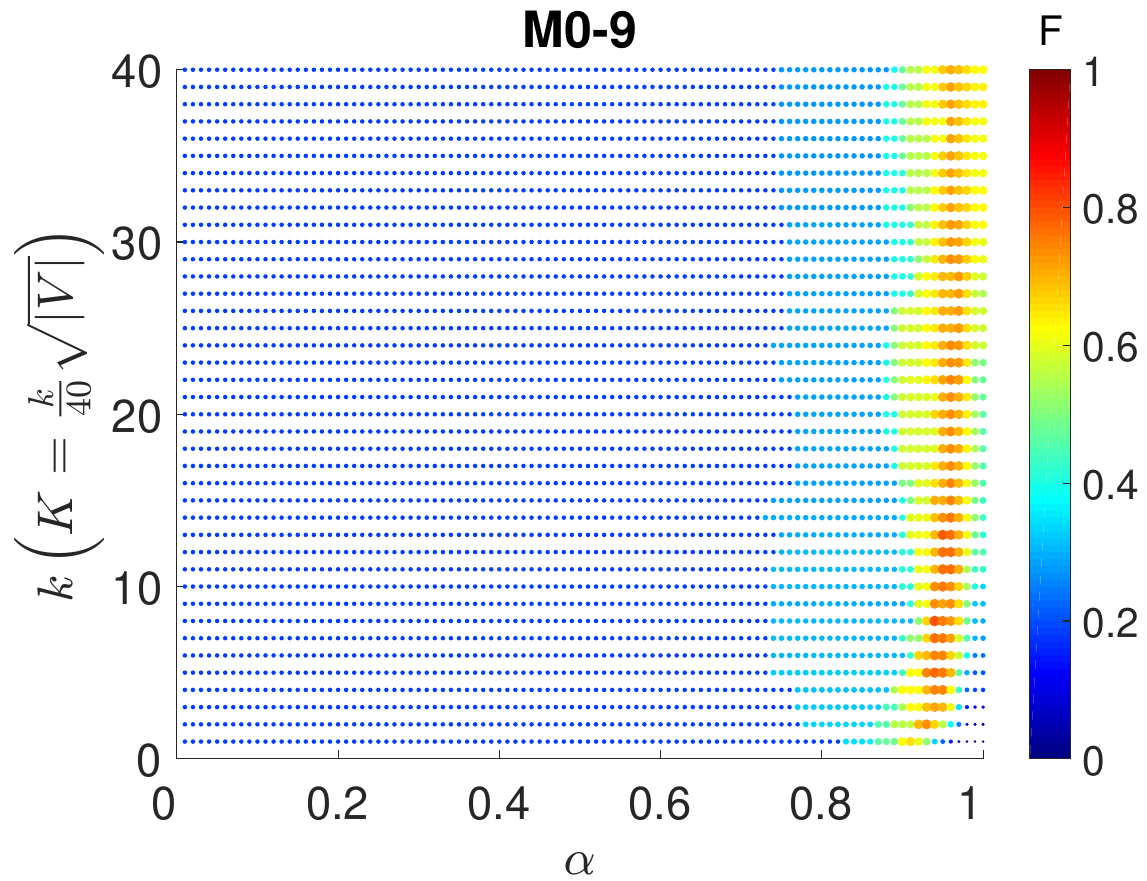}\
}\
\subfloat[face10]{
 \label{fig:paraFace10}
\includegraphics[width = 0.25\linewidth]{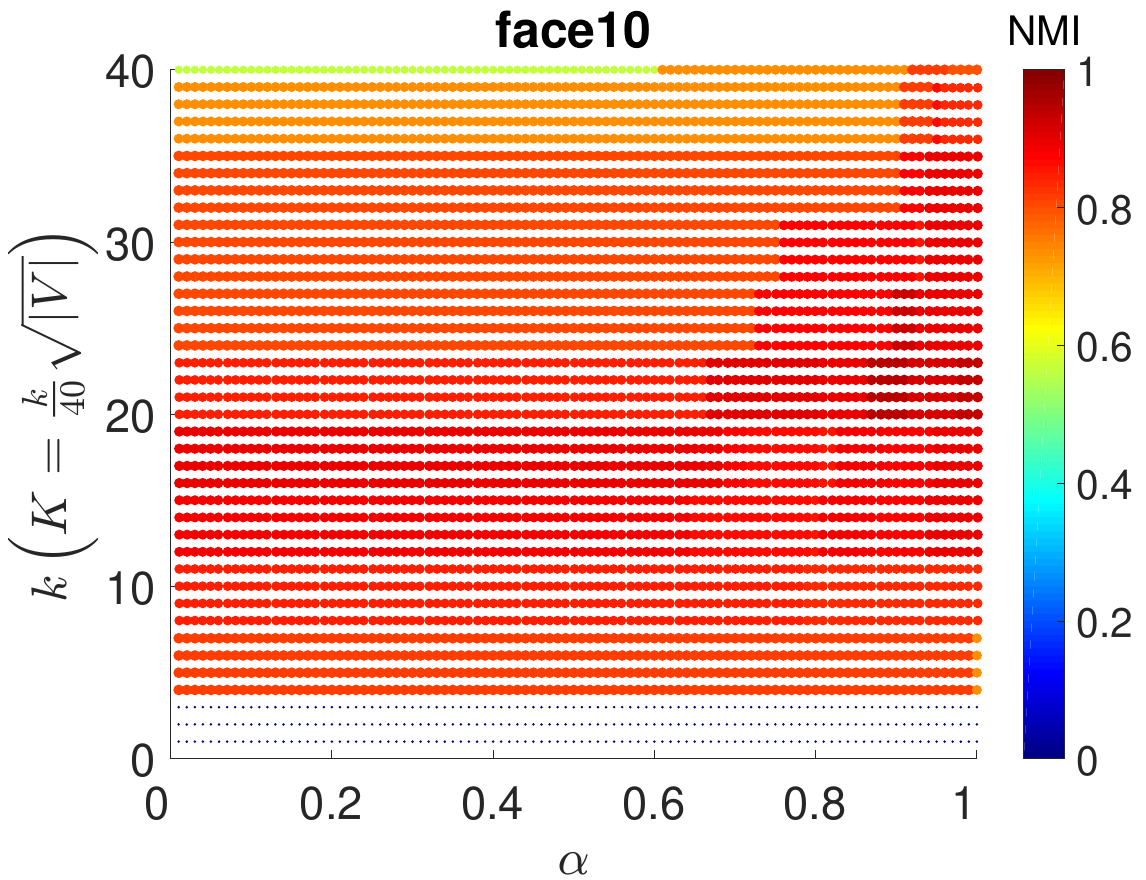}\
\includegraphics[width = 0.25\linewidth]{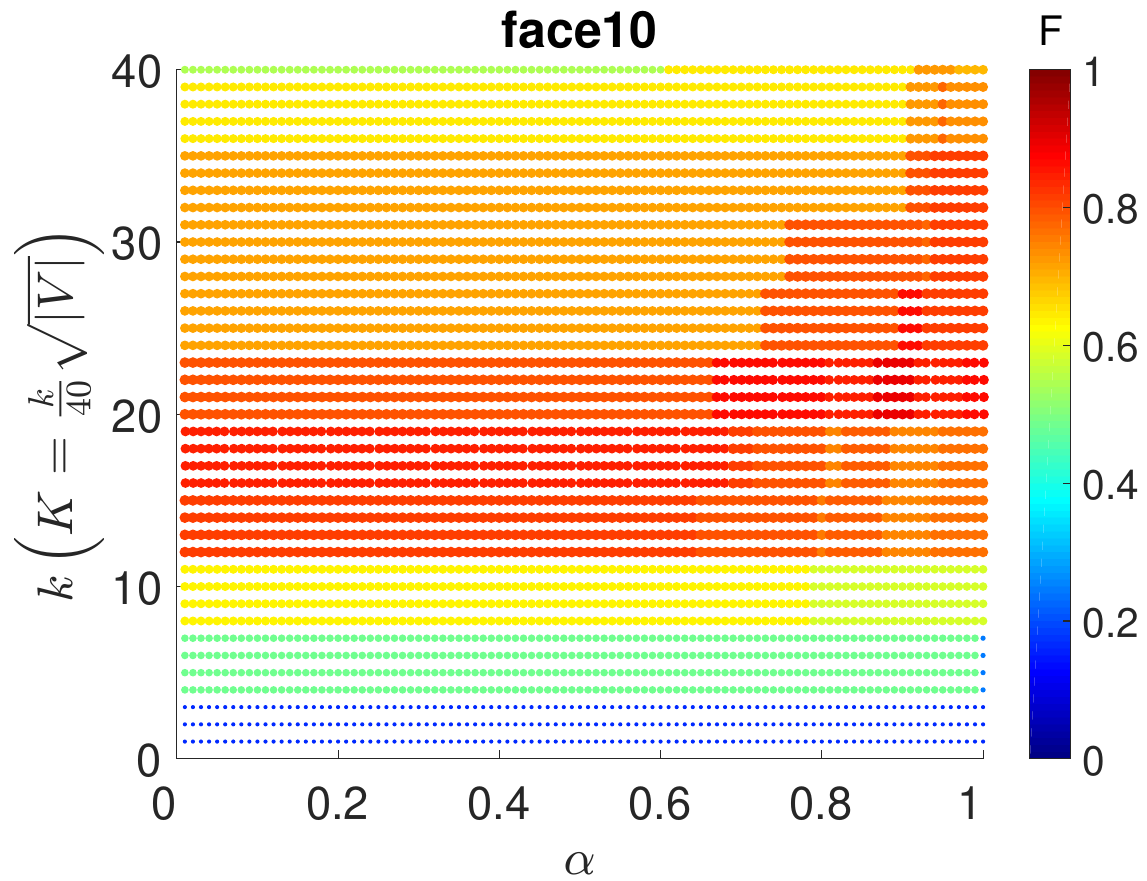}\
}
\subfloat[face40]{
  \label{fig:paraFace40}
\includegraphics[width = 0.25\linewidth]{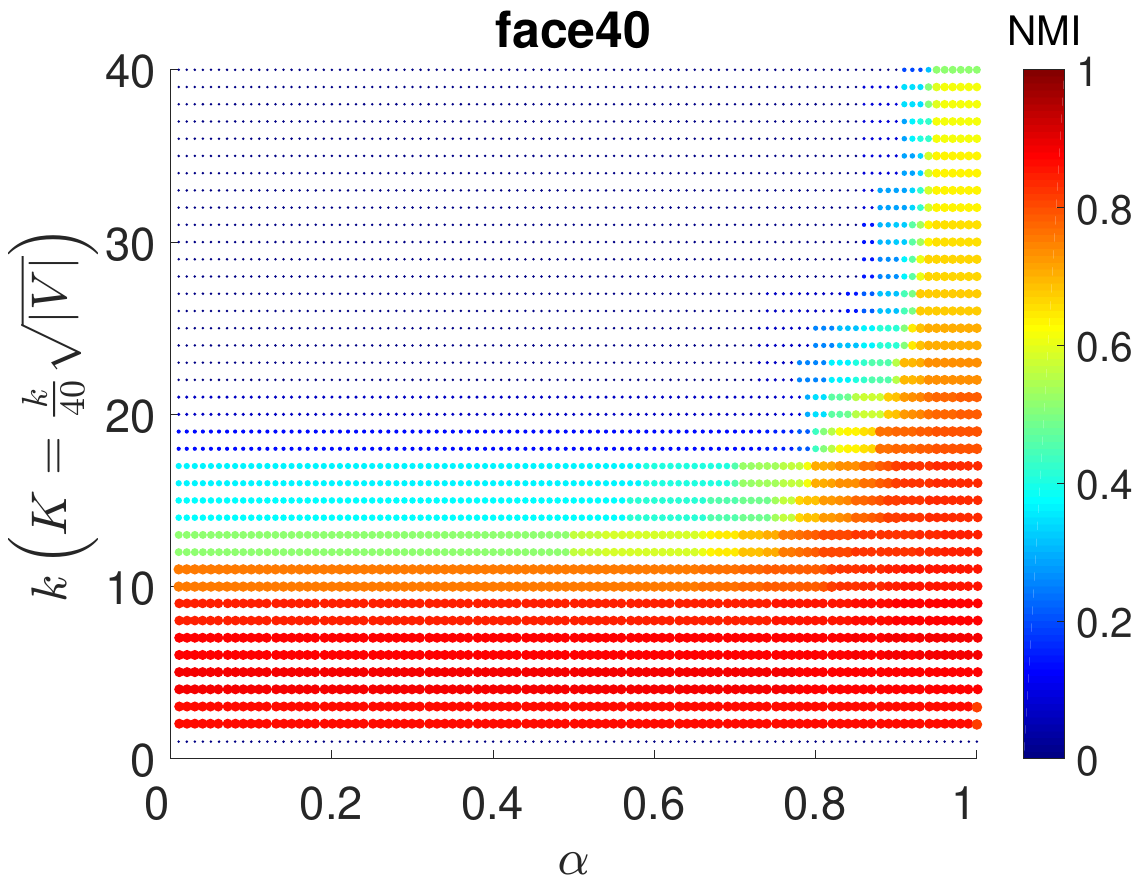}\
\includegraphics[width = 0.25\linewidth]{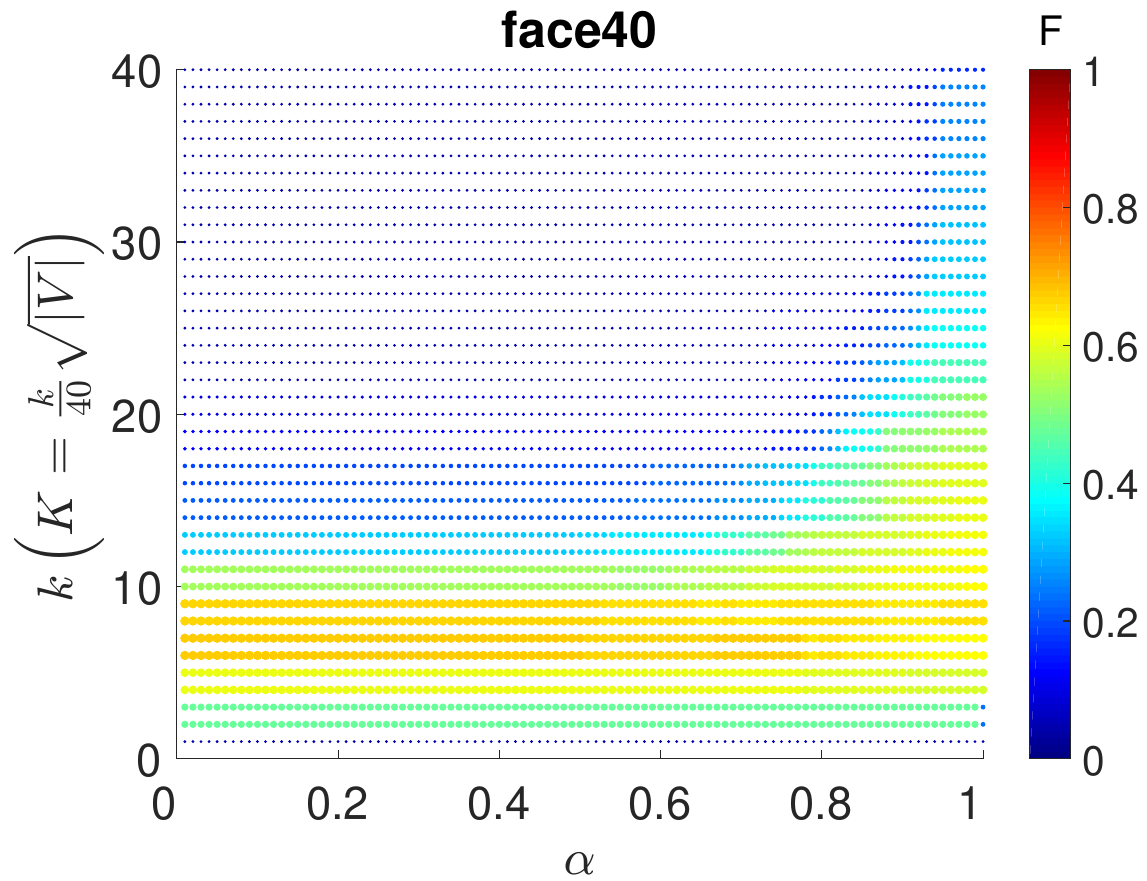}\
}\
\caption{(a)-(d) The clustering performance on 2-D synthetic datasets as a function of $K$ and $\alpha$. (e)-(j) The clustering performance on MNIST training set as a function of $K$ and $\alpha$. (k)-(l) The clustering performance on Olivetti face database as a function of $K$ and $\alpha$.}
\end{minipage}
\label{fig:paraSynt}
\end{figure}

For parameter $\alpha$, as discussed in Section \ref{sec:FJDD}, only a few values will lead to different results. To validate this claim, we search $\alpha$ in the range $[0, 1]$ with a step size of 0.01 to make a better showcase and a evident staircase can be observed from the presentation. To evaluate the clustering results, both NMI and F metrics are used. Next, we present impacts of parameters $K$ and $\alpha$ on NMI and F values for all datasets.

For synthetic datasets, due to the lack of ground truth for S2, S3, and S4, we label them according to the results shown in Figure \ref{fig:RECOMEsyn} as they agree with our intuitive understanding. As shown in Figure \ref{fig:paraS1}, for all the four datasets, both NMI and F approach one when $K$ is above $\sqrt{|V|}/4$ (in the figure, $k = 10$ for $K = \sqrt{V}/4)$ for appropriate $\alpha$'s, and then change very slowly despite the increase of $K$. Meanwhile, when $K$ falls in the range $[\sqrt{|V|}/4, \sqrt{|V|}]$, $\alpha$ valued greater than 0.8 will lead to desirable results. In addition, it is noted that, compared with S3 and S4, the maximum NMI and F values can be attained easier for S1 and S2. This is due to the fact that clusters in both S1 and S2 are convex and well separated, but for the other two datasets, irregular shapes and scales make it difficult to detect the true clusters. Regardless of the complex shapes and scales, RECOME can find the correct cluster numbers and output the desirable results in all cases with properly selected parameters.

Figure \ref{fig:paraM367} to Figure \ref{fig:paraM0-9} show the influences of parameters $K$ and $\alpha$ on performance for MNIST training set. We can see that, similar to the synthetic datasets, both NMI and F values plateau out as $K$ take around $\sqrt{|V|}/4$ for all 6 datasets. At the same time, when $K$ is in the interval $[\sqrt{|V|}/2,\sqrt{|V|}]$, the $\alpha$ value  in the range $[0.85, 0.95]$ gives the best performance. On the other hand, as the number of clusters increases, the clustering performance degrades. For example, NMI over 0.9 can be easily attained for M3467 but not for M0:9. This is because more clusters tend to have more complex sample distribution and more overlapping among different classes. Furthermore, compared with two-dimensional datasets, we can observe that the region in the $\alpha$-$K$ plot that reaches high NMI and F values shrinks. This is due to the fact that the MNIST training set with high dimension is far harder to be clustered well. However, RECOME still achieves better performance across almost all six datasets compared with baseline methods.

For the Olivetti Face Database, the trends NMI and F are different from that in the other datasets. Performance on face10 follows a step-wise pattern with regard to $K$ due to the small sample size (i.e., $\sqrt{|V|}=10$). For face40, both the NMI and F values remain stable when $K$ varies from $\sqrt{|V|}/40$ to $\sqrt{|V|}/4$. Then, with the increasing of $K$, the two indices drop rapidly and become quite small. This is
because the ground truth partition constitutes 40 clusters and only 10 pictures in each cluster. The statistical error of the estimated density on such a small set of pictures is large \cite{rodriguez2014clustering}. Therefore, for
datasets consisting of clusters with few objects, $K$ should be carefully tuned.

In summary, the performance of RECOME is dependent on both parameters of $K$ and $\alpha$. For most datasets, $K$ in the range of $[\sqrt{|V|}/2, \sqrt{|V|}]$ leads to a stable condition. Under this condition, the $\alpha$ value that produces a good performance falls in the range of $[0.8, 1]$. This implies that RECOME is not very sensitive to small changes in parameters $K$ and $\alpha$. Furthermore, as discussed in Section \ref{sec:FJDD}, the algorithm FJDD can quickly determine the JD set with tiny size. Thus, those who fall in $[0.8,1]$ can then be used as the candidate set for $\alpha$.

\subsection{FJDD validation}

\begin{figure}[!htb]
\centering
\subfloat[S1]{
\includegraphics[width = 0.245\linewidth]{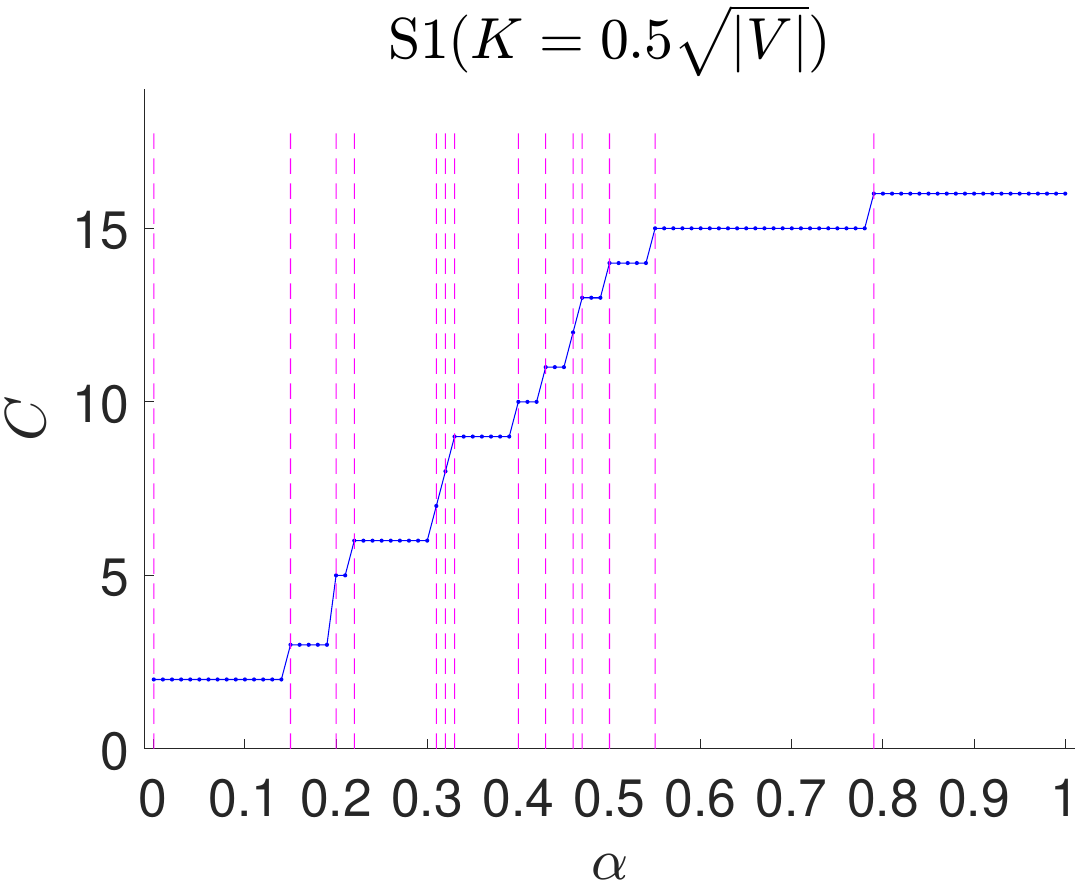}
}
\subfloat[S2]{
\includegraphics[width = 0.245\linewidth]{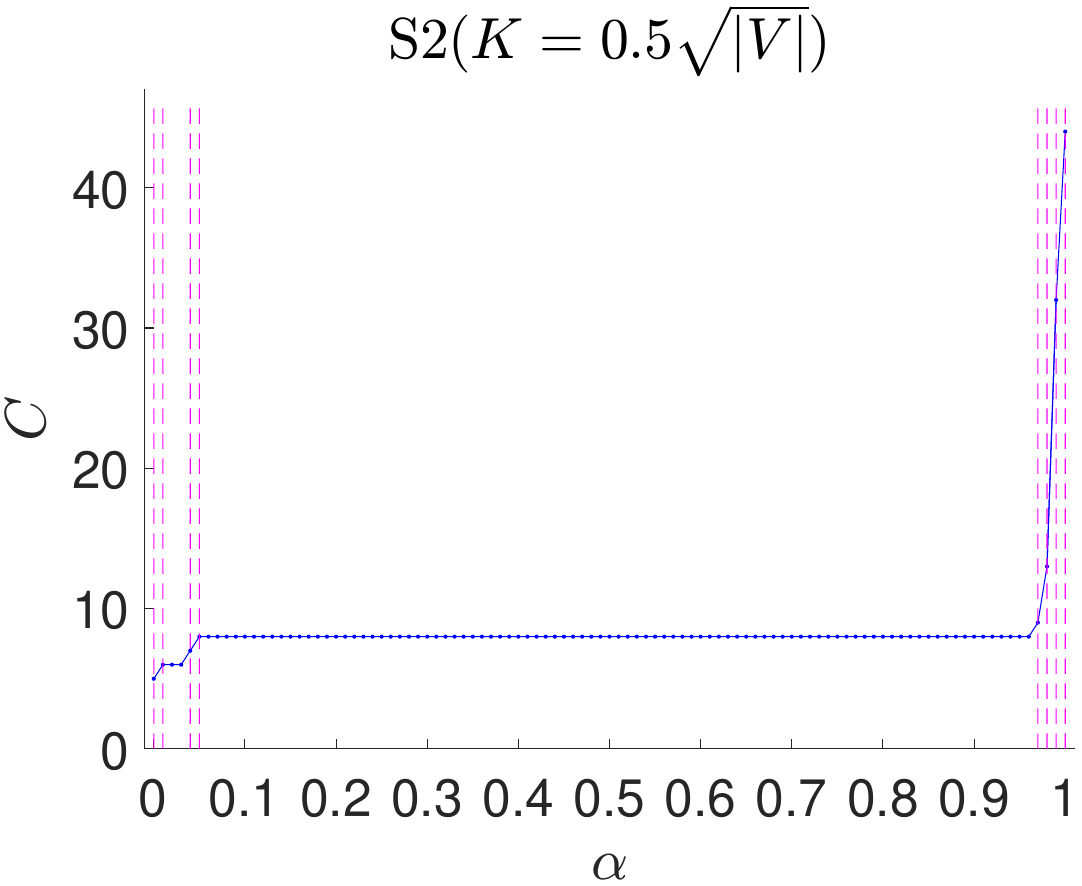}
}
\subfloat[S3]{
\includegraphics[width = 0.245\linewidth]{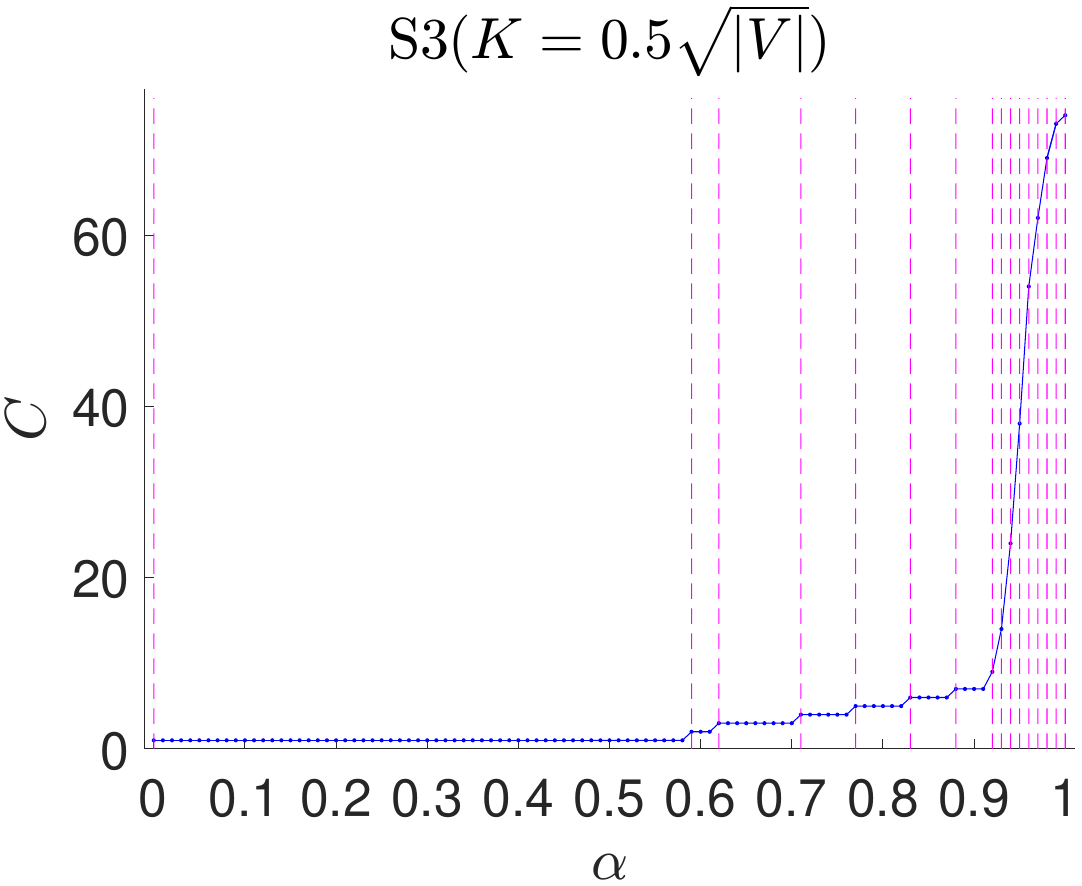}
}
\subfloat[S4]{
\includegraphics[width = 0.245\linewidth]{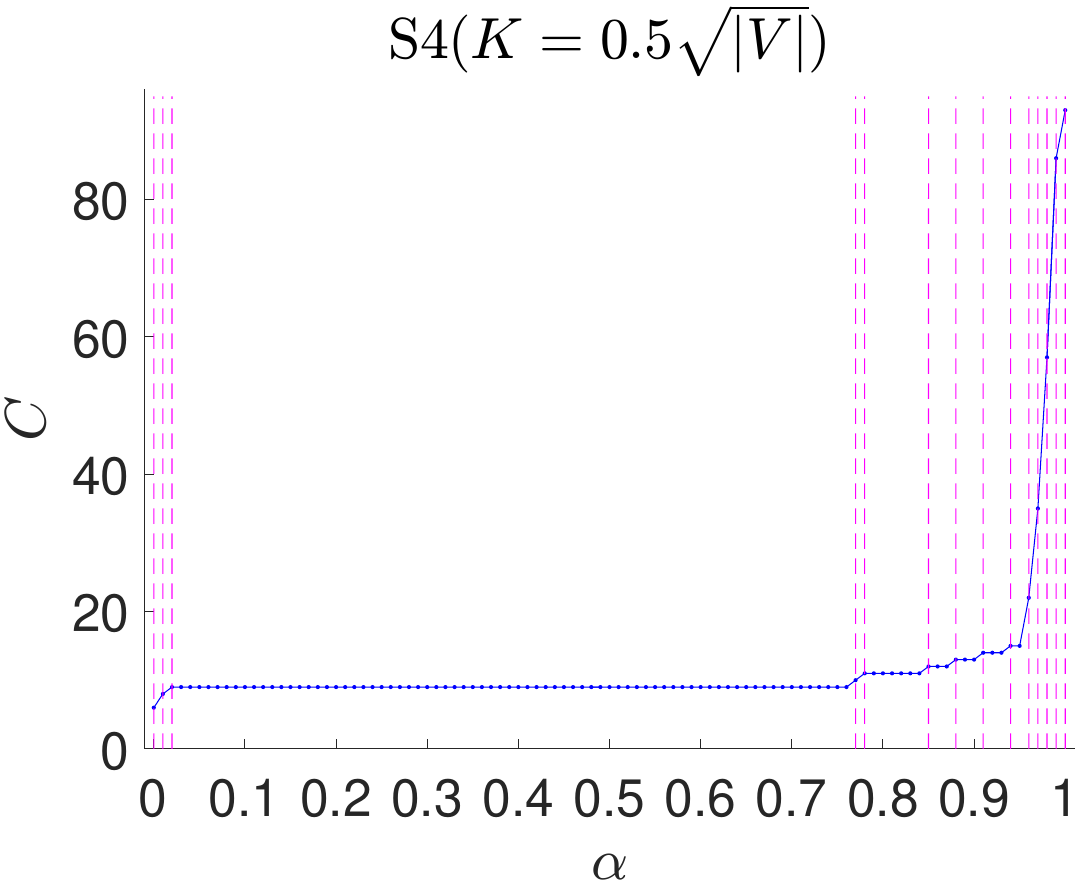}
}\
\subfloat[M367]{
\includegraphics[width = 0.245\linewidth]{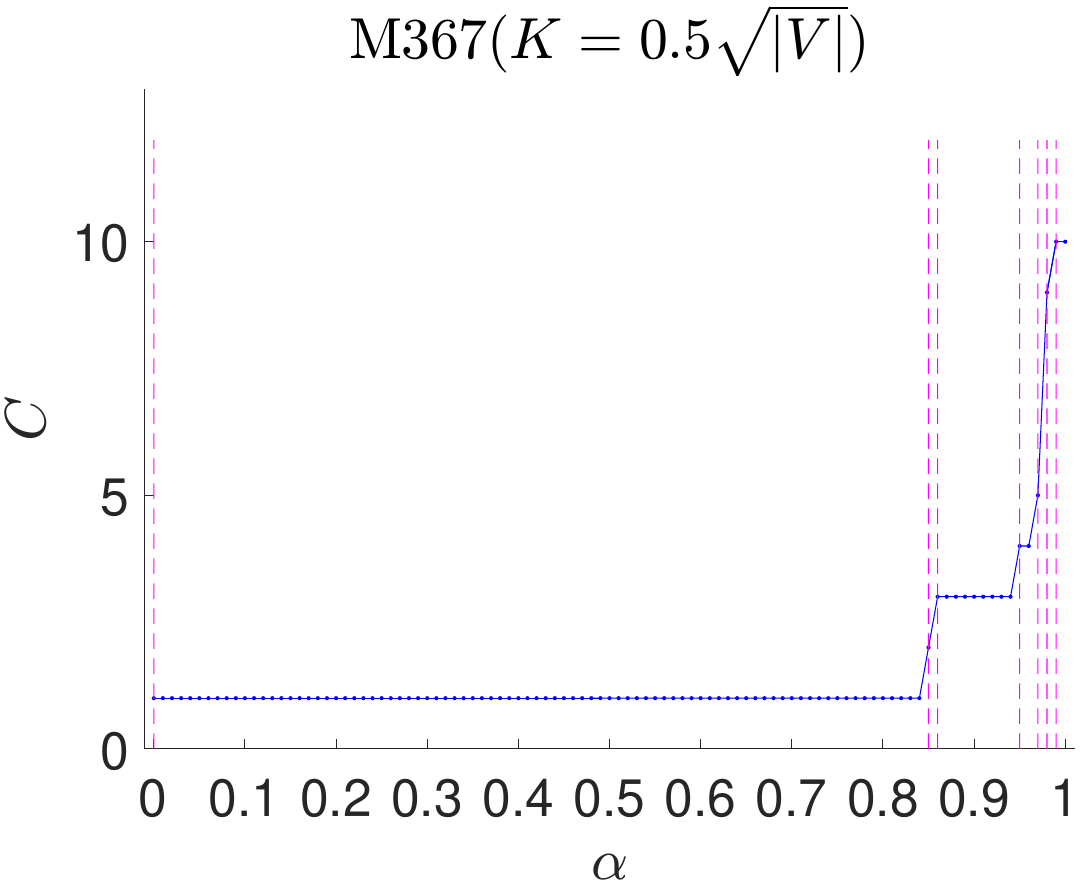}
}
\subfloat[sM3467]{
\includegraphics[width = 0.245\linewidth]{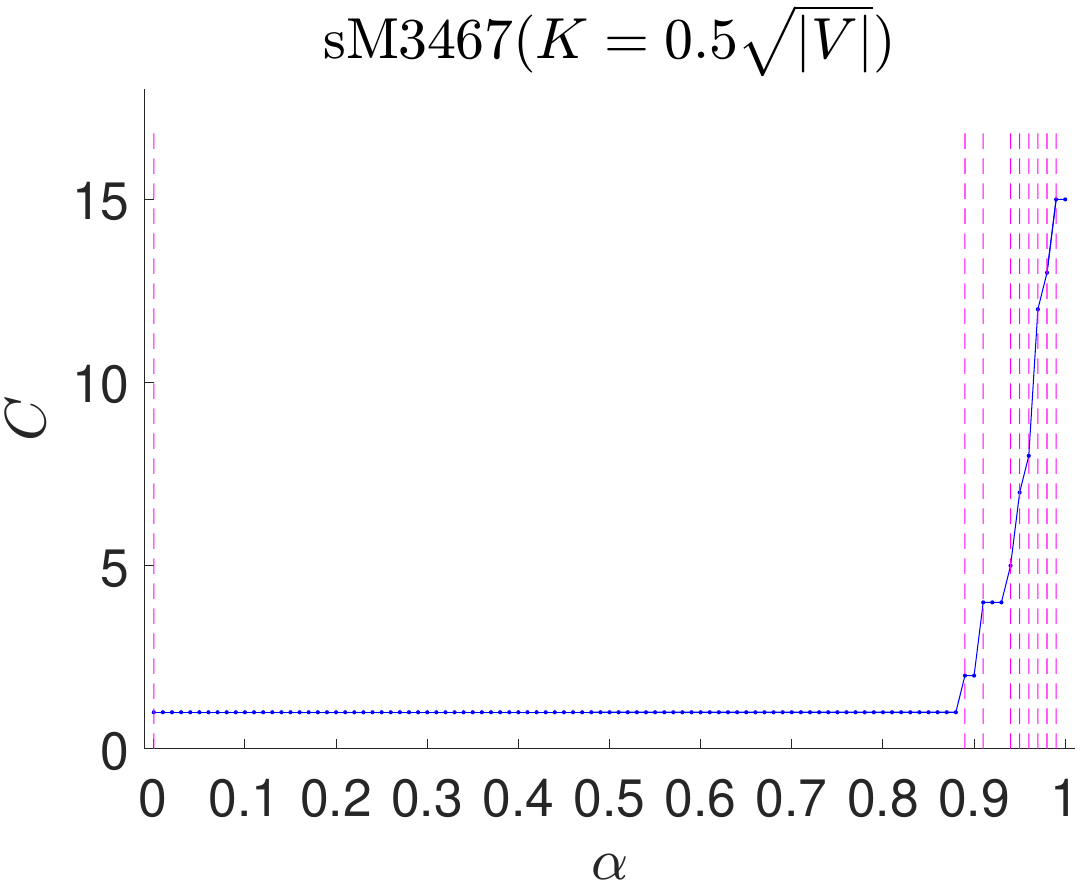}
}
\subfloat[M3467]{
\includegraphics[width = 0.245\linewidth]{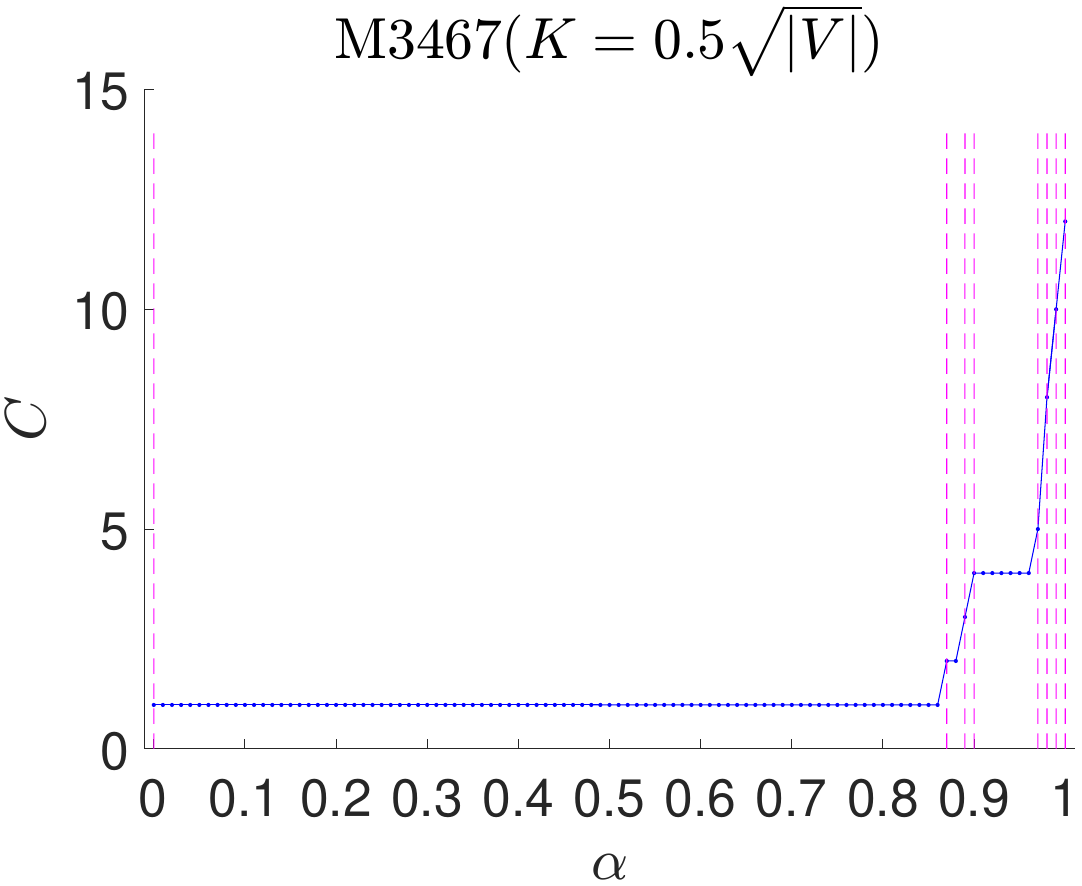}
}
\subfloat[sM0-8]{
\includegraphics[width = 0.245\linewidth]{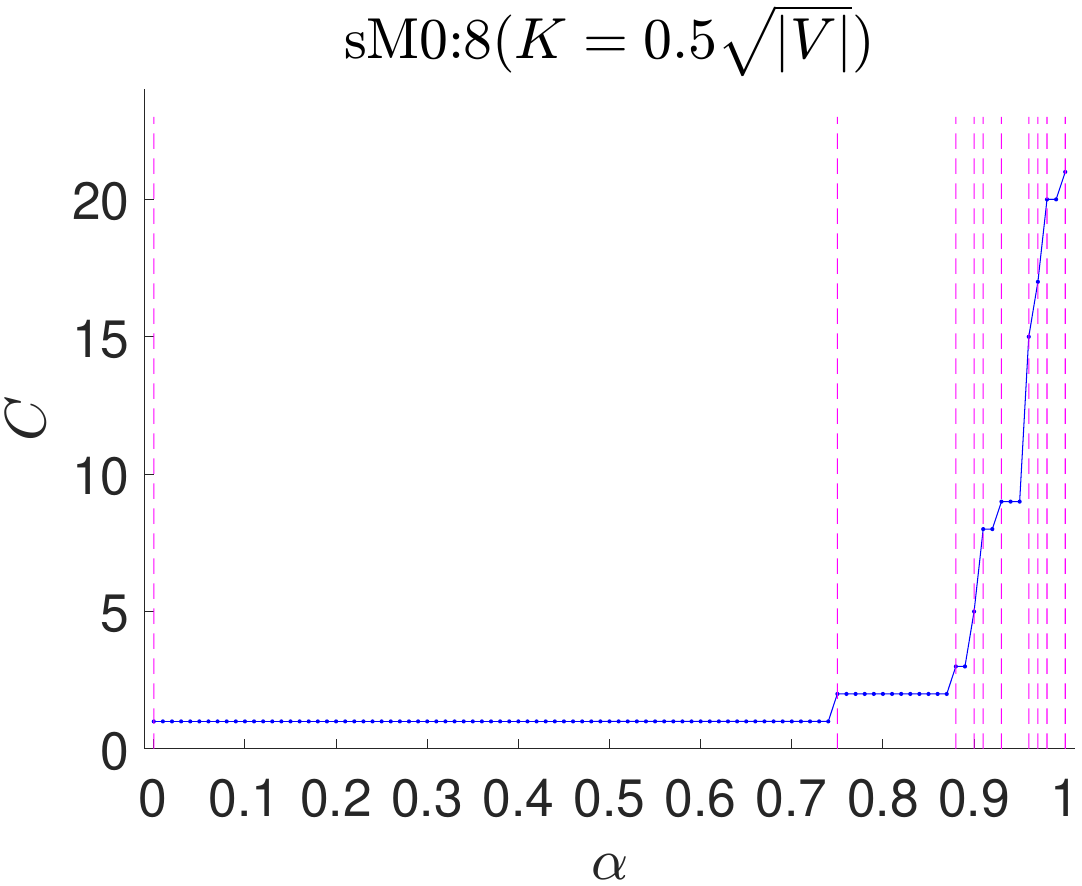}
}\
\subfloat[M0-8]{
\includegraphics[width = 0.245\linewidth]{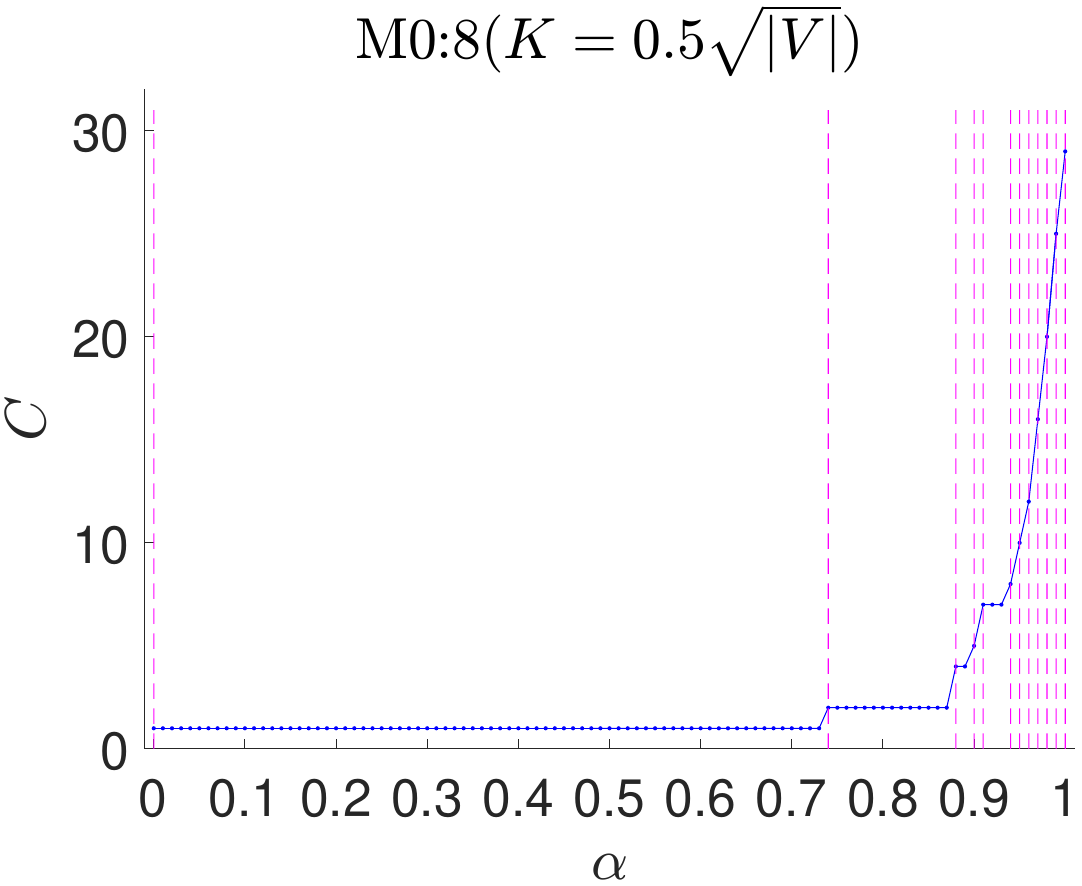}
}
\subfloat[M0-9]{
\includegraphics[width = 0.245\linewidth]{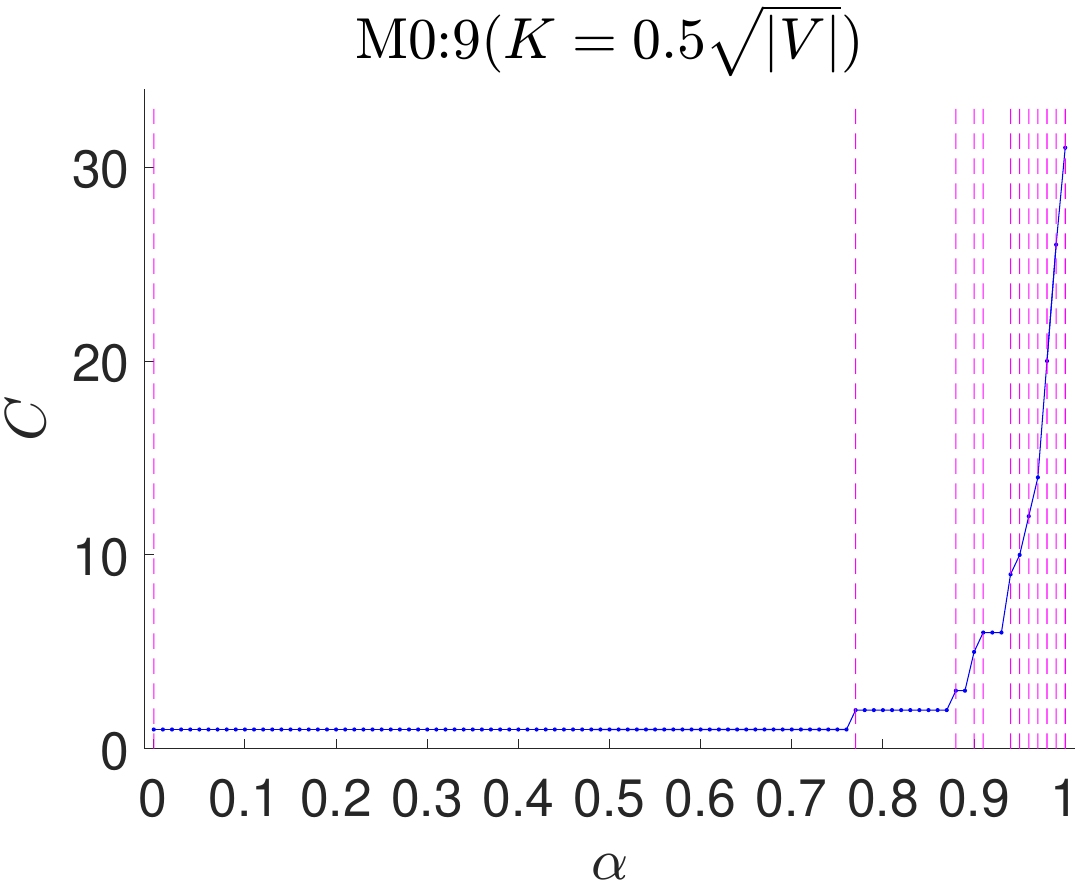}
}
\subfloat[face10]{
\includegraphics[width = 0.245\linewidth]{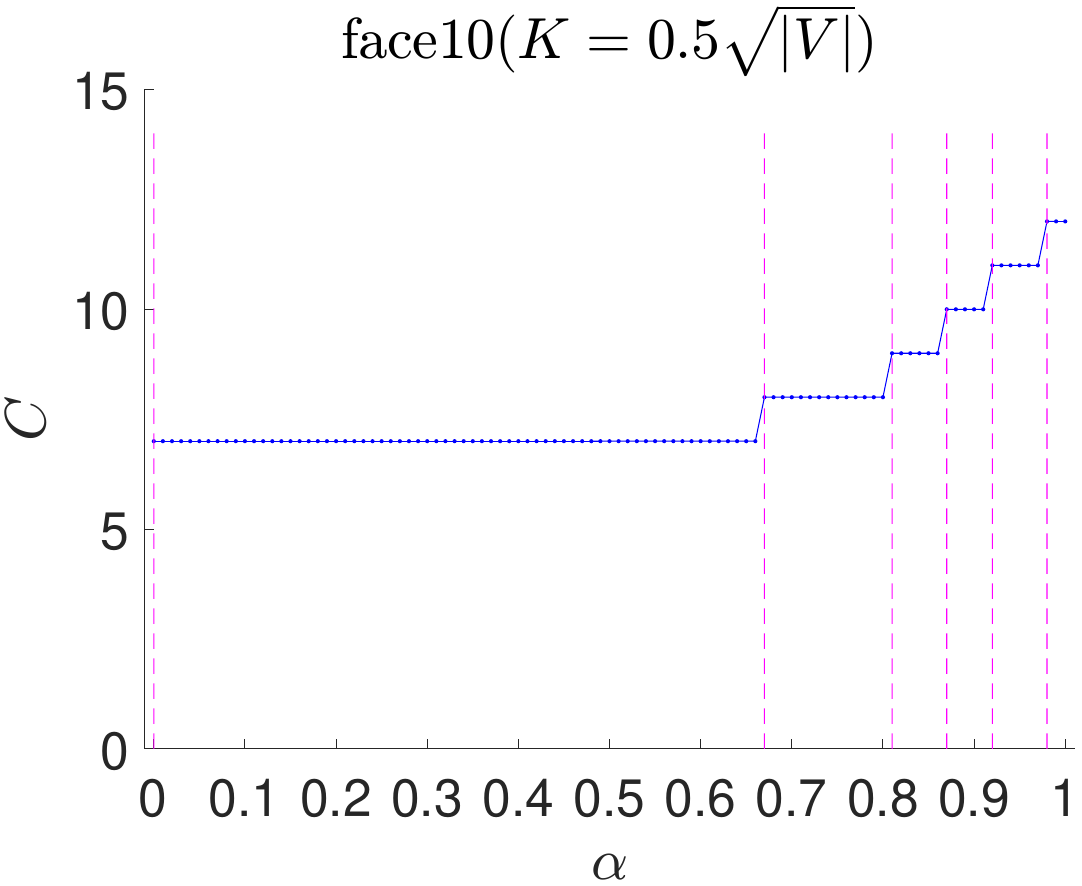}
}
\subfloat[face40]{
\includegraphics[width = 0.245\linewidth]{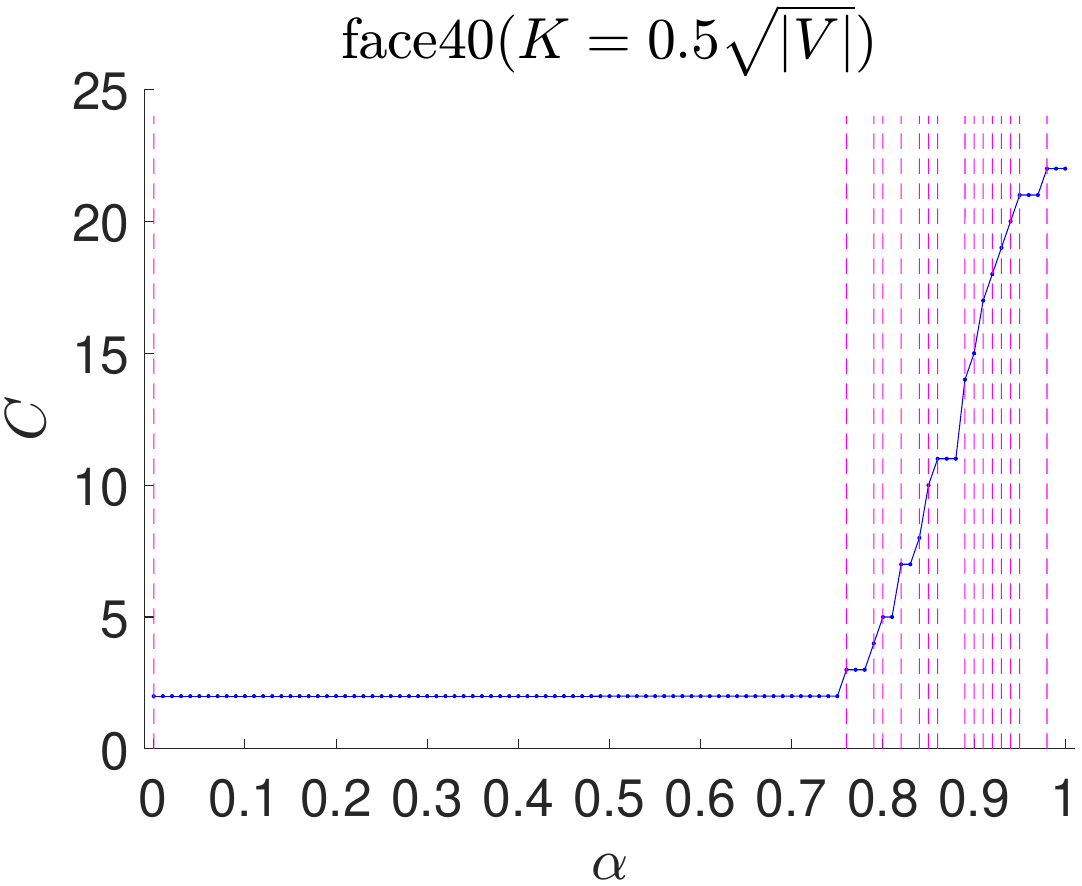}
}
\caption{The cluster number $C$ as a step function of $\alpha$ when $K=0.5\sqrt{|V|}$. The elements of JD set $L$ produced by Algorithm \ref{algo:FJDD} are marked by magenta dashed line.}
\label{fig:FJDDvali}
\end{figure}

As discussed in Section \ref{sec:FJDD}, the number of clusters produced by RECOME is a step function of $\alpha$. In this experiment, we fix $K$ at $0.5\sqrt{\vert V \vert}$ and search $\alpha$ in $[0, 1]$ with a step size 0.01. The resulting cluster number as a function of $\alpha$ is plotted in Figure~\ref{fig:FJDDvali}. It can be observed that all curves are step-wise.

We further verify the correctness of Algorithm \ref{algo:FJDD} on all datasets used in this paper with the jump points shown in the magenta dashed lines. From Figure~\ref{fig:FJDDvali}, it can be seen that FJDD extracts all jump discontinuity correctly for all datasets. Interesting, the number of jump discontinuity is less than 20 for all datasets, which means that to obtain all possible clustering results, only at most 20\% of run times is needed compared with searching for $\alpha$ in $[0, 1]$ with the step size 0.01.


\section{Conclusion}
\label{sec:conc}
In this paper, we presented a new density-based clustering method RECOME. RECOME exploits a novel density measure RNKD to detect core objects with various densities and generate atom clusters. A merging strategy based on KNN graph has been introduced to refine the atom clusters. In addition, we discovered that the number of clusters obtained by RECOME is a monotonic step function of $\alpha$, and therefore proposed an auxiliary algorithm FJDD to help end users to select parameter $\alpha$. Experiment evaluations showed that RECOME can discover clusters of different shapes, densities, and scales with parameters chosen in stable ranges. FJDD can significantly reduce the number of effective choices for the parameter $\alpha$.

In future research, we will extend the current work in three aspects. Firstly, we will design a distributed RECOME in cluster-computing frameworks (e.g. Apache Spark) to handle large volume of data. Secondly, when facing massive data, the number of jump points can grow significantly. We will investigate how to quickly identify a relevant range of $\alpha$. Lastly, it is attractive to design an automatic strategy for parameter selection to make RECOME parameter-free.


\section*{Acknowledgments}
This work is partly supported by the Fundamental Research Funds for the Central Universities (2016JBZ006), Beijing Natural Science Foundation (No. J160004), and National Science and Engineering Council, Canada.

\appendix

\section{Proof of Theorem \ref{thm:contOfNKD}}

\begin{lemma}
\label{lem:KNNIneq}
 $\forall u,v\in V$, the difference between $d_i(u)$ and  $d_i(v)$ is not greater than $d(u,v)$, i.e., $|d_i(u)-d_i(v)|\leq d(u,v)$ $(i=1,2,...,K)$.
\end{lemma}
\begin{proof}
It is equivalent to show that the following holds,
\begin{equation}
d_i(v)-d(u,v)\leq d_i(u) \leq d_i(v)+d(u,v).
\label{equ:ineqDoubSide}
\end{equation}
If $u=v$, the inequality holds trivially.
If $u\not=v$, we first show
\begin{equation}
\label{equ:ineq}
d_i(u) \leq d_i(v)+d(u,v).
\end{equation}
Assume the conclusion is false, namely, $d_i(u) > d_i(v)+d(u,v)$. Then, we can see $d_i(u)>d(u,v)$ according to the non-negativity of distance. On the other hand, we have $d_i(v)=\max\limits_{j\leq i}\{d(v,N_j(v))\}$. From triangle inequality, we have
$$
\begin{aligned}
d_i(u)&>d_i(v)+d(u,v)\\
&=\max_{j\leq i}\{d(v, N_j(v))\}+d(u,v)\\
&\geq \max_{j\leq i}\{d(u, N_j(v))\}.
\end{aligned}
$$
Denote $A=\{v, N_1(v), N_2(v),\dots, N_i(v)\}$. Therefore, $\forall w\in A$ $d_i(u)>d(u,w)$. But $A-\{u\}$ contains at least $i$ elements, a contradiction with the definition of $d_i(u)$. Thus inequality~\eqref{equ:ineq} holds.

Similar argument holds for the left-side of inequality \eqref{equ:ineqDoubSide}.
\end{proof}

Base on Lemma \ref{lem:KNNIneq}, we can prove Theorem \ref{thm:contOfNKD}.

\begin{proof}
$$
\begin{aligned}
|\rho(u)-\rho(v)|&=\theta\left|\sum_{k=1}^K{\exp\left(-\frac{d_k(u)}{\sigma}\right)}-\sum_{k=1}^K{\exp\left(-\frac{d_k(v)}{\sigma}\right)}\right|\\
&\leq\theta\sum_{k=1}^K\left|\exp\left(-\frac{d_k(u)}{\sigma}\right)-\exp\left(-\frac{d_k(v)}{\sigma}\right)\right|.\\
\end{aligned}
$$
Let $h(k)=\arg \min\limits_{w\in\{u,v\}}\{d_k(w)\}$. According to Lemma \ref{lem:KNNIneq} and the monotonicity of the $\exp$ function, we have
\begin{small}
$$
\begin{aligned}
&\theta\sum_{k=1}^K\left|\exp\left(-\frac{d_k(u)}{\sigma}\right)-\exp\left(-\frac{d_k(v)}{\sigma}\right)\right|\\
\leq&\theta\sum_{k=1}^K\left(\exp\left(-\frac{d_k(h(k))}{\sigma}\right)-\exp\left(-\frac{d_k(h(k))+d(u,v)}{\sigma}\right)\right)\\
=&\left(1-\exp\left(-\frac{d(u,v)}{\sigma}\right)\right)\theta\sum_{k=1}^K\exp\left(-\frac{d_k(h(k))}{\sigma}\right)\\
<&\left(1\!-\!\exp\left(-\!\frac{d(u,v)}{\sigma}\right)\right)\theta\sum_{k=1}^K\left(\exp\left(\!-\!\frac{d_k(u)}{\sigma}\right)\!+\!\exp\left(\!-\!\frac{d_k(v)}{\sigma}\right)\right)\\
=&\left(1-\exp\left(-\frac{d(u,v)}{\sigma}\right)\right)(\rho(u)+\rho(v)).
\end{aligned}
$$
\end{small}
Since $\rho(u)+\rho(v)>0$, the result follows.
\end{proof}
\section{Proof of Proposition \ref{prop:JDset}}
\begin{proof}
Denote $A=\{0\}\cup\{c(u,v)|u,v\in O, u\not=v\}=\{\alpha_1, \alpha_2,\dots,\alpha_t\}$, where $0\equiv\alpha_1< \alpha_2<\dots<\alpha_t$. For any $\alpha$ with $0\leq\alpha\leq1$, let $G_K^{>\alpha}$ denote the remaining graph after removing all non-core nodes with weights not larger than $\alpha$ from $G_K$. In addition, let $\text{\#}(G_K^{>\alpha})$ denote the number of components containing at least one core object in $G_K^{>\alpha}$. The key of the proof is the following equation.
\begin{equation}
\text{\#}(V,K,\alpha_i)=\text{\#}(G_K^{>\alpha_i}).
\label{equ:append1}
\end{equation}
To prove the conclusion, only two things need to be shown. First, $\forall i\not=j$, $\text{\#}(V,K, \alpha_i)\not=\text{\#}(V,K, \alpha_j)$. Second, $\forall 0\leq\alpha\leq 1$, $\exists \alpha_i\in A$ such that $\text{\#}(V,K, \alpha)=\text{\#}(V,K, \alpha_i)$.

Now we show the former. Obviously, $\text{\#}(V,K, \alpha)$ is a non-decreasing function with respect to $\alpha$, which implies that showing ``$\forall1\leq i<t, \text{\#}(V,K, \alpha_i)<\text{\#}(V,K, \alpha_{i+1})$'' is enough. Besides, for any $u,v\in O$ with $u\not=v$, ``$u$ and $v$ are located in different components in $G_K^{>\alpha_i}$'' implies ``$u$ and $v$ are located in different components in $G_K^{>\alpha_{i+1}}$''. On the other hand, from the definition of $A$, there exists a pair of core objects $u$ and $v$ such that $u$ and $v$ are connected in $G_K^{>\alpha_i}$ but disconnected in $G_K^{>\alpha_{i+1}}$. This implies ``$u$ and $v$ are located in the same component in $G_K^{>\alpha_i}$'' and ``$u$ and $v$ are located in different components in $G_K^{>\alpha_{i+1}}$''. Therefore, we have $\text{\#}(G_K^{>\alpha_i})<\text{\#}(G_K^{>\alpha_{i+1}})$. According to \eqref{equ:append1}, the conclusion follows.

Now we show the latter. For any $\alpha$ with $0\leq\alpha\leq1$, suppose $\alpha_i$ is the maximal element of $A$ that is not greater than $\alpha$. According to the definition of $A$, we have $\text{\#}(G_K^{>\alpha})=\text{\#}(G_K^{>\alpha_i})$. Thus, according to \eqref{equ:append1}, the desirable result follows.
\end{proof}

\section*{References}
\bibliography{reference}

\begin{thebibliography}{47}
\expandafter\ifx\csname natexlab\endcsname\relax\def\natexlab#1{#1}\fi
\providecommand{\url}[1]{\texttt{#1}}
\providecommand{\href}[2]{#2}
\providecommand{\path}[1]{#1}
\providecommand{\DOIprefix}{doi:}
\providecommand{\ArXivprefix}{arXiv:}
\providecommand{\URLprefix}{URL: }
\providecommand{\Pubmedprefix}{pmid:}
\providecommand{\doi}[1]{\href{http://dx.doi.org/#1}{\path{#1}}}
\providecommand{\Pubmed}[1]{\href{pmid:#1}{\path{#1}}}
\providecommand{\bibinfo}[2]{#2}
\ifx\xfnm\relax \def\xfnm[#1]{\unskip,\space#1}\fi
\bibitem[{Agrawal et~al.(1998)Agrawal, Gehrke, Gunopulos, and
  Raghavan}]{agrawal1998automatic}
\bibinfo{author}{R.~Agrawal}, \bibinfo{author}{J.~Gehrke},
  \bibinfo{author}{D.~Gunopulos}, \bibinfo{author}{P.~Raghavan},
  \bibinfo{title}{Automatic subspace clustering of high dimensional data for
  data mining applications}, volume~\bibinfo{volume}{27},
  \bibinfo{publisher}{ACM}, \bibinfo{year}{1998}.
\bibitem[{Amig車 et~al.(2009)Amig車, Gonzalo, Artiles, and Verdejo}]{Amig2009A}
\bibinfo{author}{E.~Amig車}, \bibinfo{author}{J.~Gonzalo},
  \bibinfo{author}{J.~Artiles}, \bibinfo{author}{F.~Verdejo},
\newblock \bibinfo{title}{A comparison of extrinsic clustering evaluation
  metrics based on formal constraints},
\newblock \bibinfo{journal}{Information Retrieval} \bibinfo{volume}{12}
  (\bibinfo{year}{2009}) \bibinfo{pages}{461--486}.
\bibitem[{Ankerst et~al.(1999)Ankerst, Breunig, Kriegel, and
  Sander}]{ankerst1999optics}
\bibinfo{author}{M.~Ankerst}, \bibinfo{author}{M.~M. Breunig},
  \bibinfo{author}{H.-P. Kriegel}, \bibinfo{author}{J.~Sander},
\newblock \bibinfo{title}{Optics: ordering points to identify the clustering
  structure},
\newblock in: \bibinfo{booktitle}{ACM SIGMOD}, volume~\bibinfo{volume}{28},
  \bibinfo{year}{1999}, pp. \bibinfo{pages}{49--60}.
\bibitem[{Beckmann et~al.(1990)Beckmann, Kriegel, Schneider, and
  Seeger}]{beckmann1990r}
\bibinfo{author}{N.~Beckmann}, \bibinfo{author}{H.-P. Kriegel},
  \bibinfo{author}{R.~Schneider}, \bibinfo{author}{B.~Seeger},
\newblock \bibinfo{title}{The r*-tree: an efficient and robust access method
  for points and rectangles},
\newblock in: \bibinfo{booktitle}{ACM Sigmod Record},
  volume~\bibinfo{volume}{19}, \bibinfo{organization}{Acm},
  \bibinfo{year}{1990}, pp. \bibinfo{pages}{322--331}.
\bibitem[{Breunig et~al.(2000)Breunig, Kriegel, Ng, and
  Sander}]{Breunig2000LOF}
\bibinfo{author}{M.~M. Breunig}, \bibinfo{author}{H.~P. Kriegel},
  \bibinfo{author}{R.~T. Ng}, \bibinfo{author}{J.~Sander},
\newblock \bibinfo{title}{Lof: identifying density-based local outliers},
\newblock \bibinfo{journal}{Acm Sigmod Record} \bibinfo{volume}{29}
  (\bibinfo{year}{2000}) \bibinfo{pages}{93--104}.
\bibitem[{Brown(2014)}]{Brown2014Building}
\bibinfo{author}{R.~A. Brown},
\newblock \bibinfo{title}{Building a balanced k-d tree in o(kn log n) time},
\newblock \bibinfo{journal}{CoRR} \bibinfo{volume}{abs/1410.5420}
  (\bibinfo{year}{2014}).
\bibitem[{Cassisi et~al.(2013)Cassisi, Ferro, Giugno, Pigola, and
  Pulvirenti}]{Cassisi2013Enhancing}
\bibinfo{author}{C.~Cassisi}, \bibinfo{author}{A.~Ferro},
  \bibinfo{author}{R.~Giugno}, \bibinfo{author}{G.~Pigola},
  \bibinfo{author}{A.~Pulvirenti},
\newblock \bibinfo{title}{Enhancing density-based clustering: Parameter
  reduction and outlier detection},
\newblock \bibinfo{journal}{Information Systems} \bibinfo{volume}{38}
  (\bibinfo{year}{2013}) \bibinfo{pages}{317--330}.
\bibitem[{Ert{\"o}z et~al.(2003)Ert{\"o}z, Steinbach, and
  Kumar}]{Levent2003Finding}
\bibinfo{author}{L.~Ert{\"o}z}, \bibinfo{author}{M.~Steinbach},
  \bibinfo{author}{V.~Kumar},
\newblock \bibinfo{title}{Finding clusters of different sizes, shapes, and
  densities in noisy, high dimensional data},
\newblock in: \bibinfo{booktitle}{Siam International Conference on Data Mining,
  San Francisco, Ca, Usa, May}, \bibinfo{year}{2003}.
\bibitem[{Ester et~al.(1996)Ester, Kriegel, Sander, and Xu}]{ester1996density}
\bibinfo{author}{M.~Ester}, \bibinfo{author}{H.-P. Kriegel},
  \bibinfo{author}{J.~Sander}, \bibinfo{author}{X.~Xu},
\newblock \bibinfo{title}{A density-based algorithm for discovering clusters in
  large spatial databases with noise.},
\newblock in: \bibinfo{booktitle}{ACM SIGKDD}, volume~\bibinfo{volume}{96},
  \bibinfo{year}{1996}, pp. \bibinfo{pages}{226--231}.
\bibitem[{Han et~al.(2011)Han, Kamber, and Pei}]{han2011data}
\bibinfo{author}{J.~Han}, \bibinfo{author}{M.~Kamber},
  \bibinfo{author}{J.~Pei}, \bibinfo{title}{Data mining: concepts and
  techniques}, \bibinfo{publisher}{Elsevier}, \bibinfo{year}{2011}.
\bibitem[{He et~al.(2017)He, Li, Ai, Geng, Molisch, Kristem, Zhong, and
  Yu}]{He8013075Kernel}
\bibinfo{author}{R.~He}, \bibinfo{author}{Q.~Li}, \bibinfo{author}{B.~Ai},
  \bibinfo{author}{Y.~L.~A. Geng}, \bibinfo{author}{A.~F. Molisch},
  \bibinfo{author}{V.~Kristem}, \bibinfo{author}{Z.~Zhong},
  \bibinfo{author}{J.~Yu},
\newblock \bibinfo{title}{A kernel-power-density-based algorithm for channel
  multipath components clustering},
\newblock \bibinfo{journal}{IEEE Transactions on Wireless Communications}
  \bibinfo{volume}{16} (\bibinfo{year}{2017}) \bibinfo{pages}{7138--7151}.
\bibitem[{Hinneburg and Keim(1999)}]{Hinneburg1999An}
\bibinfo{author}{A.~Hinneburg}, \bibinfo{author}{D.~A. Keim},
\newblock \bibinfo{title}{An efficient approach to clustering in large
  multimedia databases with noise},
\newblock \bibinfo{journal}{KDD} \bibinfo{volume}{98,} (\bibinfo{year}{1999}).
\bibitem[{Jain(2010)}]{jain2010data}
\bibinfo{author}{A.~K. Jain},
\newblock \bibinfo{title}{Data clustering: 50 years beyond k-means},
\newblock \bibinfo{journal}{Pattern recognition letters} \bibinfo{volume}{31}
  (\bibinfo{year}{2010}) \bibinfo{pages}{651--666}.
\bibitem[{Kaufman and Rousseeuw(2008{\natexlab{a}})}]{kaufman2008agglomerative}
\bibinfo{author}{L.~Kaufman}, \bibinfo{author}{P.~J. Rousseeuw},
\newblock \bibinfo{title}{Agglomerative nesting (program agnes)},
\newblock \bibinfo{journal}{Finding Groups in Data: An Introduction to Cluster
  Analysis}  (\bibinfo{year}{2008}{\natexlab{a}}) \bibinfo{pages}{199--252}.
\bibitem[{Kaufman and Rousseeuw(2008{\natexlab{b}})}]{kaufman2008divisive}
\bibinfo{author}{L.~Kaufman}, \bibinfo{author}{P.~J. Rousseeuw},
\newblock \bibinfo{title}{Divisive analysis (program diana)},
\newblock \bibinfo{journal}{Finding Groups in Data: An Introduction to Cluster
  Analysis}  (\bibinfo{year}{2008}{\natexlab{b}}) \bibinfo{pages}{253--279}.
\bibitem[{Kaufmann and Rousseeuw(1987)}]{Kaufmann1987Clustering}
\bibinfo{author}{L.~Kaufmann}, \bibinfo{author}{P.~J. Rousseeuw},
\newblock \bibinfo{title}{Clustering by means of medoids},
\newblock in: \bibinfo{booktitle}{Statistical Data Analysis Based on the
  L1-norm \& Related Methods}, \bibinfo{year}{1987}, pp.
  \bibinfo{pages}{405--416}.
\bibitem[{Kim and Han(2009)}]{KimH09}
\bibinfo{author}{M.~Kim}, \bibinfo{author}{J.~Han},
\newblock \bibinfo{title}{A particle-and-density based evolutionary clustering
  method for dynamic networks},
\newblock \bibinfo{journal}{{PVLDB}} \bibinfo{volume}{2} (\bibinfo{year}{2009})
  \bibinfo{pages}{622--633}.
\bibitem[{Kisore and Koteswaraiah(2016)}]{Kisore2016Improving}
\bibinfo{author}{N.~R. Kisore}, \bibinfo{author}{C.~B. Koteswaraiah},
\newblock \bibinfo{title}{Improving atm coverage area using density based
  clustering algorithm and voronoi diagrams},
\newblock \bibinfo{journal}{Information Sciences} \bibinfo{volume}{376}
  (\bibinfo{year}{2016}) \bibinfo{pages}{1--20}.
\bibitem[{Kriegel et~al.(2011)Kriegel, Kr?ger, Sander, and
  Zimek}]{Kriegel2011Density}
\bibinfo{author}{H.-P. Kriegel}, \bibinfo{author}{P.~Kr?ger},
  \bibinfo{author}{J.~Sander}, \bibinfo{author}{A.~Zimek},
\newblock \bibinfo{title}{Density-based clustering},
\newblock \bibinfo{journal}{Wiley Interdisciplinary Reviews: Data Mining and
  Knowledge Discovery} \bibinfo{volume}{1} (\bibinfo{year}{2011})
  \bibinfo{pages}{231--240}.
\bibitem[{Kubal赤k et~al.(2010)Kubal赤k, Tich?, ?indel芍?, and
  Staron}]{Kubal2010Clustering}
\bibinfo{author}{J.~Kubal赤k}, \bibinfo{author}{P.~Tich?},
  \bibinfo{author}{R.~?indel芍?}, \bibinfo{author}{R.~J. Staron},
\newblock \bibinfo{title}{Clustering methods for agent distribution
  optimization},
\newblock \bibinfo{journal}{IEEE Transactions on Systems Man and Cybernetics
  Part C} \bibinfo{volume}{40} (\bibinfo{year}{2010}) \bibinfo{pages}{78--86}.
\bibitem[{Laohakiat et~al.(2016)Laohakiat, Phimoltares, and
  Lursinsap}]{Laohakiat2016A}
\bibinfo{author}{S.~Laohakiat}, \bibinfo{author}{S.~Phimoltares},
  \bibinfo{author}{C.~Lursinsap},
\newblock \bibinfo{title}{A clustering algorithm for stream data with lda-based
  unsupervised localized dimension reduction},
\newblock \bibinfo{journal}{Information Sciences} \bibinfo{volume}{381}
  (\bibinfo{year}{2016}) \bibinfo{pages}{104每123}.
\bibitem[{Lecun et~al.(1998)Lecun, Bottou, Bengio, and
  Haffner}]{lecun1998gradient}
\bibinfo{author}{Y.~Lecun}, \bibinfo{author}{L.~Bottou},
  \bibinfo{author}{Y.~Bengio}, \bibinfo{author}{P.~Haffner},
\newblock \bibinfo{title}{Gradient-based learning applied to document
  recognition},
\newblock \bibinfo{journal}{Proceedings of the IEEE} \bibinfo{volume}{86}
  (\bibinfo{year}{1998}) \bibinfo{pages}{2278--2324}.
\bibitem[{Liang and Chen(2016)}]{Liang2016delta}
\bibinfo{author}{Z.~Liang}, \bibinfo{author}{P.~Chen},
\newblock \bibinfo{title}{Delta-density based clustering with a
  divide-and-conquer strategy: 3dc clustering},
\newblock \bibinfo{journal}{Pattern Recognition Letters} \bibinfo{volume}{73}
  (\bibinfo{year}{2016}) \bibinfo{pages}{52--59}.
\bibitem[{Liu et~al.(2007)Liu, Zhou, and Wu}]{liu2007vdbscan}
\bibinfo{author}{P.~Liu}, \bibinfo{author}{D.~Zhou}, \bibinfo{author}{N.~Wu},
\newblock \bibinfo{title}{Vdbscan: varied density based spatial clustering of
  applications with noise},
\newblock in: \bibinfo{booktitle}{2007 International conference on service
  systems and service management}, \bibinfo{organization}{IEEE},
  \bibinfo{year}{2007}, pp. \bibinfo{pages}{1--4}.
\bibitem[{Loftsgaarden and Quesenberry(1965)}]{Loftsgaarden1965A}
\bibinfo{author}{D.~O. Loftsgaarden}, \bibinfo{author}{C.~P. Quesenberry},
\newblock \bibinfo{title}{A nonparametric estimate of a multivariate density
  function},
\newblock \bibinfo{journal}{Annals of Mathematical Statistics}
  \bibinfo{volume}{36} (\bibinfo{year}{1965}) \bibinfo{pages}{1049--1051}.
\bibitem[{Loh and Yu(2015)}]{loh2015fast}
\bibinfo{author}{W.~Loh}, \bibinfo{author}{H.~Yu},
\newblock \bibinfo{title}{Fast density-based clustering through dataset
  partition using graphics processing units},
\newblock \bibinfo{journal}{Information Sciences} \bibinfo{volume}{308}
  (\bibinfo{year}{2015}) \bibinfo{pages}{94--112}.
\bibitem[{MacQueen et~al.(1967)}]{macqueen1967some}
\bibinfo{author}{J.~MacQueen}, et~al.,
\newblock \bibinfo{title}{Some methods for classification and analysis of
  multivariate observations},
\newblock in: \bibinfo{booktitle}{Proceedings of the fifth Berkeley symposium
  on mathematical statistics and probability}, volume~\bibinfo{volume}{1},
  \bibinfo{organization}{Oakland, CA, USA.}, \bibinfo{year}{1967}, pp.
  \bibinfo{pages}{281--297}.
\bibitem[{Miller and Han(2001)}]{miller2001spatial}
\bibinfo{author}{H.~Miller}, \bibinfo{author}{J.~Han},
\newblock \bibinfo{title}{Spatial clustering methods in data mining: a survey},
\newblock \bibinfo{journal}{Geographic data mining and knowledge discovery,
  Taylor and Francis}  (\bibinfo{year}{2001}).
\bibitem[{Nanda and Panda(2015)}]{nanda2015design}
\bibinfo{author}{S.~J. Nanda}, \bibinfo{author}{G.~Panda},
\newblock \bibinfo{title}{Design of computationally efficient density-based
  clustering algorithms},
\newblock \bibinfo{journal}{Data and Knowledge Engineering}
  \bibinfo{volume}{95} (\bibinfo{year}{2015}) \bibinfo{pages}{23--38}.
\bibitem[{Newman(2006)}]{newman2006modularity}
\bibinfo{author}{M.~E. Newman},
\newblock \bibinfo{title}{Modularity and community structure in networks},
\newblock \bibinfo{journal}{Proceedings of the national academy of sciences}
  \bibinfo{volume}{103} (\bibinfo{year}{2006}) \bibinfo{pages}{8577--8582}.
\bibitem[{Pei et~al.(2009)Pei, Jasra, Hand, Zhu, and Zhou}]{pei2009decode}
\bibinfo{author}{T.~Pei}, \bibinfo{author}{A.~Jasra}, \bibinfo{author}{D.~J.
  Hand}, \bibinfo{author}{A.-X. Zhu}, \bibinfo{author}{C.~Zhou},
\newblock \bibinfo{title}{Decode: a new method for discovering clusters of
  different densities in spatial data},
\newblock \bibinfo{journal}{Data Mining and Knowledge Discovery}
  \bibinfo{volume}{18} (\bibinfo{year}{2009}) \bibinfo{pages}{337--369}.
\bibitem[{Qiu and Shen(2017)}]{QIU2017102}
\bibinfo{author}{Z.~Qiu}, \bibinfo{author}{H.~Shen},
\newblock \bibinfo{title}{User clustering in a dynamic social network topic
  model for short text streams},
\newblock \bibinfo{journal}{Information Sciences} \bibinfo{volume}{414}
  (\bibinfo{year}{2017}) \bibinfo{pages}{102 -- 116}.
\bibitem[{Rodriguez and Laio(2014)}]{rodriguez2014clustering}
\bibinfo{author}{A.~Rodriguez}, \bibinfo{author}{A.~Laio},
\newblock \bibinfo{title}{Clustering by fast search and find of density peaks},
\newblock \bibinfo{journal}{Science} \bibinfo{volume}{344}
  (\bibinfo{year}{2014}) \bibinfo{pages}{1492--1496}.
\bibitem[{Samaria and Harter(1994)}]{samaria1994parameterisation}
\bibinfo{author}{F.~S. Samaria}, \bibinfo{author}{A.~C. Harter},
\newblock \bibinfo{title}{Parameterisation of a stochastic model for human face
  identification},
\newblock in: \bibinfo{booktitle}{Proceedings of 1994 IEEE Workshop on
  Applications of Computer Vision}, \bibinfo{year}{1994}, pp.
  \bibinfo{pages}{138--142}. \DOIprefix\doi{10.1109/ACV.1994.341300}.
\bibitem[{Sander et~al.(1998)Sander, Ester, Kriegel, and
  Xu}]{Sander1998Density}
\bibinfo{author}{J.~Sander}, \bibinfo{author}{M.~Ester}, \bibinfo{author}{H.~P.
  Kriegel}, \bibinfo{author}{X.~Xu},
\newblock \bibinfo{title}{Density-based clustering in spatial databases: The
  algorithm gdbscan and its applications},
\newblock \bibinfo{journal}{Data Mining and Knowledge Discovery}
  \bibinfo{volume}{2} (\bibinfo{year}{1998}) \bibinfo{pages}{169--194}.
\bibitem[{Shapira et~al.(2011)Shapira, Yuster, and Zwick}]{Shapira2011All}
\bibinfo{author}{A.~Shapira}, \bibinfo{author}{R.~Yuster},
  \bibinfo{author}{U.~Zwick},
\newblock \bibinfo{title}{All-pairs bottleneck paths in vertex weighted
  graphs},
\newblock \bibinfo{journal}{Algorithmica} \bibinfo{volume}{59}
  (\bibinfo{year}{2011}) \bibinfo{pages}{621--633}.
\bibitem[{Shi and Malik(2000)}]{Shi2000Normalized}
\bibinfo{author}{J.~Shi}, \bibinfo{author}{J.~Malik},
\newblock \bibinfo{title}{Normalized cuts and image segmentation},
\newblock \bibinfo{journal}{IEEE Trans.pattern Anal.mach.intell}
  \bibinfo{volume}{22} (\bibinfo{year}{2000}) \bibinfo{pages}{888--905}.
\bibitem[{Strehl and Ghosh(2002)}]{Strehl2002Cluster}
\bibinfo{author}{A.~Strehl}, \bibinfo{author}{J.~Ghosh},
\newblock \bibinfo{title}{Cluster ensembles --- a knowledge reuse framework for
  combining multiple partitions},
\newblock \bibinfo{journal}{Journal of Machine Learning Research}
  \bibinfo{volume}{3} (\bibinfo{year}{2002}) \bibinfo{pages}{583--617}.
\bibitem[{Terrell and Scott(1992)}]{terrell1992variable}
\bibinfo{author}{G.~R. Terrell}, \bibinfo{author}{D.~W. Scott},
\newblock \bibinfo{title}{Variable kernel density estimation},
\newblock \bibinfo{journal}{The Annals of Statistics}  (\bibinfo{year}{1992})
  \bibinfo{pages}{1236--1265}.
\bibitem[{Tran et~al.(2006)Tran, Wehrens, and Buydens}]{tran2006knn}
\bibinfo{author}{T.~N. Tran}, \bibinfo{author}{R.~Wehrens},
  \bibinfo{author}{L.~M. Buydens},
\newblock \bibinfo{title}{Knn-kernel density-based clustering for
  high-dimensional multivariate data},
\newblock \bibinfo{journal}{Computational Statistics and Data Analysis}
  \bibinfo{volume}{51} (\bibinfo{year}{2006}) \bibinfo{pages}{513--525}.
\bibitem[{Wang et~al.(2012)Wang, Lai, and Zhu}]{Wang2012Graph}
\bibinfo{author}{C.~D. Wang}, \bibinfo{author}{J.~H. Lai},
  \bibinfo{author}{J.~Y. Zhu},
\newblock \bibinfo{title}{Graph-based multiprototype competitive learning and
  its applications},
\newblock \bibinfo{journal}{IEEE Transactions on Systems Man and Cybernetics
  Part C} \bibinfo{volume}{42} (\bibinfo{year}{2012})
  \bibinfo{pages}{934--946}.
\bibitem[{Wang and Song(2016)}]{Wang2016Automatic}
\bibinfo{author}{G.~Wang}, \bibinfo{author}{Q.~Song},
\newblock \bibinfo{title}{Automatic clustering via outward statistical testing
  on density metrics},
\newblock \bibinfo{journal}{IEEE Transactions on Knowledge \& Data Engineering}
  \bibinfo{volume}{28} (\bibinfo{year}{2016}) \bibinfo{pages}{1--1}.
\bibitem[{Wang et~al.(1997)Wang, Yang, and Muntz}]{wang1997sting}
\bibinfo{author}{W.~Wang}, \bibinfo{author}{J.~Yang}, \bibinfo{author}{R.~R.
  Muntz},
\newblock \bibinfo{title}{Sting: A statistical information grid approach to
  spatial data mining},
\newblock in: \bibinfo{booktitle}{VLDB}, volume~\bibinfo{volume}{97},
  \bibinfo{year}{1997}, pp. \bibinfo{pages}{186--195}.
\bibitem[{Xu et~al.(2007)Xu, Yuruk, Feng, and Schweiger}]{Xu2007SCAN}
\bibinfo{author}{X.~Xu}, \bibinfo{author}{N.~Yuruk}, \bibinfo{author}{Z.~Feng},
  \bibinfo{author}{T.~A.~J. Schweiger},
\newblock \bibinfo{title}{Scan: a structural clustering algorithm for
  networks},
\newblock in: \bibinfo{booktitle}{ACM SIGKDD International Conference on
  Knowledge Discovery and Data Mining}, \bibinfo{year}{2007}, pp.
  \bibinfo{pages}{824--833}.
\bibitem[{Ye et~al.(2003)Ye, Gao, and Zeng}]{Qixiang2003Multimedia}
\bibinfo{author}{Q.~Ye}, \bibinfo{author}{W.~Gao}, \bibinfo{author}{W.~Zeng},
\newblock \bibinfo{title}{Color image segmentation using density-based
  clustering},
\newblock in: \bibinfo{booktitle}{Multimedia and Expo, 2003. ICME '03.
  Proceedings. 2003 International Conference on}, volume~\bibinfo{volume}{2},
  \bibinfo{year}{2003}, pp. \bibinfo{pages}{II--401--4 vol.2}.
  \DOIprefix\doi{10.1109/ICME.2003.1221638}.
\bibitem[{Yin and Wang(2014)}]{Yin2014A}
\bibinfo{author}{J.~Yin}, \bibinfo{author}{J.~Wang},
\newblock \bibinfo{title}{A dirichlet multinomial mixture model-based approach
  for short text clustering},
\newblock in: \bibinfo{booktitle}{ACM SIGKDD}, \bibinfo{year}{2014}, pp.
  \bibinfo{pages}{233--242}.
\bibitem[{Zhang et~al.(2016)Zhang, Zhai, Zhang, and Li}]{zhang2016spectral}
\bibinfo{author}{H.~Zhang}, \bibinfo{author}{H.~Zhai},
  \bibinfo{author}{L.~Zhang}, \bibinfo{author}{P.~Li},
\newblock \bibinfo{title}{Spectral--spatial sparse subspace clustering for
  hyperspectral remote sensing images},
\newblock \bibinfo{journal}{IEEE Transactions on Geoscience and Remote Sensing}
  \bibinfo{volume}{54} (\bibinfo{year}{2016}) \bibinfo{pages}{3672--3684}.

\end{thebibliography}
\end{document}